\input psfig.tex
\catcode `\@=11 

\def\@version{1.3}
\def\@verdate{28.11.1992}


%
%
%
%
%
%

\font\fiverm=cmr5
\font\fivei=cmmi5	\skewchar\fivei='177
\font\fivesy=cmsy5	\skewchar\fivesy='60
\font\fivebf=cmbx5

\font\sevenrm=cmr7
\font\seveni=cmmi7	\skewchar\seveni='177
\font\sevensy=cmsy7	\skewchar\sevensy='60
\font\sevenbf=cmbx7

\font\eightrm=cmr8
\font\eightbf=cmbx8
\font\eightit=cmti8
\font\eighti=cmmi8			\skewchar\eighti='177
\font\eightmib=cmmib10 at 8pt	\skewchar\eightmib='177
\font\eightsy=cmsy8			\skewchar\eightsy='60
\font\eightsyb=cmbsy10 at 8pt	\skewchar\eightsyb='60
\font\eightsl=cmsl8
\font\eighttt=cmtt8			\hyphenchar\eighttt=-1
\font\eightcsc=cmcsc10 at 8pt
\font\eightsf=cmss8

\font\ninerm=cmr9
\font\ninebf=cmbx9
\font\nineit=cmti9
\font\ninei=cmmi9			\skewchar\ninei='177
\font\ninemib=cmmib10 at 9pt	\skewchar\ninemib='177
\font\ninesy=cmsy9			\skewchar\ninesy='60
\font\ninesyb=cmbsy10 at 9pt	\skewchar\ninesyb='60
\font\ninesl=cmsl9
\font\ninett=cmtt9			\hyphenchar\ninett=-1
\font\ninecsc=cmcsc10 at 9pt
\font\ninesf=cmss9

\font\tenrm=cmr10
\font\tenbf=cmbx10
\font\tenit=cmti10
\font\teni=cmmi10		\skewchar\teni='177
\font\tenmib=cmmib10	\skewchar\tenmib='177
\font\tensy=cmsy10		\skewchar\tensy='60
\font\tensyb=cmbsy10	\skewchar\tensyb='60
\font\tenex=cmex10
\font\tensl=cmsl10
\font\tentt=cmtt10		\hyphenchar\tentt=-1
\font\tencsc=cmcsc10
\font\tensf=cmss10

\font\elevenrm=cmr10 scaled \magstephalf
\font\elevenbf=cmbx10 scaled \magstephalf
\font\elevenit=cmti10 scaled \magstephalf
\font\eleveni=cmmi10 scaled \magstephalf	\skewchar\eleveni='177
\font\elevenmib=cmmib10 scaled \magstephalf	\skewchar\elevenmib='177
\font\elevensy=cmsy10 scaled \magstephalf	\skewchar\elevensy='60
\font\elevensyb=cmbsy10 scaled \magstephalf	\skewchar\elevensyb='60
\font\elevensl=cmsl10 scaled \magstephalf
\font\eleventt=cmtt10 scaled \magstephalf	\hyphenchar\eleventt=-1
\font\elevencsc=cmcsc10 scaled \magstephalf
\font\elevensf=cmss10 scaled \magstephalf

\font\fourteenrm=cmr10 scaled \magstep2
\font\fourteenbf=cmbx10 scaled \magstep2
\font\fourteenit=cmti10 scaled \magstep2
\font\fourteeni=cmmi10 scaled \magstep2		\skewchar\fourteeni='177
\font\fourteenmib=cmmib10 scaled \magstep2	\skewchar\fourteenmib='177
\font\fourteensy=cmsy10 scaled \magstep2	\skewchar\fourteensy='60
\font\fourteensyb=cmbsy10 scaled \magstep2	\skewchar\fourteensyb='60
\font\fourteensl=cmsl10 scaled \magstep2
\font\fourteentt=cmtt10 scaled \magstep2	\hyphenchar\fourteentt=-1
\font\fourteencsc=cmcsc10 scaled \magstep2
\font\fourteensf=cmss10 scaled \magstep2

\font\seventeenrm=cmr10 scaled \magstep3
\font\seventeenbf=cmbx10 scaled \magstep3
\font\seventeenit=cmti10 scaled \magstep3
\font\seventeeni=cmmi10 scaled \magstep3	\skewchar\seventeeni='177
\font\seventeenmib=cmmib10 scaled \magstep3	\skewchar\seventeenmib='177
\font\seventeensy=cmsy10 scaled \magstep3	\skewchar\seventeensy='60
\font\seventeensyb=cmbsy10 scaled \magstep3	\skewchar\seventeensyb='60
\font\seventeensl=cmsl10 scaled \magstep3
\font\seventeentt=cmtt10 scaled \magstep3	\hyphenchar\seventeentt=-1
\font\seventeencsc=cmcsc10 scaled \magstep3
\font\seventeensf=cmss10 scaled \magstep3

\def\@typeface{Computer Modern} 

\def\hexnumber@#1{\ifnum#1<10 \number#1\else
 \ifnum#1=10 A\else\ifnum#1=11 B\else\ifnum#1=12 C\else
 \ifnum#1=13 D\else\ifnum#1=14 E\else\ifnum#1=15 F\fi\fi\fi\fi\fi\fi\fi}

\def\mib{\hexnumber@\mibfam}
\def\syb{\hexnumber@\sybfam}

\def\makestrut{%
  \setbox\strutbox=\hbox{%
    \vrule height.7\baselineskip depth.3\baselineskip width 0pt}%
}

\def\bls#1{%
  \normalbaselineskip=#1%
  \normalbaselines%
  \makestrut%
}

%

\newfam\mibfam 
\newfam\sybfam 
\newfam\scfam  
\newfam\sffam  

\def\em{\ifdim\fontdimen1\font>0 \rm\else\it\fi}

\textfont3=\tenex
\scriptfont3=\tenex
\scriptscriptfont3=\tenex

\def\eightpoint{
  \def\rm{\fam0\eightrm}%
  \textfont0=\eightrm \scriptfont0=\sevenrm \scriptscriptfont0=\fiverm%
  \textfont1=\eighti  \scriptfont1=\seveni  \scriptscriptfont1=\fivei%
  \textfont2=\eightsy \scriptfont2=\sevensy \scriptscriptfont2=\fivesy%
  \textfont\itfam=\eightit\def\it{\fam\itfam\eightit}%
  \textfont\bffam=\eightbf%
    \scriptfont\bffam=\sevenbf%
      \scriptscriptfont\bffam=\fivebf%
  \def\bf{\fam\bffam\eightbf}%
  \textfont\slfam=\eightsl\def\sl{\fam\slfam\eightsl}%
  \textfont\ttfam=\eighttt\def\tt{\fam\ttfam\eighttt}%
  \textfont\scfam=\eightcsc\def\sc{\fam\scfam\eightcsc}%
  \textfont\sffam=\eightsf\def\sf{\fam\sffam\eightsf}%
  \textfont\mibfam=\eightmib%
  \textfont\sybfam=\eightsyb%
  \bls{10pt}%
}

\def\ninepoint{
  \def\rm{\fam0\ninerm}%
  \textfont0=\ninerm \scriptfont0=\sevenrm \scriptscriptfont0=\fiverm%
  \textfont1=\ninei  \scriptfont1=\seveni  \scriptscriptfont1=\fivei%
  \textfont2=\ninesy \scriptfont2=\sevensy \scriptscriptfont2=\fivesy%
  \textfont\itfam=\nineit\def\it{\fam\itfam\nineit}%
  \textfont\bffam=\ninebf%
    \scriptfont\bffam=\sevenbf%
      \scriptscriptfont\bffam=\fivebf%
  \def\bf{\fam\bffam\ninebf}%
  \textfont\slfam=\ninesl\def\sl{\fam\slfam\ninesl}%
  \textfont\ttfam=\ninett\def\tt{\fam\ttfam\ninett}%
  \textfont\scfam=\ninecsc\def\sc{\fam\scfam\ninecsc}%
  \textfont\sffam=\ninesf\def\sf{\fam\sffam\ninesf}%
  \textfont\mibfam=\ninemib%
  \textfont\sybfam=\ninesyb%
  \bls{12pt}%
}

\def\tenpoint{
  \def\rm{\fam0\tenrm}%
  \textfont0=\tenrm \scriptfont0=\sevenrm \scriptscriptfont0=\fiverm%
  \textfont1=\teni  \scriptfont1=\seveni  \scriptscriptfont1=\fivei%
  \textfont2=\tensy \scriptfont2=\sevensy \scriptscriptfont2=\fivesy%
  \textfont\itfam=\tenit\def\it{\fam\itfam\tenit}%
  \textfont\bffam=\tenbf%
    \scriptfont\bffam=\sevenbf%
      \scriptscriptfont\bffam=\fivebf%
  \def\bf{\fam\bffam\tenbf}%
  \textfont\slfam=\tensl\def\sl{\fam\slfam\tensl}%
  \textfont\ttfam=\tentt\def\tt{\fam\ttfam\tentt}%
  \textfont\scfam=\tencsc\def\sc{\fam\scfam\tencsc}%
  \textfont\sffam=\tensf\def\sf{\fam\sffam\tensf}%
  \textfont\mibfam=\tenmib%
  \textfont\sybfam=\tensyb%
  \bls{12pt}%
}

\def\elevenpoint{
  \def\rm{\fam0\elevenrm}%
  \textfont0=\elevenrm \scriptfont0=\eightrm \scriptscriptfont0=\fiverm%
  \textfont1=\eleveni  \scriptfont1=\eighti  \scriptscriptfont1=\fivei%
  \textfont2=\elevensy \scriptfont2=\eightsy \scriptscriptfont2=\fivesy%
  \textfont\itfam=\elevenit\def\it{\fam\itfam\elevenit}%
  \textfont\bffam=\elevenbf%
    \scriptfont\bffam=\eightbf%
      \scriptscriptfont\bffam=\fivebf%
  \def\bf{\fam\bffam\elevenbf}%
  \textfont\slfam=\elevensl\def\sl{\fam\slfam\elevensl}%
  \textfont\ttfam=\eleventt\def\tt{\fam\ttfam\eleventt}%
  \textfont\scfam=\elevencsc\def\sc{\fam\scfam\elevencsc}%
  \textfont\sffam=\elevensf\def\sf{\fam\sffam\elevensf}%
  \textfont\mibfam=\elevenmib%
  \textfont\sybfam=\elevensyb%
  \bls{13pt}%
}

\def\fourteenpoint{
  \def\rm{\fam0\fourteenrm}%
  \textfont0\fourteenrm  \scriptfont0\tenrm  \scriptscriptfont0\sevenrm%
  \textfont1\fourteeni   \scriptfont1\teni   \scriptscriptfont1\seveni%
  \textfont2\fourteensy  \scriptfont2\tensy  \scriptscriptfont2\sevensy%
  \textfont\itfam=\fourteenit\def\it{\fam\itfam\fourteenit}%
  \textfont\bffam=\fourteenbf%
    \scriptfont\bffam=\tenbf%
      \scriptscriptfont\bffam=\sevenbf%
  \def\bf{\fam\bffam\fourteenbf}%
  \textfont\slfam=\fourteensl\def\sl{\fam\slfam\fourteensl}%
  \textfont\ttfam=\fourteentt\def\tt{\fam\ttfam\fourteentt}%
  \textfont\scfam=\fourteencsc\def\sc{\fam\scfam\fourteencsc}%
  \textfont\sffam=\fourteensf\def\sf{\fam\sffam\fourteensf}%
  \textfont\mibfam=\fourteenmib%
  \textfont\sybfam=\fourteensyb%
  \bls{17pt}%
}

\def\seventeenpoint{
  \def\rm{\fam0\seventeenrm}%
  \textfont0\seventeenrm  \scriptfont0\elevenrm  \scriptscriptfont0\ninerm%
  \textfont1\seventeeni   \scriptfont1\eleveni   \scriptscriptfont1\ninei%
  \textfont2\seventeensy  \scriptfont2\elevensy  \scriptscriptfont2\ninesy%
  \textfont\itfam=\seventeenit\def\it{\fam\itfam\seventeenit}%
  \textfont\bffam=\seventeenbf%
    \scriptfont\bffam=\elevenbf%
      \scriptscriptfont\bffam=\ninebf%
  \def\bf{\fam\bffam\seventeenbf}%
  \textfont\slfam=\seventeensl\def\sl{\fam\slfam\seventeensl}%
  \textfont\ttfam=\seventeentt\def\tt{\fam\ttfam\seventeentt}%
  \textfont\scfam=\seventeencsc\def\sc{\fam\scfam\seventeencsc}%
  \textfont\sffam=\seventeensf\def\sf{\fam\sffam\seventeensf}%
  \textfont\mibfam=\seventeenmib%
  \textfont\sybfam=\seventeensyb%
  \bls{20pt}%
}

\lineskip=1pt      \normallineskip=\lineskip
\lineskiplimit=0pt \normallineskiplimit=\lineskiplimit




\def\Nulle{0}  
\def\Aue{1}    
\def\Afe{2}    
\def\Sue{4}    
\def\Hae{5}    
\def\Hbe{6}    
\def\Hce{7}    
\def\Hde{8}    
\def\Kwe{9}    
\def\Txe{10}   
\def\Lie{11}   
\def\Bbe{12}   


\newdimen\DimenA
\newbox\BoxA

\newcount\LastMac \LastMac=\Nulle
\newcount\HeaderNumber \HeaderNumber=0
\newcount\DefaultHeader \DefaultHeader=\HeaderNumber
\newskip\Indent

\newskip\half      \half=5.5pt plus 1.5pt minus 2.25pt
\newskip\one       \one=11pt plus 3pt minus 5.5pt
\newskip\onehalf   \onehalf=16.5pt plus 5.5pt minus 8.25pt
\newskip\two       \two=22pt plus 5.5pt minus 11pt

\def\Half{\vskip-\lastskip\vskip\half}
\def\One{\vskip-\lastskip\vskip\one}
\def\OneHalf{\vskip-\lastskip\vskip\onehalf}
\def\Two{\vskip-\lastskip\vskip\two}


\def\rTenPT{10pt plus \Feathering}

\def\TenPT{10pt plus \Feathering} 
\def\ElevenPT{11pt plus \Feathering}

\def\Raggedright{
 \rightskip=0pt plus \hsize
}

\def\Fullout{
\rightskip=0pt
}

\def\Hang#1#2{
 \hangindent=#1
 \hangafter=#2
}

\def\EveryMac{
 \Fullout
 \everypar{}
}



\def\title#1{
 \EveryMac
 \LastMac=\Nulle
 \global\HeaderNumber=0
 \global\DefaultHeader=1
 \vbox to 1pc{\vss}
 \seventeenpoint
 \Raggedright
 \noindent \bf #1
}

\def\author#1{
 \EveryMac
 \ifnum\LastMac=\Afe \OneHalf
  \else \Two
 \fi
 \LastMac=\Aue
 \fourteenpoint
 \Raggedright
 \noindent \rm #1\par
 \vskip 3pt\relax
}

\def\affiliation#1{
 \EveryMac
 \LastMac=\Afe
 \eightpoint\bls{\TenPT}
 \Raggedright
 \noindent \it #1\par
}

\def\abstract{%
 \EveryMac
 \Two
 \LastMac=\Sue
 \everypar{\Hang{11pc}{0}}
 \noindent\ninebf ABSTRACT\par
 \tenpoint\bls{\ElevenPT}
 \Fullout
 \noindent\rm
}

\def\keywords{
 \EveryMac
 \Half
 \LastMac=\Kwe
 \everypar{\Hang{11pc}{0}}
 \tenpoint\bls{\ElevenPT}
 \Fullout
 \noindent\hbox{\bf Key words:\ }
 \rm
}


\def\maketitle{%
  \Two%
  \EndOpening%
  \MakePage%
}


\def\pageoffset#1#2{\hoffset=#1\relax\voffset=#2\relax}


\def\Autonumber{
 \global\AutoNumbertrue  
}

\newif\ifAutoNumber \AutoNumberfalse
\newcount\Sec        
\newcount\SecSec
\newcount\SecSecSec

\Sec=0

\def\:{\let\@sptoken= } \:  
\def\:{\@xifnch} \expandafter\def\: {\futurelet\@tempc\@ifnch}

\def\@ifnextchar#1#2#3{%
  \let\@tempMACe #1%
  \def\@tempMACa{#2}%
  \def\@tempMACb{#3}%
  \futurelet \@tempMACc\@ifnch%
}

\def\@ifnch{%
\ifx \@tempMACc \@sptoken%
  \let\@tempMACd\@xifnch%
\else%
  \ifx \@tempMACc \@tempMACe%
    \let\@tempMACd\@tempMACa%
  \else%
    \let\@tempMACd\@tempMACb%
  \fi%
\fi%
\@tempMACd%
}

\def\@ifstar#1#2{\@ifnextchar *{\def\@tempMACa*{#1}\@tempMACa}{#2}}

\def\section{\@ifstar{\@ssection}{\@section}}

\def\@section#1{
 \EveryMac
 \Two
 \LastMac=\Hae
 \ninepoint\bls{\ElevenPT}
 \bf
 \Raggedright
 \ifAutoNumber
  \advance\Sec by 1
  \noindent\number\Sec\hskip 1pc \uppercase{#1}
  \SecSec=0
 \else
  \noindent \uppercase{#1}
 \fi
 \nobreak
}

\def\@ssection#1{
 \EveryMac
 \ifnum\LastMac=\Hae \Half
  \else \OneHalf
 \fi
 \LastMac=\Hae
 \tenpoint\bls{\ElevenPT}
 \bf
 \Raggedright
 \noindent\uppercase{#1}
}

\def\subsection#1{
 \EveryMac
 \ifnum\LastMac=\Hae \Half
  \else \OneHalf
 \fi
 \LastMac=\Hbe
 \tenpoint\bls{\ElevenPT}
 \bf
 \Raggedright
 \ifAutoNumber
  \advance\SecSec by 1
  \noindent\number\Sec.\number\SecSec
  \hskip 1pc #1
  \SecSecSec=0
 \else
  \noindent #1
 \fi
 \nobreak
}

\def\subsubsection#1{
 \EveryMac
 \ifnum\LastMac=\Hbe \Half
  \else \OneHalf
 \fi
 \LastMac=\Hce
 \ninepoint\bls{\ElevenPT}
 \it
 \Raggedright
 \ifAutoNumber
  \advance\SecSecSec by 1
  \noindent\number\Sec.\number\SecSec.\number\SecSecSec
  \hskip 1pc #1
 \else
  \noindent #1
 \fi
 \nobreak
}

\def\paragraph#1{
 \EveryMac
 \One
 \LastMac=\Hde
 \ninepoint\bls{\ElevenPT}
 \noindent \it #1
 \rm
}


\def\tx{
 \EveryMac
 \ifnum\LastMac=\Lie \Half\fi
 \ifnum\LastMac=\Hae \nobreak\Half\fi
 \ifnum\LastMac=\Hbe \nobreak\Half\fi
 \ifnum\LastMac=\Hce \nobreak\Half\fi
 \ifnum\LastMac=\Lie \else \noindent\fi
 \LastMac=\Txe
 \ninepoint\bls{\ElevenPT}
 \rm
}


\def\item{
 \par
 \EveryMac
 \ifnum\LastMac=\Lie
  \else \Half
 \fi
 \LastMac=\Lie
 \ninepoint\bls{\ElevenPT}
 \rm
}


\def\bibitem{
 \par
 \EveryMac
 \ifnum\LastMac=\Bbe
  \else \Half
 \fi
 \LastMac=\Bbe
 \Hang{1.5em}{1}
 \eightpoint\bls{\TenPT}
 \Raggedright
 \noindent \rm
}


\newtoks\CatchLine

\def\@journal{Mon.\ Not.\ R.\ Astron.\ Soc.\ }  
\def\@pubyear{1993}        
\def\@pagerange{000--000}  
\def\@volume{000}          
\def\@microfiche{}         %

\def\pubyear#1{\gdef\@pubyear{#1}\@makecatchline}
\def\pagerange#1{\gdef\@pagerange{#1}\@makecatchline}
\def\volume#1{\gdef\@volume{#1}\@makecatchline}
\def\microfiche#1{\gdef\@microfiche{and Microfiche\ #1}\@makecatchline}

\def\@makecatchline{%
  \global\CatchLine{%
    {\rm \@journal {\bf \@volume},\ \@pagerange\ (\@pubyear)\ \@microfiche}}%
}

\@makecatchline 

\newtoks\LeftHeader
\def\shortauthor#1{
 \global\LeftHeader{#1}
}

\newtoks\RightHeader
\def\shorttitle#1{
 \global\RightHeader{#1}
}

\def\PageHead{
 \EveryMac
 \ifnum\HeaderNumber=1 \Pagehead
  \else \Catchline
 \fi
}

\def\Catchline{%
 \vbox to 0pt{\vskip-22.5pt
  \hbox to \PageWidth{\vbox to8.5pt{}\noindent
  \eightpoint\the\CatchLine\hfill}\vss}
 \nointerlineskip
}

\def\Pagehead{%
 \ifodd\pageno
   \vbox to 0pt{\vskip-22.5pt
   \hbox to \PageWidth{\vbox to8.5pt{}\elevenpoint\it\noindent
    \hfill\the\RightHeader\hskip1.5em\rm\folio}\vss}
 \else
   \vbox to 0pt{\vskip-22.5pt
   \hbox to \PageWidth{\vbox to8.5pt{}\elevenpoint\rm\noindent
   \folio\hskip1.5em\it\the\LeftHeader\hfill}\vss}
 \fi
 \nointerlineskip
}

\def\PageFoot{} 

\def\authorcomment#1{%
  \gdef\PageFoot{%
    \nointerlineskip%
    \vbox to 22pt{\vfil%
      \hbox to \PageWidth{\elevenpoint\rm\noindent \hfil #1 \hfil}}%
  }%
}

\everydisplay{\displaysetup}

\newif\ifeqno
\newif\ifleqno

\def\displaysetup#1$${%
 \displaytest#1\eqno\eqno\displaytest
}

\def\displaytest#1\eqno#2\eqno#3\displaytest{%
 \if!#3!\ldisplaytest#1\leqno\leqno\ldisplaytest
 \else\eqnotrue\leqnofalse\def\eqn{#2}\def\eq{#1}\fi
 \generaldisplay$$}

\def\ldisplaytest#1\leqno#2\leqno#3\ldisplaytest{%
 \def\eq{#1}%
 \if!#3!\eqnofalse\else\eqnotrue\leqnotrue
  \def\eqn{#2}\fi}

\def\generaldisplay{%
\ifeqno \ifleqno 
   \hbox to \hsize{\noindent
     $\displaystyle\eq$\hfil$\displaystyle\eqn$}
  \else
    \hbox to \hsize{\noindent
     $\displaystyle\eq$\hfil$\displaystyle\eqn$}
  \fi
 \else
 \hbox to \hsize{\vbox{\noindent
  $\displaystyle\eq$\hfil}}
 \fi
}

\def\@notice{%
  \par\Two%
  \bls{12pt}%
  \noindent\tenrm This paper has been produced using the Blackwell
                  Scientific Publications \TeX\ macros.%
}

\outer\def\bye{\@notice\par\vfill\supereject\end}

\everyjob{%
  \Warn{Monthly notices of the RAS journal style (\@typeface)\space
        v\@version,\space \@verdate.}\Warn{}%
}




\newif\if@debug \@debugfalse  

\def\Print#1{\if@debug\immediate\write16{#1}\else \fi}
\def\Warn#1{\immediate\write16{#1}}
\def\wlog#1{}

\newcount\Iteration 

\newif\ifFigureBoxes  
\FigureBoxestrue

\def\Single{0} \def\Double{1}                 
\def\Figure{0} \def\Table{1}                  

\def\InStack{0}  
\def\InZoneA{1}
\def\InZoneB{2}
\def\InZoneC{3}

\newcount\TEMPCOUNT 
\newdimen\TEMPDIMEN 
\newbox\TEMPBOX     
\newbox\VOIDBOX     

\newcount\LengthOfStack 
\newcount\MaxItems      
\newcount\StackPointer
\newcount\Point         
\newcount\NextFigure    
\newcount\NextTable     
\newcount\NextItem      

\newcount\StatusStack   
\newcount\NumStack      
\newcount\TypeStack     
\newcount\SpanStack     
\newcount\BoxStack      

\newcount\ItemSTATUS    
\newcount\ItemNUMBER    
\newcount\ItemTYPE      
\newcount\ItemSPAN      
\newbox\ItemBOX         
\newdimen\ItemSIZE      

\newdimen\PageHeight    
\newdimen\TextLeading   
\newdimen\Feathering    
\newcount\LinesPerPage  
\newdimen\ColumnWidth   
\newdimen\ColumnGap     
\newdimen\PageWidth     
\newdimen\BodgeHeight   
\newcount\Leading       

\newdimen\ZoneBSize  
\newdimen\TextSize   
\newbox\ZoneABOX     
\newbox\ZoneBBOX     
\newbox\ZoneCBOX     

\newif\ifFirstSingleItem
\newif\ifFirstZoneA
\newif\ifMakePageInComplete
\newif\ifMoreFigures \MoreFiguresfalse 
\newif\ifMoreTables  \MoreTablesfalse  

\newif\ifFigInZoneB 
\newif\ifFigInZoneC 
\newif\ifTabInZoneB 
\newif\ifTabInZoneC

\newif\ifZoneAFullPage

\newbox\MidBOX    
\newbox\LeftBOX
\newbox\RightBOX
\newbox\PageBOX   

\newif\ifLeftCOL  
\LeftCOLtrue

\newdimen\ZoneBAdjust

\newcount\ItemFits
\def\Yes{1}
\def\No{2}




\MaxItems=15
\NextFigure=0        
\NextTable=1

\BodgeHeight=6pt
\TextLeading=11pt    
\Leading=11
\Feathering=0pt      
\LinesPerPage=61     
\topskip=\TextLeading
\ColumnWidth=20pc    
\ColumnGap=2pc       

\def\ItemSep{\vskip \TextLeading plus \TextLeading minus 4pt}

\FigureBoxesfalse 

\parskip=0pt
\parindent=18pt
\widowpenalty=0
\clubpenalty=10000
\tolerance=1500
\hbadness=1500
\abovedisplayskip=6pt plus 2pt minus 2pt
\belowdisplayskip=6pt plus 2pt minus 2pt
\abovedisplayshortskip=6pt plus 2pt minus 2pt
\belowdisplayshortskip=6pt plus 2pt minus 2pt

\PageHeight=\TextLeading 
\multiply\PageHeight by \LinesPerPage
\advance\PageHeight by \topskip

\PageWidth=2\ColumnWidth
\advance\PageWidth by \ColumnGap




\newcount\DUMMY \StatusStack=\allocationnumber
\newcount\DUMMY \newcount\DUMMY \newcount\DUMMY 
\newcount\DUMMY \newcount\DUMMY \newcount\DUMMY 
\newcount\DUMMY \newcount\DUMMY \newcount\DUMMY
\newcount\DUMMY \newcount\DUMMY \newcount\DUMMY 
\newcount\DUMMY \newcount\DUMMY \newcount\DUMMY

\newcount\DUMMY \NumStack=\allocationnumber
\newcount\DUMMY \newcount\DUMMY \newcount\DUMMY 
\newcount\DUMMY \newcount\DUMMY \newcount\DUMMY 
\newcount\DUMMY \newcount\DUMMY \newcount\DUMMY 
\newcount\DUMMY \newcount\DUMMY \newcount\DUMMY 
\newcount\DUMMY \newcount\DUMMY \newcount\DUMMY

\newcount\DUMMY \TypeStack=\allocationnumber
\newcount\DUMMY \newcount\DUMMY \newcount\DUMMY 
\newcount\DUMMY \newcount\DUMMY \newcount\DUMMY 
\newcount\DUMMY \newcount\DUMMY \newcount\DUMMY 
\newcount\DUMMY \newcount\DUMMY \newcount\DUMMY 
\newcount\DUMMY \newcount\DUMMY \newcount\DUMMY

\newcount\DUMMY \SpanStack=\allocationnumber
\newcount\DUMMY \newcount\DUMMY \newcount\DUMMY 
\newcount\DUMMY \newcount\DUMMY \newcount\DUMMY 
\newcount\DUMMY \newcount\DUMMY \newcount\DUMMY 
\newcount\DUMMY \newcount\DUMMY \newcount\DUMMY 
\newcount\DUMMY \newcount\DUMMY \newcount\DUMMY

\newbox\DUMMY   \BoxStack=\allocationnumber
\newbox\DUMMY   \newbox\DUMMY \newbox\DUMMY 
\newbox\DUMMY   \newbox\DUMMY \newbox\DUMMY 
\newbox\DUMMY   \newbox\DUMMY \newbox\DUMMY 
\newbox\DUMMY   \newbox\DUMMY \newbox\DUMMY 
\newbox\DUMMY   \newbox\DUMMY \newbox\DUMMY

\def\wlog{\immediate\write-1}


\def\GetItemAll#1{%
 \GetItemSTATUS{#1}
 \GetItemNUMBER{#1}
 \GetItemTYPE{#1}
 \GetItemSPAN{#1}
 \GetItemBOX{#1}
}

\def\GetItemSTATUS#1{%
 \Point=\StatusStack
 \advance\Point by #1
 \global\ItemSTATUS=\count\Point
}

\def\GetItemNUMBER#1{%
 \Point=\NumStack
 \advance\Point by #1
 \global\ItemNUMBER=\count\Point
}

\def\GetItemTYPE#1{%
 \Point=\TypeStack
 \advance\Point by #1
 \global\ItemTYPE=\count\Point
}

\def\GetItemSPAN#1{%
 \Point\SpanStack
 \advance\Point by #1
 \global\ItemSPAN=\count\Point
}

\def\GetItemBOX#1{%
 \Point=\BoxStack
 \advance\Point by #1
 \global\setbox\ItemBOX=\vbox{\copy\Point}
 \global\ItemSIZE=\ht\ItemBOX
 \global\advance\ItemSIZE by \dp\ItemBOX
 \TEMPCOUNT=\ItemSIZE
 \divide\TEMPCOUNT by \Leading
 \divide\TEMPCOUNT by 65536
 \advance\TEMPCOUNT by 1
 \ItemSIZE=\TEMPCOUNT pt
 \global\multiply\ItemSIZE by \Leading
}


\def\JoinStack{%
 \ifnum\LengthOfStack=\MaxItems 
  \Warn{WARNING: Stack is full...some items will be lost!}
 \else
  \Point=\StatusStack
  \advance\Point by \LengthOfStack
  \global\count\Point=\ItemSTATUS
  \Point=\NumStack
  \advance\Point by \LengthOfStack
  \global\count\Point=\ItemNUMBER
  \Point=\TypeStack
  \advance\Point by \LengthOfStack
  \global\count\Point=\ItemTYPE
  \Point\SpanStack
  \advance\Point by \LengthOfStack
  \global\count\Point=\ItemSPAN
  \Point=\BoxStack
  \advance\Point by \LengthOfStack
  \global\setbox\Point=\vbox{\copy\ItemBOX}
  \global\advance\LengthOfStack by 1
  \ifnum\ItemTYPE=\Figure 
   \global\MoreFigurestrue
  \else
   \global\MoreTablestrue
  \fi
 \fi
}


\def\LeaveStack#1{%
 {\Iteration=#1
 \loop
 \ifnum\Iteration<\LengthOfStack
  \advance\Iteration by 1
  \GetItemSTATUS{\Iteration}
   \advance\Point by -1
   \global\count\Point=\ItemSTATUS
  \GetItemNUMBER{\Iteration}
   \advance\Point by -1
   \global\count\Point=\ItemNUMBER
  \GetItemTYPE{\Iteration}
   \advance\Point by -1
   \global\count\Point=\ItemTYPE
  \GetItemSPAN{\Iteration}
   \advance\Point by -1
   \global\count\Point=\ItemSPAN
  \GetItemBOX{\Iteration}
   \advance\Point by -1
   \global\setbox\Point=\vbox{\copy\ItemBOX}
 \repeat}
 \global\advance\LengthOfStack by -1
}


\newif\ifStackNotClean

\def\CleanStack{%
 \StackNotCleantrue
 {\Iteration=0
  \loop
   \ifStackNotClean
    \GetItemSTATUS{\Iteration}
    \ifnum\ItemSTATUS=\InStack
     \advance\Iteration by 1
     \else
      \LeaveStack{\Iteration}
    \fi
   \ifnum\LengthOfStack<\Iteration
    \StackNotCleanfalse
   \fi
 \repeat}
}


\def\FindItem#1#2{%
 \global\StackPointer=-1 
 {\Iteration=0
  \loop
  \ifnum\Iteration<\LengthOfStack
   \GetItemSTATUS{\Iteration}
   \ifnum\ItemSTATUS=\InStack
    \GetItemTYPE{\Iteration}
    \ifnum\ItemTYPE=#1
     \GetItemNUMBER{\Iteration}
     \ifnum\ItemNUMBER=#2
      \global\StackPointer=\Iteration
      \Iteration=\LengthOfStack 
     \fi
    \fi
   \fi
  \advance\Iteration by 1
 \repeat}
}


\def\FindNext{%
 \global\StackPointer=-1 
 {\Iteration=0
  \loop
  \ifnum\Iteration<\LengthOfStack
   \GetItemSTATUS{\Iteration}
   \ifnum\ItemSTATUS=\InStack
    \GetItemTYPE{\Iteration}
   \ifnum\ItemTYPE=\Figure
    \ifMoreFigures
      \global\NextItem=\Figure
      \global\StackPointer=\Iteration
      \Iteration=\LengthOfStack 
    \fi
   \fi
   \ifnum\ItemTYPE=\Table
    \ifMoreTables
      \global\NextItem=\Table
      \global\StackPointer=\Iteration
      \Iteration=\LengthOfStack 
    \fi
   \fi
  \fi
  \advance\Iteration by 1
 \repeat}
}


\def\ChangeStatus#1#2{%
 \Point=\StatusStack
 \advance\Point by #1
 \global\count\Point=#2
}



\def\Zone{\InZoneA}

\ZoneBAdjust=0pt

\def\MakePage{
 \global\ZoneBSize=\PageHeight
 \global\TextSize=\ZoneBSize
 \global\ZoneAFullPagefalse
 \global\topskip=\TextLeading
 \MakePageInCompletetrue
 \MoreFigurestrue
 \MoreTablestrue
 \FigInZoneBfalse
 \FigInZoneCfalse
 \TabInZoneBfalse
 \TabInZoneCfalse
 \global\FirstSingleItemtrue
 \global\FirstZoneAtrue
 \global\setbox\ZoneABOX=\box\VOIDBOX
 \global\setbox\ZoneBBOX=\box\VOIDBOX
 \global\setbox\ZoneCBOX=\box\VOIDBOX
 \loop
  \ifMakePageInComplete
 \FindNext
 \ifnum\StackPointer=-1
  \NextItem=-1
  \MoreFiguresfalse
  \MoreTablesfalse
 \fi
 \ifnum\NextItem=\Figure
   \FindItem{\Figure}{\NextFigure}
   \ifnum\StackPointer=-1 \global\MoreFiguresfalse
   \else
    \GetItemSPAN{\StackPointer}
    \ifnum\ItemSPAN=\Single \def\Zone{\InZoneB}\relax
     \ifFigInZoneC \global\MoreFiguresfalse\fi
    \else
     \def\Zone{\InZoneA}
     \ifFigInZoneB \def\Zone{\InZoneC}\fi
    \fi
   \fi
   \ifMoreFigures\Print{}\FigureItems\fi
 \fi
\ifnum\NextItem=\Table
   \FindItem{\Table}{\NextTable}
   \ifnum\StackPointer=-1 \global\MoreTablesfalse
   \else
    \GetItemSPAN{\StackPointer}
    \ifnum\ItemSPAN=\Single\relax
     \ifTabInZoneC \global\MoreTablesfalse\fi
    \else
     \def\Zone{\InZoneA}
     \ifTabInZoneB \def\Zone{\InZoneC}\fi
    \fi
   \fi
   \ifMoreTables\Print{}\TableItems\fi
 \fi
   \MakePageInCompletefalse 
   \ifMoreFigures\MakePageInCompletetrue\fi
   \ifMoreTables\MakePageInCompletetrue\fi
 \repeat
 \ifZoneAFullPage
  \global\TextSize=0pt
  \global\ZoneBSize=0pt
  \global\vsize=0pt\relax
  \global\topskip=0pt\relax
  \vbox to 0pt{\vss}
  \eject
 \else
 \global\advance\ZoneBSize by -\ZoneBAdjust
 \global\vsize=\ZoneBSize
 \global\hsize=\ColumnWidth
 \global\ZoneBAdjust=0pt
 \ifdim\TextSize<23pt
 \Warn{}
 \Warn{* Making column fall short: TextSize=\the\TextSize *}
 \vskip-\lastskip\eject\fi
 \fi
}

\def\MakeRightCol{
 \global\TextSize=\ZoneBSize
 \MakePageInCompletetrue
 \MoreFigurestrue
 \MoreTablestrue
 \global\FirstSingleItemtrue
 \global\setbox\ZoneBBOX=\box\VOIDBOX
 \def\Zone{\InZoneB}
 \loop
  \ifMakePageInComplete
 \FindNext
 \ifnum\StackPointer=-1
  \NextItem=-1
  \MoreFiguresfalse
  \MoreTablesfalse
 \fi
 \ifnum\NextItem=\Figure
   \FindItem{\Figure}{\NextFigure}
   \ifnum\StackPointer=-1 \MoreFiguresfalse
   \else
    \GetItemSPAN{\StackPointer}
    \ifnum\ItemSPAN=\Double\relax
     \MoreFiguresfalse\fi
   \fi
   \ifMoreFigures\Print{}\FigureItems\fi
 \fi
 \ifnum\NextItem=\Table
   \FindItem{\Table}{\NextTable}
   \ifnum\StackPointer=-1 \MoreTablesfalse
   \else
    \GetItemSPAN{\StackPointer}
    \ifnum\ItemSPAN=\Double\relax
     \MoreTablesfalse\fi
   \fi
   \ifMoreTables\Print{}\TableItems\fi
 \fi
   \MakePageInCompletefalse 
   \ifMoreFigures\MakePageInCompletetrue\fi
   \ifMoreTables\MakePageInCompletetrue\fi
 \repeat
 \ifZoneAFullPage
  \global\TextSize=0pt
  \global\ZoneBSize=0pt
  \global\vsize=0pt\relax
  \global\topskip=0pt\relax
  \vbox to 0pt{\vss}
  \eject
 \else
 \global\vsize=\ZoneBSize
 \global\hsize=\ColumnWidth
 \ifdim\TextSize<23pt
 \Warn{}
 \Warn{* Making column fall short: TextSize=\the\TextSize *}
 \vskip-\lastskip\eject\fi
\fi
}

\def\FigureItems{
 \Print{Considering...}
 \ShowItem{\StackPointer}
 \GetItemBOX{\StackPointer} 
 \GetItemSPAN{\StackPointer}
  \CheckFitInZone 
  \ifnum\ItemFits=\Yes
   \ifnum\ItemSPAN=\Single
     \ChangeStatus{\StackPointer}{\InZoneB} 
     \global\FigInZoneBtrue
     \ifFirstSingleItem
      \hbox{}\vskip-\BodgeHeight
     \global\advance\ItemSIZE by \TextLeading
     \fi
     \unvbox\ItemBOX\ItemSep
     \global\FirstSingleItemfalse
     \global\advance\TextSize by -\ItemSIZE
     \global\advance\TextSize by -\TextLeading
   \else
    \ifFirstZoneA
     \global\advance\ItemSIZE by \TextLeading
     \global\FirstZoneAfalse\fi
    \global\advance\TextSize by -\ItemSIZE
    \global\advance\TextSize by -\TextLeading
    \global\advance\ZoneBSize by -\ItemSIZE
    \global\advance\ZoneBSize by -\TextLeading
    \ifFigInZoneB\relax
     \else
     \ifdim\TextSize<3\TextLeading
     \global\ZoneAFullPagetrue
     \fi
    \fi
    \ChangeStatus{\StackPointer}{\Zone}
    \ifnum\Zone=\InZoneC \global\FigInZoneCtrue\fi
  \fi
   \Print{TextSize=\the\TextSize}
   \Print{ZoneBSize=\the\ZoneBSize}
  \global\advance\NextFigure by 1
   \Print{This figure has been placed.}
  \else
   \Print{No space available for this figure...holding over.}
   \Print{}
   \global\MoreFiguresfalse
  \fi
}

\def\TableItems{
 \Print{Considering...}
 \ShowItem{\StackPointer}
 \GetItemBOX{\StackPointer} 
 \GetItemSPAN{\StackPointer}
  \CheckFitInZone 
  \ifnum\ItemFits=\Yes
   \ifnum\ItemSPAN=\Single
    \ChangeStatus{\StackPointer}{\InZoneB}
     \global\TabInZoneBtrue
     \ifFirstSingleItem
      \hbox{}\vskip-\BodgeHeight
     \global\advance\ItemSIZE by \TextLeading
     \fi
     \unvbox\ItemBOX\ItemSep
     \global\FirstSingleItemfalse
     \global\advance\TextSize by -\ItemSIZE
     \global\advance\TextSize by -\TextLeading
   \else
    \ifFirstZoneA
    \global\advance\ItemSIZE by \TextLeading
    \global\FirstZoneAfalse\fi
    \global\advance\TextSize by -\ItemSIZE
    \global\advance\TextSize by -\TextLeading
    \global\advance\ZoneBSize by -\ItemSIZE
    \global\advance\ZoneBSize by -\TextLeading
    \ifFigInZoneB\relax
     \else
     \ifdim\TextSize<3\TextLeading
     \global\ZoneAFullPagetrue
     \fi
    \fi
    \ChangeStatus{\StackPointer}{\Zone}
    \ifnum\Zone=\InZoneC \global\TabInZoneCtrue\fi
   \fi
  \global\advance\NextTable by 1
   \Print{This table has been placed.}
  \else
  \Print{No space available for this table...holding over.}
   \Print{}
   \global\MoreTablesfalse
  \fi
}


\def\CheckFitInZone{%
{\advance\TextSize by -\ItemSIZE
 \advance\TextSize by -\TextLeading
 \ifFirstSingleItem
  \advance\TextSize by \TextLeading
 \fi
 \ifnum\Zone=\InZoneA\relax
  \else \advance\TextSize by -\ZoneBAdjust
 \fi
 \ifdim\TextSize<3\TextLeading \global\ItemFits=\No
 \else \global\ItemFits=\Yes\fi}
}

\def\BF#1#2{
 \ItemSTATUS=\InStack
 \ItemNUMBER=#1
 \ItemTYPE=\Figure
 \if#2S \ItemSPAN=\Single
  \else \ItemSPAN=\Double
 \fi
 \setbox\ItemBOX=\vbox{}
}

\def\BT#1#2{
 \ItemSTATUS=\InStack
 \ItemNUMBER=#1
 \ItemTYPE=\Table
 \if#2S \ItemSPAN=\Single
  \else \ItemSPAN=\Double
 \fi
 \setbox\ItemBOX=\vbox{}
}

\def\BeginOpening{%
 \hsize=\PageWidth
 \global\setbox\ItemBOX=\vbox\bgroup
}

\let\begintopmatter=\BeginOpening  

\def\EndOpening{%
 \egroup
 \ItemNUMBER=0
 \ItemTYPE=\Figure
 \ItemSPAN=\Double
 \ItemSTATUS=\InStack
 \JoinStack
}


\newbox\tmpbox

\def\FC#1#2#3#4{%
  \ItemSTATUS=\InStack
  \ItemNUMBER=#1
  \ItemTYPE=\Figure
  \if#2S
    \ItemSPAN=\Single \TEMPDIMEN=\ColumnWidth
  \else
    \ItemSPAN=\Double \TEMPDIMEN=\PageWidth
  \fi
  {\hsize=\TEMPDIMEN
   \global\setbox\ItemBOX=\vbox{%
     \ifFigureBoxes
       \B{\TEMPDIMEN}{#3}
     \else
       \vbox to #3{\vfil}%
     \fi%
     \eightpoint\rm\bls{\rTenPT}%
     \vskip 5.5pt plus 6pt%
     \setbox\tmpbox=\vbox{#4\par}%
     \ifdim\ht\tmpbox>10pt 
       \noindent #4\par%
     \else
       \hbox to \hsize{\hfil #4\hfil}%
     \fi%
   }%
  }%
  \JoinStack%
  \Print{Processing source for figure {\the\ItemNUMBER}}%
}

\let\figure=\FC  

\def\TH#1#2#3#4{%
 \ItemSTATUS=\InStack
 \ItemNUMBER=#1
 \ItemTYPE=\Table
 \if#2S \ItemSPAN=\Single \TEMPDIMEN=\ColumnWidth
  \else \ItemSPAN=\Double \TEMPDIMEN=\PageWidth
 \fi
{\hsize=\TEMPDIMEN
\eightpoint\bls{\rTenPT}\rm
\global\setbox\ItemBOX=\vbox{\noindent#3\vskip 5.5pt plus5.5pt\noindent#4}}
 \JoinStack
 \Print{Processing source for table {\the\ItemNUMBER}}
}

\let\table=\TH  

\def\UnloadZoneA{%
\FirstZoneAtrue
 \Iteration=0
  \loop
   \ifnum\Iteration<\LengthOfStack
    \GetItemSTATUS{\Iteration}
    \ifnum\ItemSTATUS=\InZoneA
     \GetItemBOX{\Iteration}
     \ifFirstZoneA \vbox to \BodgeHeight{\vfil}%
     \FirstZoneAfalse\fi
     \unvbox\ItemBOX\ItemSep
     \LeaveStack{\Iteration}
     \else
     \advance\Iteration by 1
   \fi
 \repeat
}

\def\UnloadZoneC{%
\Iteration=0
  \loop
   \ifnum\Iteration<\LengthOfStack
    \GetItemSTATUS{\Iteration}
    \ifnum\ItemSTATUS=\InZoneC
     \GetItemBOX{\Iteration}
     \ItemSep\unvbox\ItemBOX
     \LeaveStack{\Iteration}
     \else
     \advance\Iteration by 1
   \fi
 \repeat
}


\def\ShowItem#1{
  {\GetItemAll{#1}
  \Print{\the#1:
  {TYPE=\ifnum\ItemTYPE=\Figure Figure\else Table\fi}
  {NUMBER=\the\ItemNUMBER}
  {SPAN=\ifnum\ItemSPAN=\Single Single\else Double\fi}
  {SIZE=\the\ItemSIZE}}}
}

\def\ShowStack{%
 \Print{}
 \Print{LengthOfStack = \the\LengthOfStack}
 \ifnum\LengthOfStack=0 \Print{Stack is empty}\fi
 \Iteration=0
 \loop
 \ifnum\Iteration<\LengthOfStack
  \ShowItem{\Iteration}
  \advance\Iteration by 1
 \repeat
}

\def\B#1#2{%
\hbox{\vrule\kern-0.4pt\vbox to #2{%
\hrule width #1\vfill\hrule}\kern-0.4pt\vrule}
}

\def\Ref#1{\begingroup\global\setbox\TEMPBOX=\vbox{\hsize=2in\noindent#1}\endgroup
\ht1=0pt\dp1=0pt\wd1=0pt\vadjust{\vtop to 0pt{\advance
\hsize0.5pc\kern-10pt\moveright\hsize\box\TEMPBOX\vss}}}

\def\MarkRef#1{\leavevmode\thinspace\hbox{\vrule\vtop
{\vbox{\hrule\kern1pt\hbox{\vphantom{\rm/}\thinspace{\rm#1}%
\thinspace}}\kern1pt\hrule}\vrule}\thinspace}%


\output{%
 \ifLeftCOL
  \global\setbox\LeftBOX=\vbox to \ZoneBSize{\box255\unvbox\ZoneBBOX}
  \global\LeftCOLfalse
  \MakeRightCol
 \else
  \setbox\RightBOX=\vbox to \ZoneBSize{\box255\unvbox\ZoneBBOX}
  \setbox\MidBOX=\hbox{\box\LeftBOX\hskip\ColumnGap\box\RightBOX}
  \setbox\PageBOX=\vbox to \PageHeight{%
  \UnloadZoneA\box\MidBOX\UnloadZoneC}
  \shipout\vbox{\PageHead\box\PageBOX\PageFoot}
  \global\advance\pageno by 1
  \global\HeaderNumber=\DefaultHeader
  \global\LeftCOLtrue
  \CleanStack
  \MakePage
 \fi
}


\catcode `\@=12 


\pageoffset{-2pc}{0pc}
\Autonumber
\begintopmatter
\title{HST Observations of X-ray-selected AGN}
\author{David Schade$^1$, B.J.Boyle$^2$, \& Michael Letawsky$^1$}
\affiliation{$^1$ Dominion Astrophysical Observatory, 5071 West Saanich 
Road, Victoria, V8X 4M6, Canada}
\affiliation{$^2$ Anglo-Australian Observatory, PO Box 296, Epping, 
NSW 1710, Australia}

\shortauthor{D.J.Schade, B.J.Boyle, M.Letawsky}
\shorttitle{HST observations of AGN}

\abstract 

We report on the initial results of a comprehensive Hubble Space
Telescope (HST) snapshot imaging survey of 76 low redshift ($z < 0.15$)
X-ray-selected active galactic nuclei (AGN) in the Einstein Extended
Medium Sensitivity Survey. This survey is expected to show no bias
with respect to host galaxy types and so is arguably one of the best
available samples with HST imaging for the study of the host
galaxies. The HST observations in the F814W band are complemented by
deeper ground-based observations in the $B$ and $R$ bands for all AGN. The
absolute magnitudes for AGN in this sample lie in the range $-24
< M_{B{\rm (AB)}} < -18$, bracketing the extrapolated break in the QSO
luminosity function ($M_{B{\rm (AB)}}= -22.3$) at these low redshifts. We
find a weak correlation between the luminosity of the host galaxy and
the central AGN. We find no convincing cases of an AGN with no
detectable host galaxy, although the faintest host galaxies of
moderately luminous AGN do extend as faint as $M_{B{\rm (AB)}} = - 18$. We
find no evidence for strong interaction/merger activity in any of the
AGN in this sample. The median ratio of AGN to host galaxy luminosity
($L_{\rm AGN}/L_{\rm Host}=0.2$) is lower than previously observed, although the
observed scatter is large. Approximately 55 per cent of these radio-quiet AGN
have host galaxies that are fit best by a `bulge-only' model (or
alternatively are classified visually as ellipticals/S0 galaxies)
confirming the result by McLure et al.\ that radio-quiet AGN are not
exclusively found in spiral galaxies. A comparison with the Autofib
field galaxy survey shows that the morphological type distribution is
skewed toward earlier types than a field galaxy sample drawn at random
with the same distribution of luminosities. This is consistent with
the observation that the luminosity of the host galaxies is higher by
$0.75 \pm 0.25\,$mag than a matching sample drawn from the Autofib
survey. Given the bias toward early-type galaxies, the AGN host
galaxies are consistent with a luminosity and size distribution
identical to normal galaxies. In every respect these galaxies are
intermediate in their properties between the large, luminous host
galaxies found around high luminosity AGN in the local Universe and
the fainter host galaxies identified around lower luminosity Seyfert
galaxies. These results suggest that, with the exception of a bias
toward early spectral types, host galaxies of AGN are drawn at random
from the overall galaxy population with the nuclear properties
governed (weakly) by spheroid mass.

\keywords X-rays: general -- galaxies: active -- quasars: general
\maketitle

\section{Introduction}\tx

The properties of the host galaxies of active galactic nuclei (AGN)
and QSOs play a fundamental role in our understanding of the AGN
phenomenon.  The size, luminosity and structure of the host galaxy can
provide valuable clues to the origin and fuelling of AGN (e.g. Smith
\& Heckman 1990).  Ground-based optical imaging studies of low
redshift AGN over the past 20 years (see e.g.  Adams 1977, Simkin, Su
and Schwarz 1980, Smith et al.\ 1986, MacKenty 1990, Zitelli et
al. 1993, Kontilainen \& Ward 1994) have been limited by the spatial
resolution attainable from the ground; 1$\,$arcsec $\equiv
2.8$h$_{50}^{-1}\,$kpc at $z=0.1$, and no strong concensus has been
been reached over the general properties of AGN host galaxies from
such studies (see e.g. V\`eron-Cetty and Woltjer 1990).

More recently, near infra-red imaging studies of AGN have yielded a
clearer picture (McLeod \& Reike 1994, Dunlop et al.\ 1993, Taylor et
al. 1996).  Infra-red studies, of course, not only benefit from the
improved seeing in the $H$ and $K$ bands, but also because of the
increased dominance of the red host galaxy against the blue AGN.  Such
studies reveal that powerful AGN inhabit luminous ($L>L^*$) and
massive ($r_{1/2}>10\,$kpc) galaxies.  Furthermore, these studies
find that radio-loud AGN are found exclusively in early-type galaxies.
Taylor et al.\ (1996) also found early-type galaxies acting as hosts
for almost half of the radio-quiet AGN in their sample, challenging
the existing orthodoxy that radio-quiet AGN are predominantly found in
spiral galaxies.

With the excellent imaging performance provided by the COSTAR-corrected
optics, a number of QSO host galaxy studies have recently been carried 
out with the HST (Bachall et al.\ 1997, Boyce et al.\ 1997, 
McLure et al.\ 1999). These studies each contain typically 15--20 QSOs
and confirm many of the earlier 
results obtained in the infra-red. The predominantly bright ($M_B<-23$) QSOs
imaged by the HST appear to lie in bright ($L>L^*$) with large radii  
($r_{1/2}>10\,$kpc).  McClure et al.\ (1999) also confirm their previous
finding that a signficant fraction (90 per cent) of the radio-quiet QSOs imaged 
have elliptical hosts.   

An extremely comprehensive HST imaging study of 256 AGN and starbursts
has also been carried out by Malkan, Gorjian \& Tam (1998). This study
has focussed on much lower redshift ($z < 0.035$) and consequently lower
luminosity AGN. This survey also confirms the tendency for a significant
fraction of broad-line radio-quiet AGN to reside in earlier type galaxies. In contrast to studies of
bright AGN, few of the AGN host galaxies in this study show direct evidence for
interactions or recent merger activity.

Despite careful selection of the object sample, all these studies
have relied heavily on existing heterogeneous compilations of 
QSO catalogues (see e.g. V\`eron and V\`eron-Cetty 1997) on which to base
their initial target list.  In particular they have focussed on
luminous optically-selected or radio-selected QSOs, where strong selection 
effects may favour particular, and possibly non-representative types of QSOs.
For example, the Palomar-Green (PG) survey (Green, Schmidt \& Liebert 1986) 
is the source of many of QSOs used in the studies above, yet it is strongly
biassed toward star-like images in the original photographically-identified 
sample.

Imaging surveys of radio-selected AGN also avoid the problem
associated with optical selection biases, but such objects only comprise
$\sim5\,$per cent of all AGN (Peacock, Miller \&
Mead 1986) and so inferences drawn from such samples are limited to a
small fraction of the AGN population.

With limitations for both optically-selected and radio-selected
AGN samples, the increasing availability of complete, X-ray-selected
samples of AGN offers an alternative method to study of AGN host
galaxies.  Unlike radio samples, X-ray AGN do form a representative
sample of all AGN; there being few, if any, X-ray-quiet AGN (Avni \&
Tanenbaum 1986).  In addition, X-ray flux limited samples with {\it
complete} optical identification suffer from none of the inherent
biases towards dominant nuclei or peculiar morphological types present
in existing optically-selected samples of low redshift AGN.

X-ray-selected samples of AGN have been studied in the past;
Kontilainen \& Ward (1994) carried out ground-based optical and
near-infra-red imaging of 31 AGN in the $2-10\,$keV sample of
Piccinotti et al.\ (1982), and Malkan, Margon \& Chanan (1984) obtained optical
images for 24 AGN selected from the $0.3-3.5\,$keV Einstein Medium
Sensitivity Survey (EMSS, Stocke et al.\ 1991).  The AGN studied by
Kontilainen \& Ward (1994) were heavily weighted towards extremely low
redshifts ($z<0.015$) and thus were of low luminosity
($M_B>-21$).  Conversely, the Malkan et al.\ survey comprised a wide
range of much higher redshift objects ($0.1<z<1.8$), although the
ground-based imaging gave inconclusive results for the nine AGN with
$z>0.4$ and limited results on the properties of the host galaxy
assciated with the lower redshift AGN.

Nevertheless, the EMSS is an extremely powerful sample of AGN
to use.  With near-complete optical identification (94 per cent), it
does not suffer from any strong optical biases. Over 95 per cent of
the sample is radio-quiet, including all of the AGN with $z<0.2$. For
$z<0.15$ the spatial resolution of the HST is ideally suited to the
study of the innermost regions ($<400$h$_{50}^{-1}\,$pc) 
of the host galaxy.

We therefore initiated an imaging campaign with HST to obtain snapshot
F814W observations of approximately 100 AGN in the EMSS with
$0.03<z<0.15$.  The magnitude range spanned by these AGN is
$-24<M_{B{\rm (AB) }}<-18$, straddling the predicted `break' luminosity
($M_{B{\rm (AB) }}\sim-22.3$) in the AGN luminosity function (LF) at these
redshifts (Boyle et al.\ 1988). 

An important aspect to this programme is that we also have
ground-based imaging in the $B$ and $R$ passbands from the 1-m Jacobus
Kapteyn Telescope (JKT) and 40-inch telescope operated by the
Mount Stromlo and Siding Spring Observatories (MSSSO) 
to complement the HST observations.  Although
the ground-based images were only taken in moderate seeing conditions
(1$\,$arcsec -- 3$\,$arcsec) they are complementary to the HST data,
permitting us to model the host galaxy accurately well beyond the
central regions, to surface brightness levels ($B_{\mu} =
26\,$mag$\,$arcsec$^{-2}$) unattainable with the snapshot HST
observations.

In this paper we report on the results obtained from the 76 AGN imaged
in this programme. We describe the HST and ground-based observations in 
section 2. In section 3 we discuss the fitting procedure
used, including the technique of 2-dimensional
profile fitting used to extract information on the AGN host galaxy. We
present our results in section 4, comparing the properties of AGN host
galaxies derived from this study with those obtained from previous
observations. We summarise our conclusions in section 5.

\section{Data}\tx

\subsection{The AGN sample}\tx

The AGN sample used in this imaging study were selected from the EMSS
(Stocke et al.\  1991).  Over
830 X-ray sources were identified in the EMSS, of which 420 were
classified as AGN, i.e., as having broad emission lines.  The EMSS was
selected in the `soft' X-ray band $0.3-3.5\,$keV, with a mean flux
limit of $S(0.3-3.5\,{\rm keV)}\sim 10^{-13}\,$ erg$\,$s$^{-1}\,$cm$^{-2}$.

We selected 80 low redshift ($z<0.15$) AGN for our imaging
study. Our imaging campaign began with observations made at the 1-m
JKT and so our sample was initially
defined to be those low redshift EMSS AGN that were observable from La
Palma. However, the subsequent sucess of our HST snapshot proposal led
us to expand the sample to include a further 13 EMSS QSOs at southern
declinations.  Follow-up ground-based observations for these QSOs was
carried out on the MSSSO 40-inch telescope.  
The ground-based studies and the HST
imaging campaigns were largely carried out in parallel over the period
1993 -- 1998.  The unpredictability of both the weather in the
ground-based observations and the sequence of images obtained in
snapshot mode meant that it was impossible to maintain an exact
correspondance between the AGN imaged in the ground-based and HST
programs.

We obtained a total of 76 snapshot images with the HST. These AGN form
the basis of the sample analysed in this paper. We have some form of
ground-based $B$ or $R$ imaging data for 69 of these AGN; of these 11 have
only B-band imaging and 2 have R-band imaging only. Positions,
redshifts and observational details for all AGN are listed in Table
1. Positions and redshifts were taken from the revised EMSS catalogue
published by Maccacaro et al.\ (1995).

Fig.\ 1 illustrates the region of the AGN absolute magnitude-redshift
plane sampled by this study. In this diagram we have plotted the
catalogued redshift against total (nuclear + host) $M_{B{\rm (AB)}}$ magnitudes
for each AGN in the sample. The magnitudes were derived from the HST
and ground-based images using the fitting procedures described
below. The AGN span a range $-23.6 < M_{B{\rm (AB)}} < -18.5$, with a
median luminosity $M_{B{\rm (AB)}} < -21.5$

All magnitudes given in the present paper are in the AB system. For
the ground-based observations, we adopted the following
transformations from the Landolt system: $B_{\rm AB} = B - 0.17$ and
$R_{\rm AB} = R + 0.05$.  Throughout this paper we use $H_0 = 50{\rm
h}_{50}\,$km$\,$s$^{-1}$Mpc$^{-1}$, $\Omega_{\rm M}=1$,
$\Omega_{\Lambda}=0$.

\figure{1}{S}{0mm}{
\psfig{figure=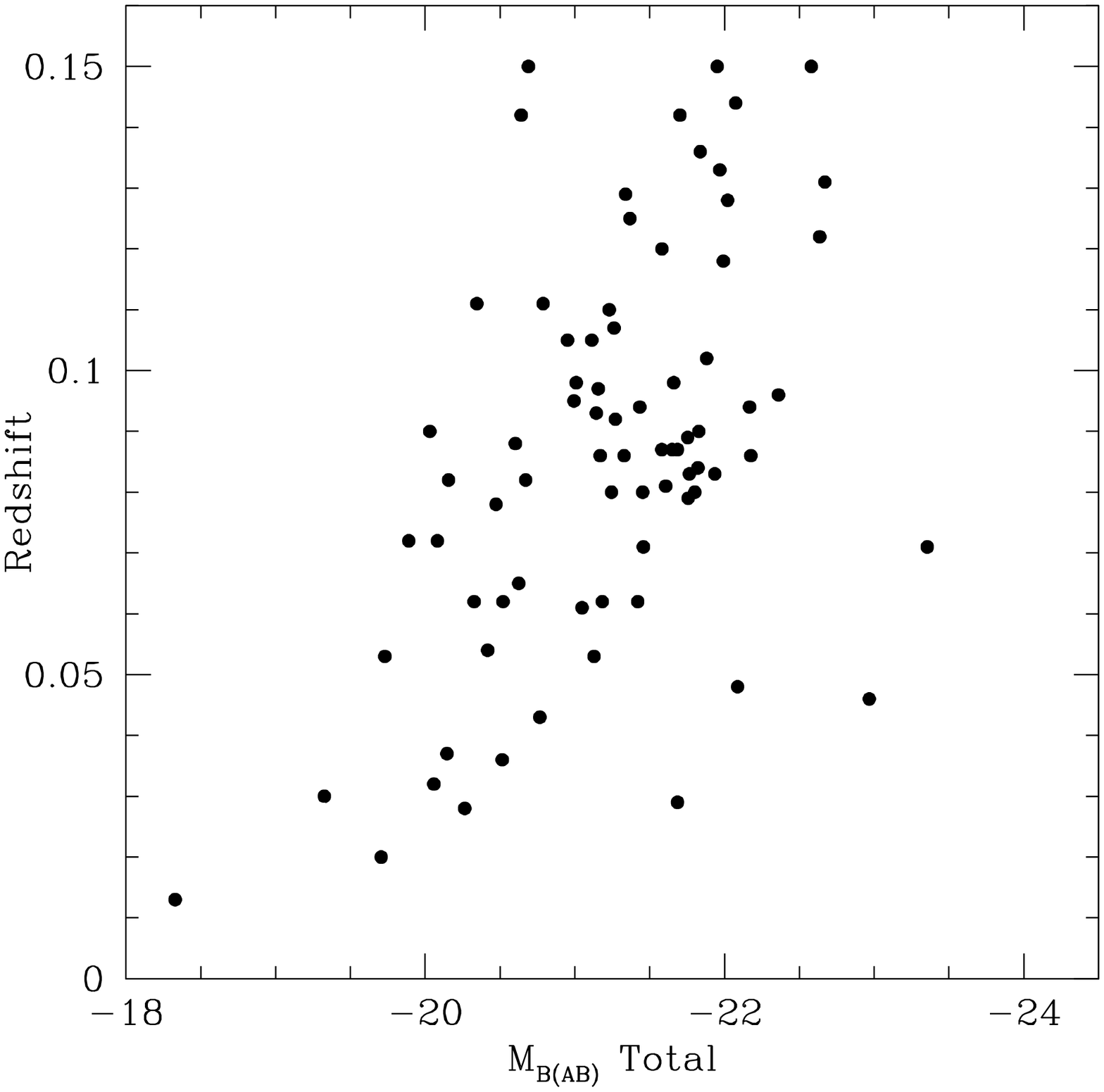,width=3.2in}\break
\noindent\bf Figure 1. \rm Absolute Magnitude {\it v.} redshift for the 
76 AGN in this sample.}

Eight of the EMSS AGN in our data set were classified by Stocke et
al.\ (1991) as uncertain or ambiguous AGN, usually because the
identification spectra lack sufficient signal-to-noise to determine
the presence of broad Balmer emission lines. More detailed
spectroscopy of `ambiguous' EMSS AGN by Boyle et al.\ (1995) has shown
that these sources are a mix of AGN (Seyfert 1.5 -- 2) and star-forming
galaxies. The only ambiguous AGN in this data set that was also
studied by Boyle et al.\ (1995) is MS1334.6+0351 which was classified
by Boyle et al.\ (1995) as a Seyfert 1.5 on the basis of its broad H$\alpha$
emission line. For this analysis, MS1334.6+0351 was therefore
classified as a bona fide AGN, whereas the remainder of these objects
were treated as uncertain AGN.
\subsection{HST observations}\tx

HST WFPC2 observations were obtained for all 76 AGN listed in Table 1
as part of a Cycle 6 snapshot program. The observations used in this
analysis were obtained over the period May 1996 to January 1999; the
date of each snapshot observation is given in Table 1. The
observations were carried out in the F814W ($I$) passband, chosen to
assist in the detection of the redder host galaxy components over the
bluer nucleus. Each snapshot observation lasted for 600$\,$sec,
comprising three separate 200-sec exposures. In each case, the AGN was
imaged at the centre of the Planetary Camera (PC), with a pixel scale
of 0.0455 arcsec$\,$pixel$^{-1}$ thus maximising the resolution attainable. 
For a few of the brightest objects in the sample, a short (10 sec) exposure 
was also taken to obtain an unsaturated image of the nuclear component. 
Fig.\ 2 shows the central 30 arcsec $\times$ 30 arcsec region of the 
reduced HST image for each AGN in the sample.

The data were processed using standard STSDAS pipelines including
flat-fielding and bias-subtraction. The three images of each object
were obtained using an integer-shift dithering pattern. The images
were combined using these offsets and taking the median of the images
which resulted in the removal of warm pixels and cosmic
rays. Photometric zeropoints were obtained using the header
information and converting to the AB system $m_{\rm AB} = -2.5\log f_{\nu} 
+48.6$.  Saturation was dealt with by replacing saturated pixels with
re-scaled unsaturated pixels from the 10-sec integrations where
available. Otherwise saturated pixels were defined as unusable and
were ignored in the fitting process.

\figure{2}{D}{0mm}{
\psfig{figure=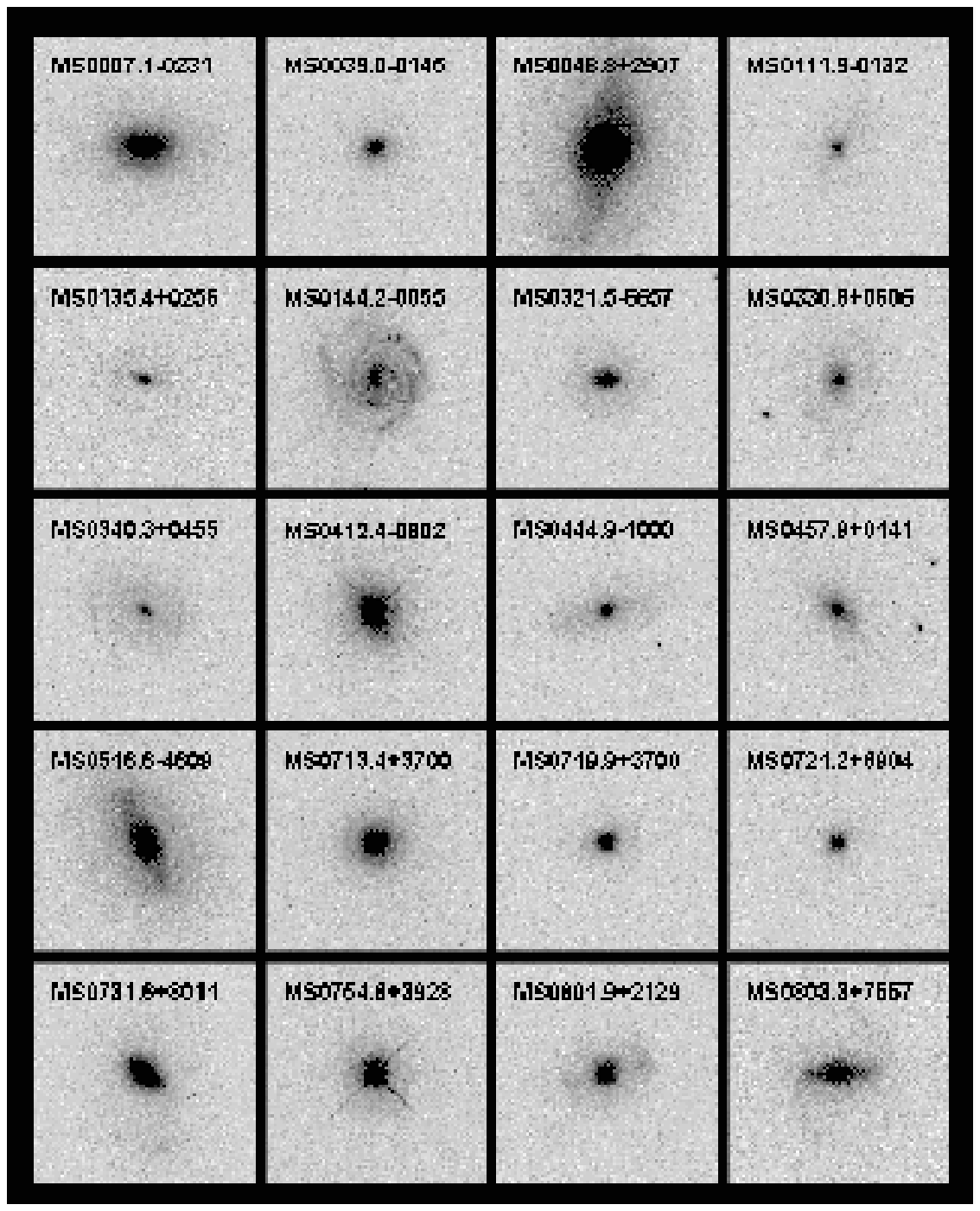,width=6.9in}\break
\noindent\bf Figure 2 \rm Reduced HST snapshot image for each EMSS AGN
in the survey.  Each image is 30 arcsec $\times$ 30 arcsec}
\figure{3}{D}{0mm}{
\psfig{figure=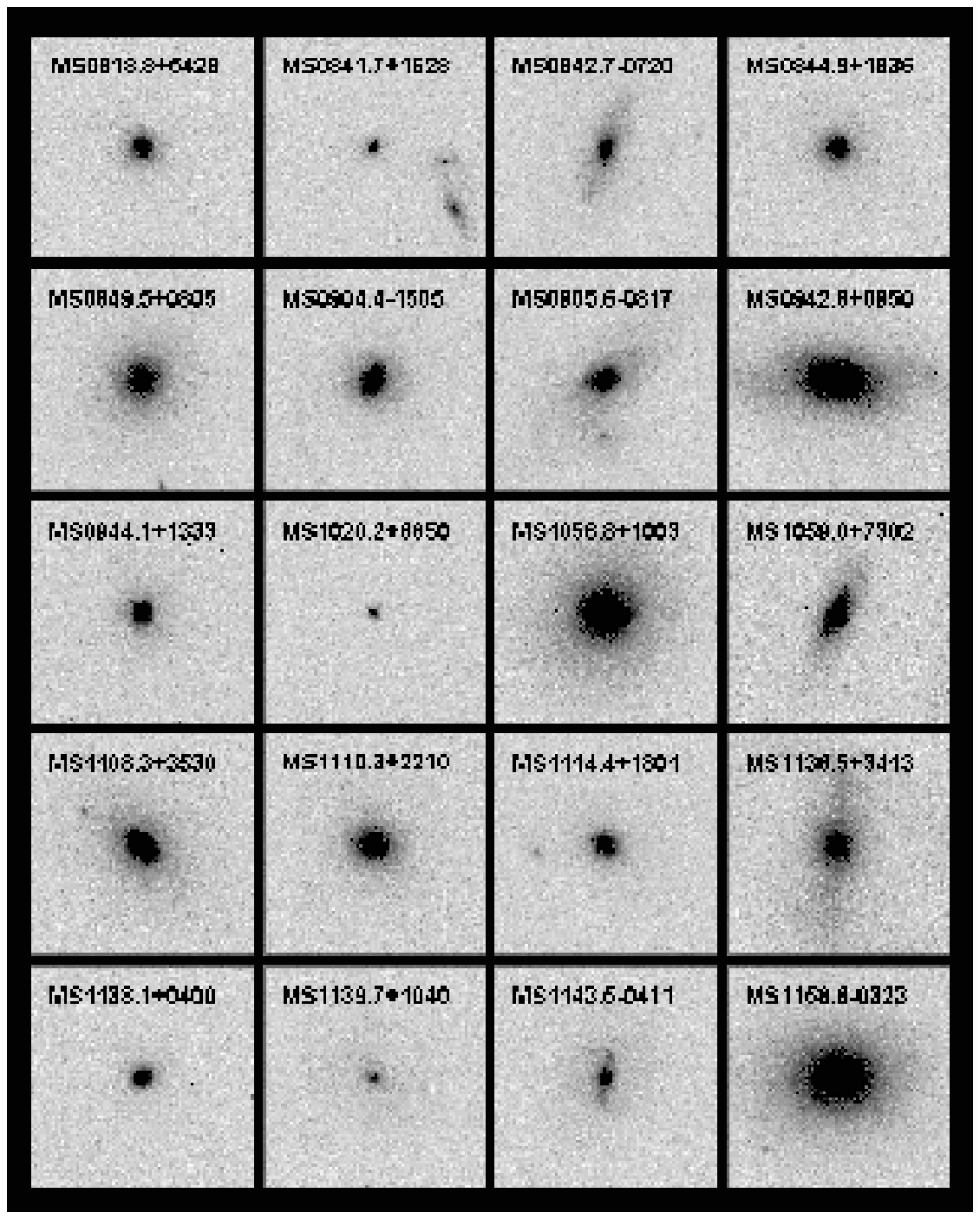,width=6.9in}\break
\noindent\bf Figure 2 \rm contd.}
\figure{4}{D}{0mm}{
\psfig{figure=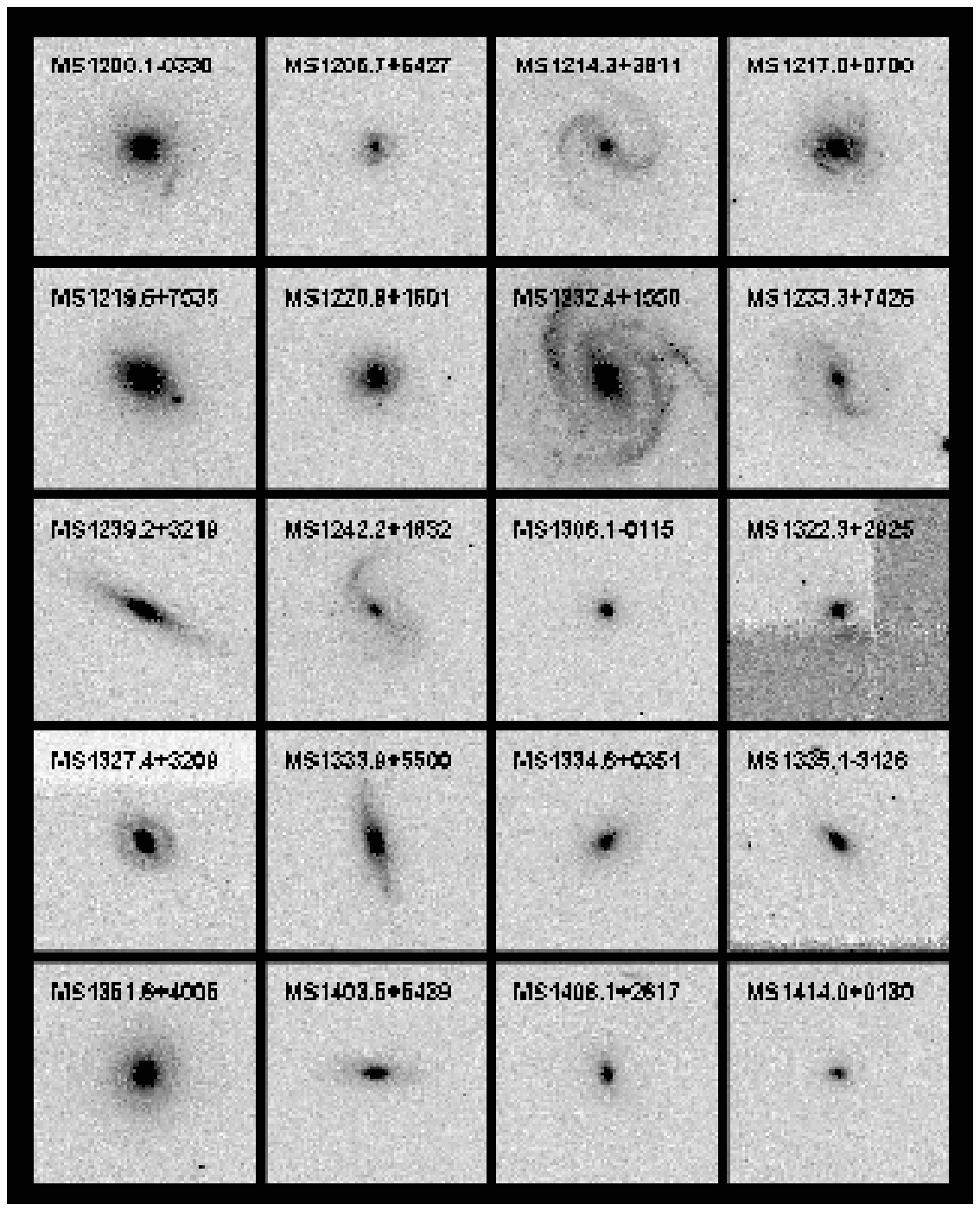,width=6.9in}\break
\noindent\bf Figure 2 \rm contd.}
\figure{5}{D}{0mm}{
\psfig{figure=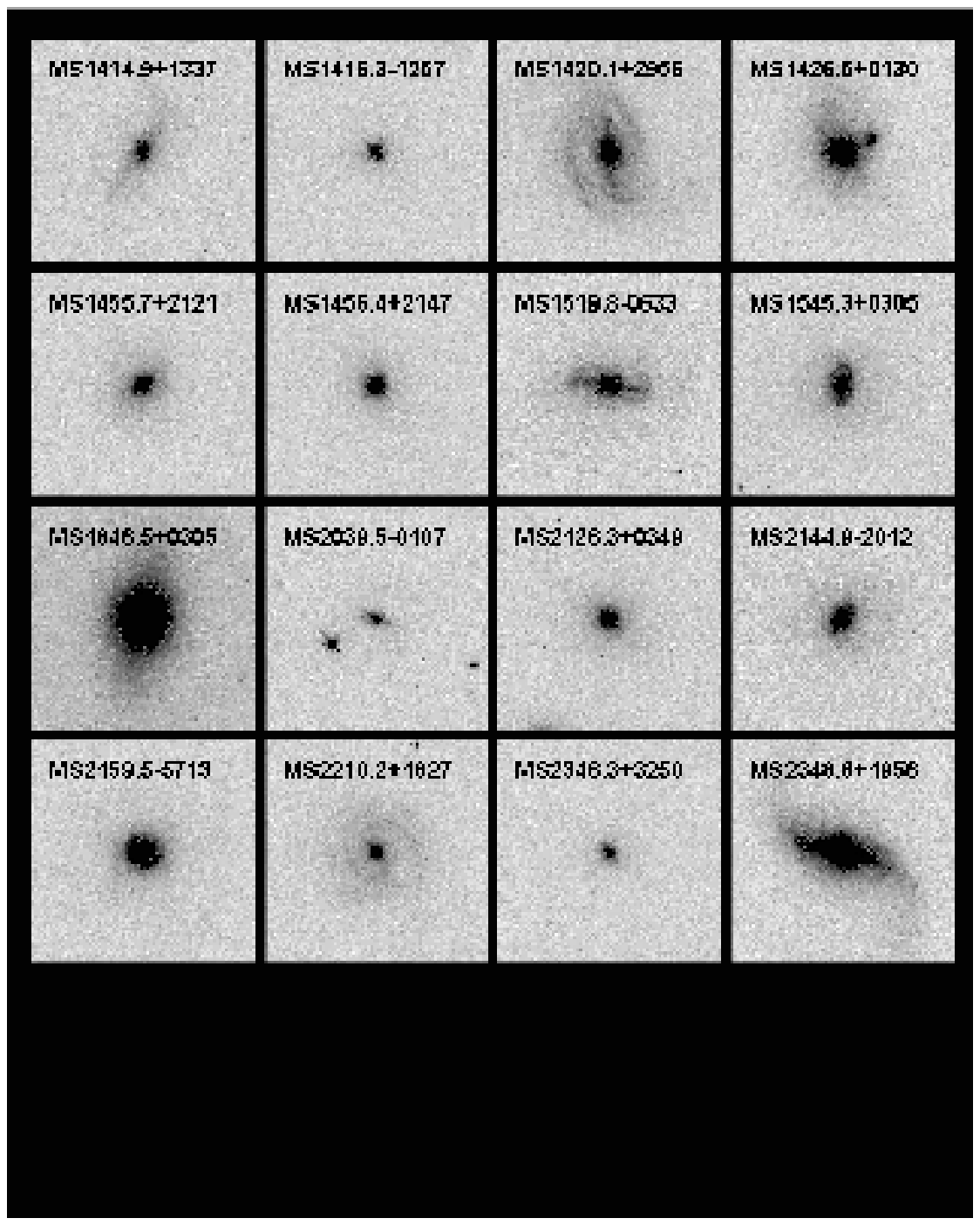,width=6.9in}\break
\noindent\bf Figure 2 \rm contd.}

\subsection{Ground-based imaging}\tx

Ground-based observations in the $B$ and $R$ passbands were carried
out on the 1-m JKT and the 40-inch MSSSO telescope. Harris $B$ and $R$
filter sets were used on both telescopes. The JKT observations were
made over three observing seasons: 1993 January 18-24, 1994 January
4-10 and 1995 January 4-10. The observations in 1993 and 1994 were
made with a $700\times 500$-pixel GEC chip
($0.35\,$arcsec$\,$pixel$^{-1}$) and the 1995 observations were made
with a $1024^2$-pixel Tektronix chip ($0.31\,$arcsec$\,$pixel$^{-1}$).
An equivalent of nine nights' data were obtained over the three runs,
with the 1994 run providing the best conditions. $B$-band observations
were carried out exclusively in 1993 and 1994, with $R$-band
observations made in 1995.

\table{1}{D}{\noindent\bf Table 1.\ \rm AGN observed in the imaging survey}
{\tabskip=1em plus 2em minus .7em 
\halign to\hsize{#\hfil&#\hfil&\hfil#&\hfil#&\hfil#&\hfil#&\hfil#&\hfil#
&\hfil#\hfil&\hfil#\hfil&\hfil#\hfil
&\hfil#\hfil&\hfil#\hfil&\hfil#\hfil&\hfil#\hfil\cr
\noalign{\medskip} 
\noalign{\hrule}
\noalign{\smallskip} 
\hfil Name \hfil &\hfil z\hfil &\multispan6\hfil RA (1950) Dec\hfil&
\multispan3\hfil$B$\hfil&
\multispan3\hfil$R$\hfil\cr 
&&h&m&\hfil s\hfil&$^\circ$&$'$&$''$&
Telescope&Exposure&FWHM&Telescope&Exposure&FWHM&Date of\cr
&&&&&&&&&(secs)&(arcsecs)&&(secs)&(arcsecs)&HST obs\cr
\noalign{\smallskip} \noalign{\hrule}  \noalign{\medskip}
MS0007.1$-$0231&0.087&0&7&5.8&$-$2&31&18&MSSSO&$4800$&2.9&MSSSO&$1200$&4.3&10/1/97\cr
MS0039.0$-$0145$\dag$&0.11&0&39&03.4&$-$1&45&40&---&---&---&---&---&---&20/7/96\cr
MS0048.8$+$2907&0.036&0&48&53.1&29&7&55&JKT&$3600$&2.7&JKT&$900$&2.3&29/9/96\cr
MS0111.9$-$0132&0.120&1&11&54.2&$-$1&32&25&JKT&$4800$&1.3&---&---&---&19/01/99\cr
MS0135.4$+$0256&0.150&1&35&29.2&2&55&36&JKT&$3600$&1.9&JKT&$900$&2.7&14/11/96\cr
MS0144.2$-$0055&0.080&1&44&11.3&$-$0&55&41&---&---&---&---&---&---&20/7/96\cr
MS0321.5$-$6657&0.093&3&21&41.0&$-$66&57&44&MSSSO&$2400$&2.6&---&---&---&26/3/97\cr
MS0330.8$+$0606&0.105&3&30&52.9&6&6&38&JKT&$4800$&1.6&JKT&$900$&1.3&19/7/96\cr
MS0340.3$+$0455$\dag$&0.097&3&40&16.8&4&55&38&---&---&---&---&---&---&16/3/97\cr
MS0412.4$-$0802&0.037&4&12&27.3&$-$8&3&8&JKT&$3600$&2.8&JKT&$1200$&3.0&12/7/96\cr
MS0444.9$-$1000$\dag$&0.095&4&44&53.6&$-$10&0&51&JKT&$3600$&1.2&JKT&$900$&2.8&17/4/97\cr
MS0457.9$+$0141&0.128&4&57&56.8&1&41&49&JKT&$3600$&1.3&JKT&$900$&1.7&13/4/97\cr
MS0516.6$-$4609&0.048&5&16&37.7&$-$46&09&19&&---&---&---&---&---&25/4/97\cr
MS0713.4$+$3700&0.122&7&13&29.5&36&59&58&JKT&$3600$&1.2&JKT&$900$&1.7&18/3/97\cr
MS0719.9$+$7100&0.125&7&19&58.2&71&0&14&JKT&$3600$&1.8&JKT&$1800$&2.3&31/5/96\cr
MS0721.2$+$6904&0.111&7&21&14.5&69&3&49&JKT&$3600$&1.5&JKT&$900$&2.7&5/6/96\cr
MS0731.6$+$8011&0.087&7&31&45.4&80&10&44&---&---&---&---&---&---&13/7/96\cr
MS0754.6$+$3928&0.096&7&54&38.7&39&28&36&JKT&$2050$&1.2&JKT&$900$&1.8&3/4/97\cr
MS0801.9$+$2129&0.118&8&1&56.6&21&29&24&JKT&$8400$&1.5&JKT&$2700$&2.4&21/10/96\cr
MS0803.3$+$7557&0.094&8&3&20.1&75&57&39&JKT&$7200$&1.7&JKT&$900$&2.9&24/3/97\cr
MS0818.8$+$5428&0.086&8&18&46.8&54&28&14&JKT&$3600$&1.8&JKT&$1800$&3.0&21/3/97\cr
MS0841.7$+$1628&0.150&8&41&40.4&16&27&49&JKT&$3600$&1.1&JKT&$1800$&2.1&7/4/97\cr
MS0842.7$-$0720&0.144&8&42&43.1&$-$7&21&8&JKT&$3600$&1.7&JKT&$900$&2.8&17/4/97\cr
MS0844.9$+$1836&0.086&8&44&57.9&18&35&44&JKT&$6000$&1.1&JKT&$900$&2.1&12/4/97\cr
MS0849.5$+$0805&0.063&8&49&34.6&8&4&56&---&---&---&JKT&$900$&2.3&26/4/97\cr
MS0904.4$-$1505&0.054&9&4&26.6&$-$15&5&38&JKT&$2400$&1.9&JKT&$900$&2.7&19/4/97\cr
MS0905.6$-$0817&0.071&9&5&38.4&$-$8&17&39&JKT&$3600$&1.6&JKT&$900$&2.2&12/4/97\cr
MS0942.8$+$0950$\dag$&0.013&9&42&49.3&9&50&0&JKT&$1200$&1.8&JKT&$900$&2.2&19/10/96\cr
MS0944.1$+$1333&0.131&9&44&10.1&13&33&47&JKT&$3600$&0.9&JKT&$900$&2.0&31/3/97\cr
MS1020.2$+$6850&0.078&10&20&19.1&68&50&10&JKT&$2750$&2.1&JKT&$900$&2.8&24/3/97\cr
MS1058.8$+$1003$\dag$&0.028&10&58&49.8&10&3&27&JKT&$3600$&1.7&JKT&$900$&2.4&26/3/97\cr
MS1059.0$+$7302&0.089&10&59&7.7&73&2&47&JKT&$3600$&3.1&JKT&$900$&2.9&14/9/96\cr
MS1108.3$+$3530&0.061&11&8&20.8&35&30&8&JKT&$3600$&1.8&JKT&$900$&2.6&2/4/97\cr
MS1110.3$+$2210$\dag$&0.030&11&10&19.0&22&10&52&JKT&$3600$&2.2&JKT&$900$&2.3&8/4/97\cr
MS1114.4$+$1801$\dag$&0.092&11&14&25.8&18&1&3&JKT&$4800$&1.3&JKT&$900$&1.8&7/4/97\cr
MS1136.5$+$3413&0.032&11&36&36.0&34&12&27&JKT&$3600$&1.8&JKT&$900$&2.0&25/3/97\cr
MS1138.1$+$0400&0.098&11&38&7.3&4&0&44&JKT&$3600$&1.1&JKT&$900$&1.8&3/12/96\cr
MS1139.7$+$1040&0.150&11&39&42.0&10&40&16&JKT&$3600$&1.4&JKT&$900$&1.8&2/2/97\cr
MS1143.5$-$0411&0.133&11&43&30.5&$-$4&11&21&---&---&---&---&---&---&21/12/96\cr
MS1158.6$-$0323&0.020&11&58&40.6&$-$3&23&58&JKT&$3600$&3.4&JKT&$1200$&3.1&30/3/97\cr
MS1200.1$-$0330&0.065&12&0&11.5&$-$3&30&40&JKT&$3600$&2.9&JKT&$900$&1.9&30/12/96\cr
MS1205.7$+$6427&0.105&12&5&45.1&64&27&27&JKT&$3600$&1.5&JKT&$900$&2.3&25/3/97\cr
MS1214.3$+$3811&0.062&12&14&21.8&38&11&16&JKT&$2400$&2.1&JKT&$900$&2.1&21/10/96\cr
MS1217.0$+$0700&0.080&12&16&58.0&7&0&12&JKT&$2400$&1.1&JKT&$900$&1.9&11/4/97\cr
MS1219.6$+$7535&0.070&12&19&33.8&75&35&16&JKT&$3600$&1.9&JKT&$900$&2.3&10/3/97\cr
MS1220.9$+$1601&0.081&12&20&59.0&16&1&43&JKT&$3600$&1.5&JKT&$900$&1.8&12/4/97\cr
MS1232.4$+$1550&0.046&12&32&25.0&15&50&26&JKT&$3600$&1.7&JKT&$900$&2.1&8/4/97\cr
MS1233.3$+$7426&0.084&12&33&18.4&74&26&37&JKT&$4800$&2.1&JKT&$900$&2.2&8/12/98\cr
MS1239.2$+$3219&0.053&12&39&19.5&32&19&21&JKT&$3600$&1.8&JKT&$900$&2.2&28/10/96\cr
MS1242.2$+$1632&0.087&12&42&11.6&16&32&33&JKT&$2400$&1.1&JKT&$900$&2.5&29/3/97\cr
MS1306.1$-$0115&0.111&13&6&11.4&$-$1&14&55&JKT&$4800$&1.0&JKT&$900$&2.3&8/4/97\cr
MS1322.3$+$2925&0.072&13&22&19.8&29&25&51&JKT&$2400$&1.7&---&---&---&16/8/96\cr
MS1327.4$+$3209&0.093&13&27&27.8&32&9&8&JKT&$4800$&1.8&JKT&$900$&2.0&8/5/97\cr
MS1333.9$+$5500&0.107&13&33&56.6&55&0&5&JKT&$3600$&1.3&JKT&$900$&1.7&16/10/96\cr
MS1334.6$+$0351$\ddag$&0.136&13&34&37.8&3&51&11&MSSSO&$2400$&2.5&MSSSO&$2400$&3.2&7/4/97\cr
\noalign{\medskip}
\noalign{\hrule}
\noalign{\smallskip}
\noalign{\dag Uncertain/ambiguous ID from Stocke et al. (1991)}
\noalign{\medskip}
}}

\table{2}{D}{\noindent\bf Table 1 \rm contd.}
{\tabskip=1em plus 2em minus .7em 
\halign to\hsize{#\hfil&#\hfil&\hfil#&\hfil#&\hfil#&\hfil#&\hfil#&\hfil#
&\hfil#\hfil&\hfil#\hfil&\hfil#\hfil
&\hfil#\hfil&\hfil#\hfil&\hfil#\hfil&\hfil#\hfil\cr
\noalign{\medskip} 
\noalign{\hrule}
\noalign{\smallskip} 
\hfil Name\hfil&\hfil z \hfil&\multispan6\hfil RA (1950) Dec\hfil&
\multispan3\hfil$B$\hfil&
\multispan3\hfil$R$\hfil\cr 
&&h&m&\hfil s\hfil&$^\circ$&$'$&$''$&
Telescope&Exposure&FWHM&Telescope&Exposure&FWHM&Date of\cr
&&&&&&&&&(secs)&(arcsecs)&&(secs)&(arcsecs)&HST obs\cr
\noalign{\smallskip} \noalign{\hrule}  \noalign{\medskip}
MS1335.1$-$3128&0.082&13&35&10.3&$-$31&28&42&MSSSO&$2400$&2.8&MSSSO&$1200$&3.5&28/8/96\cr
MS1351.6$+$4005&0.062&13&51&38.8&40&5&44&JKT&$2400$&1.8&---&---&---&4/7/96\cr
MS1403.5$+$5439&0.082&14&3&30.4&54&39&15&JKT&$2700$&1.3&JKT&$900$&2.0&17/3/97\cr
MS1408.1$+$2617&0.072&14&8&8.8&26&17&35&JKT&$4800$&1.1&JKT&$900$&1.8&24/8/96\cr
MS1414.0$+$0130&0.142&14&14&07.1&1&30&18&MSSSO&$2400$&3.1&---&---&---&29/3/97\cr
MS1414.9$+$1337&0.088&14&14&58.0&13&37&18&JKT&$3600$&1.5&JKT&$900$&1.9&2/9/96\cr
MS1416.3$-$1257&0.129&14&16&21.3&$-$12&56&58&MSSSO&$2700$&4.3&---&---&---&7/4/97\cr
MS1420.1$+$2956&0.053&14&20&8.7&29&56&30&JKT&$2400$&1.9&JKT&$900$&3.5&16/3/97\cr
MS1426.5$+$0130&0.086&14&26&33.9&1&30&27&---&---&---&---&---&---&29/3/97\cr
MS1455.7$+$2121&0.080&14&55&44.4&21&21&53&JKT&$3600$&1.6&JKT&$900$&2.0&19/3/97\cr
MS1456.4$+$2147&0.062&14&56&27.6&21&48&5&JKT&$3600$&1.9&---&---&---&29/3/97\cr
MS1519.8$-$0633&0.083&15&19&49.0&$-$6&34&1&MSSSO&$4800$&3.7&MSSSO&$1200$&3.5&11/3/97\cr
MS1545.3$+$0305&0.098&15&45&21.2&3&5&1&MSSSO&$2400$&3.1&---&---&---&5/9/96\cr
MS1846.5$-$7857&0.029&18&46&33.4&$-$78&57&33&MSSSO&$1200$&3.7&MSSSO&$1800$&4.4&24/4/97\cr
MS2039.5$-$0107&0.142&20&39&30.8&$-$01&08&06&MSSSO&$2400$&2.5&---&---&---&29/3/97\cr
MS2128.3$+$0349&0.094&21&28&21.8&3&49&18&MSSSO&$2400$&2.8&---&---&---&4/4/97\cr
MS2144.9$-$2019&0.102&21&44&57.3&$-$20&12&08&MSSSO&$9600$&2.7&---&---&---&21/7/96\cr
MS2159.5$-$1050&0.083&21&59&30.3&$-$57&14&09&---&---&---&---&---&---&11/3/97\cr
MS2210.2$+$1827&0.079&22&10&13.6&18&27&34&JKT&$2200$&2.2&JKT&$900$&2.3&30/12/96\cr
MS2348.3$+$3250$\dag$&0.090&23&48&21.4&32&51&08&---&---&---&JKT&$900$&$2.5$&17/5/97\cr
MS2348.6$+$1956&0.045&23&48&41.3&19&57&2&JKT&$3600$&2.2&JKT&$2700$&2.7&27/1/97\cr
\noalign{\medskip}
\noalign{\hrule}
\noalign{\smallskip}
\noalign{\dag  Uncertain/ambiguous ID from Stocke et al. (1991)}
\noalign{\ddag Uncertain/ambiguous ID from Stocke et al. (1991), confirmed
as AGN by Boyle et al. (1995)}
\noalign{\medskip}
}}

Additional imaging for the southern QSOs in our target list was
obtained with the MSSSO 40-inch telescope equipped with a $1024^2$
Tektronix CCD ($0.25\,$arcsec$\,$pixel$^{-1}$) on the nights of 1997
July 28-31. Only 1.5 nights' data in poor seeing ($>2\,$arcsec) were
obtained.

Table 1 gives the total integration time and median seeing of the
images for each AGN, together with the telescope used to obtained the
ground-based images. Observations of each AGN were split into a number
of short exposures, typically $1200\,$sec for $B$-band observations
and $900\,$sec for $R$-band observations.

As far as possible during the observing runs, we attempted to match
the prevailing seeing conditions with the redshift of the AGN
currently being observed, thus preserving, as far as possible, a
constant physical size for the resolution. Observations of some AGN
were repeated until a lower FWHM ($< 2\,$arcsec) was achieved. Since the
groundbased images were primarily used to provide information on the
larger angular scales (e.g. bulge and particularly disk) where the HST
images provide less information, good seeing was not considered as
important as image depth. Some images with very poor seeing still
provided useful constraints at large angular scales on the fitting
process, improving the overall 3-component fit.

We reduced the CCD frames using the IRAF package at the Cambridge
{\ninerm STARLINK} node and at the DAO. Standard techniques were used to
bias-correct and flat-field the data. We created flat-fields for each
night by averaging sky-limited data frames from different AGN fields,
after bright stars had been masked out and other deviant points
rejected using a 3oe clipping algorithm.

Based on 3--4 Landolt (1992) standard star sequences we observed each
night, we were able to obtain a zeropoint consistent to $\pm 2$ per
cent on each night on which observations were carried out.

\subsection{HST and ground-based imaging}\tx

A feature of this analysis is the complementary information provided
by the ground- and space-based images. The ground-based observations
provide low-resolution data that is suitable for defining the extended
disk component (even with poor seeing) while the HST observations
provide the information on small spatial scales required to deconvolve
the strongly peaked ($r^{1/4}$) galaxy bulge and point source
contributions. The relative levels of signal-to-noise are such that
the HST data are only moderately effective at characterising the host
galaxy properties (because of their 600-sec integration times) on
large scales. The low-resolution ground-based imaging provides
effective constraints at large radii but has little power to
discriminate between bulge and point-source components.

The choice of filters reinforces the role of each dataset. We chose
the F814W imaging from HST to emphasize the redder bulge component
relative to the bluer nuclear source. By the same token, the bluer $B$
and $R$ filters in the groundbased observations provide important colour
information on the outer regions of the galaxy.

\section{Profile Fitting}\tx

\subsection{Method}\tx

To derive the observed parameters for the different AGN components, we
performed a simultaneous three-component parametric model fit to the
$B$, $R$ and $I$ images for each AGN in the sample. The components
fitted were a point source, an exponential disk, and a de Vaucouleurs
$r^{1/4}$ bulge.  In this procedure the specified model was
transformed into the observational space of each dataset using the
pixel scale, detector orientation, filter, and point spread function
(PSF) appropriate for each image.

For the ground-based images, the PSF was derived from several
(typically five) bright stars in the same image that contained the
AGN. The PSFs were defined and managed using the DAOPHOT package
(Stetson 1987) within IRAF. The core of ground-based observations was
fitted with a gaussian function and the residuals were retained as a
lookup table.

For the HST images, the PSF was fitted using a Lorentzian function
plus a look-up table of residuals. Sampling errors are severe with HST
and simulations showed that these can be substantially reduced in our
fitting procedure by using DAOPHOT to construct the PSF for each
observation centered at the same position (with respect to the pixel
grid) as the observed object before the fitting procedure begins.

Very few of the HST observations had suitable PSF stars on the PC
chip. Therefore, a single PSF constructed from several bright stars in
a star cluster observations was used to fit all of the galaxies. We
were able to check the adopted PSF against seven stellar PSFs observed
in this sample, where a PSF star was present near the AGN. After
processing the stellar PSFs in the same manner as done for the fitting
procedure, the FWHM for the PSFs showed a full range of 0.11 PC pixels
($0.005\,$arcsec). As a result the normalisation of the PSF is not
exact, leading to a photometric errors at 2 -- 3 per cent level when
measured with a 0.05 arcsec aperture.

Details of the fitting procedure are described by Schade et al.\ (1996). 
The bulge component is characterised by:
$$I_B(r_B)=I_B(0){\rm exp} \biggl[ -7.67 \biggl({r_B\over r_e}\biggr)^{0.25} \biggr]$$
and the disk component by:
$$I_D(r_D)=I_D(0) {\rm exp} \biggl({r_D\over h }\biggr)$$

where $I(0)$ is the central surface brightness, $r_e$ is the bulge
effective (or half-light) radius, and $h$ is the disk scale
length. The point source is simply a scaled version of the
PSF and is assumed to be coincident with the galaxy center (we
found no case where this assumption failed).

Given the position of the galaxy center $(x_c,y_c)$, then at a
position $(x,y)$, $dx=x-x_c$ and $dy=y-y_c$, $dx_B=dx*{\rm
cos}(\theta_B) +dy*{\rm sin}(\theta_B)$ $dy_B=\bigl(-dx*{\rm
sin}(\theta_B) +dy*{\rm cos}(\theta_B)\bigr)/ar_B$ and $r_B^2=dx_B^2 +
dy_B^2$ where $\theta_B$ is the position angle of the major axis of
the bulge component and $ar_B$ is the axial ratio (minor/major) of the
bulge. A similar equation holds for the disk component. The position
angles of the two components are allowed to vary independently.

Since the colors of the bulge, disk, and point-source components are
expected to be different, the normalisation of the components in each
passband were also allowed to vary independently in the fit. However,
the structural parameters e.g. orientation, axial ratio and scale
length were held fixed for each component across the different
passbands.

This gives a maximum of 17 free parameters for each fit, the ($x,y$)
position of the image centre, the relative normalisation of the point
source, bulge and disk component in each passband, plus the axial
ratio, orientation and scale length of both the bulge and disk
components.

For each $BRI$ image set, the relative rotation of the detectors used
was determined to better than one degree prior to the fitting
procedure by comparison of images in each passband. The ground-based
frames had rotations near 90 or 180 degrees from each other (i.e. one
of the detector axes was always aligned within a few degrees of north)
whereas the HST rotation varied continuously and was determined from
the position angle of the V3 axis given in the WFPC2 image header.

The fitting was done by minimising $\chi^2$ using a modified
Levenberg-Marquardt algorithm. The fitting was typically done over a
radius of six arcseconds on both the ground-based and HST images with
some variation for individual objects where necessary. The
point-source probability was derived using an F-test comparing the
value of $\chi^2$ for the best-fit model and the model that fit best
without a point source.  

We used a relatively large radius in the fitting process primarily to
ensure the galaxy model went to zero at large radii.  As a result, a
significant number of sky pixels that are effectively perfectly fit by the
model are included in the calculation of the reduced $\chi^2$.  For
some fits, this may bias our estimates of reduced $\chi^2$ value towards 
lower values. 

The fitting procedure is difficult and complex because the general
models have three concentric components (bulge, disk, point source)
which may, for some parameter values, be similar in shape and
size. Thus there is a high degree of correlation between these
parameters. In other words, there may be long, flat-bottomed valleys
in the $\chi^2$ surface where various combinations of bulge, disk and point
source are equally good fits. The correlations will be greatly reduced
when the galaxy components are much larger that the point-spread
function and/or when the galaxy components have axial ratios much
different from unity. In order to ensure that we fit models that are
minimal in the sense that they contain the smallest number of
components consistent with a good fit to the data, we performed fits
of pure disk, pure bulge, bulge-plus-disk, disk-plus-point,
bulge-plus-point, bulge-plus-disk-plus-point, and pure point-source
models. The models and residuals were examined and the minimal model
that was a good fit was accepted.

\figure{6}{S}{0mm}{
\psfig{figure=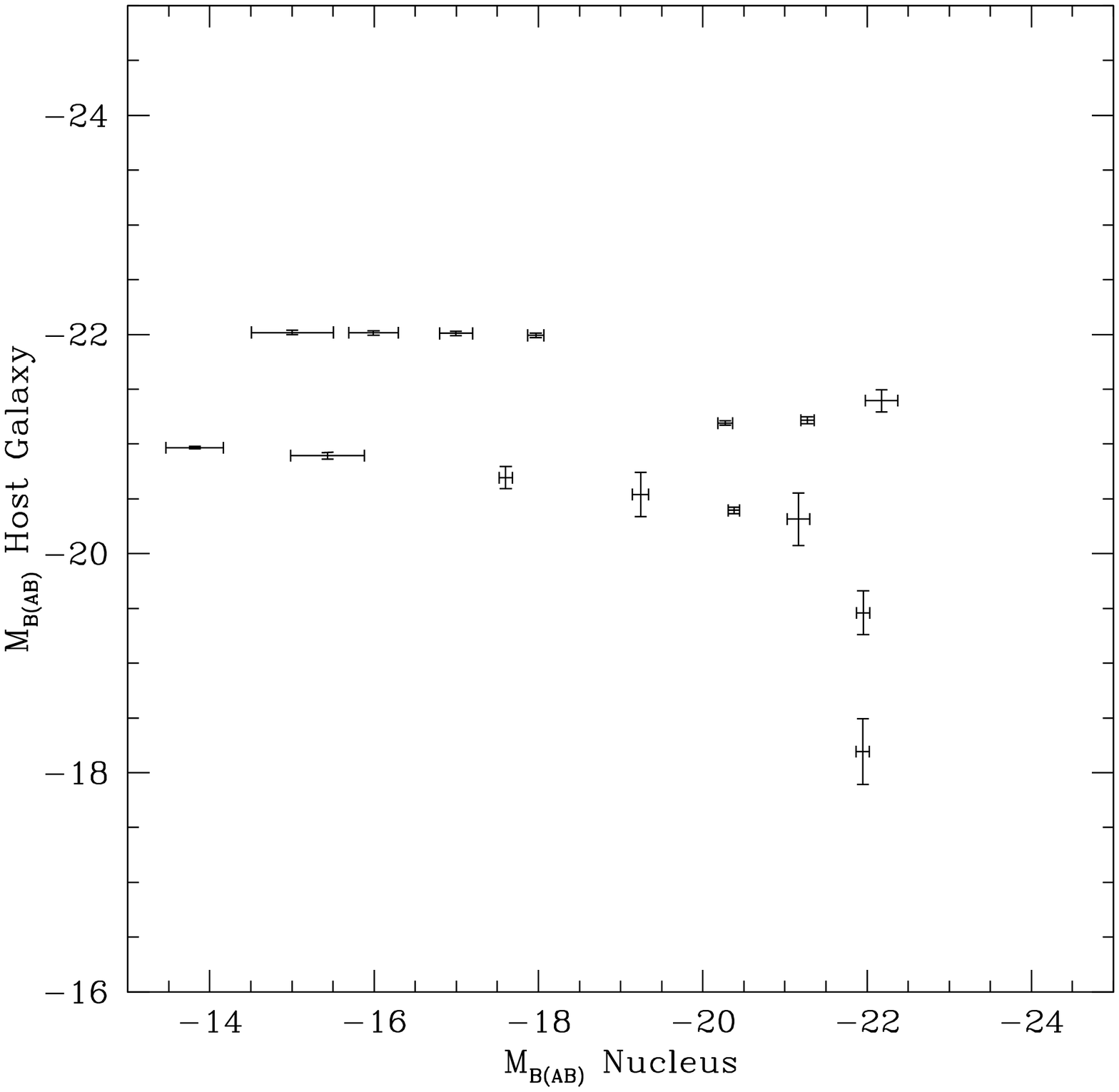,width=3.2in}\break
\noindent\bf Figure 3 \rm The errors in $M_{B{\rm (AB)}}$ for nuclear and 
bulge compoments derived from the fits to simulated galaxies. Errors 
include both statistical
errors of the fitting process and systematic errors due to the entire
processing and fitting procedure (see text). All of the simulations
are restricted to bulge-plus-point source models.}

\figure{7}{S}{0mm}{
\psfig{figure=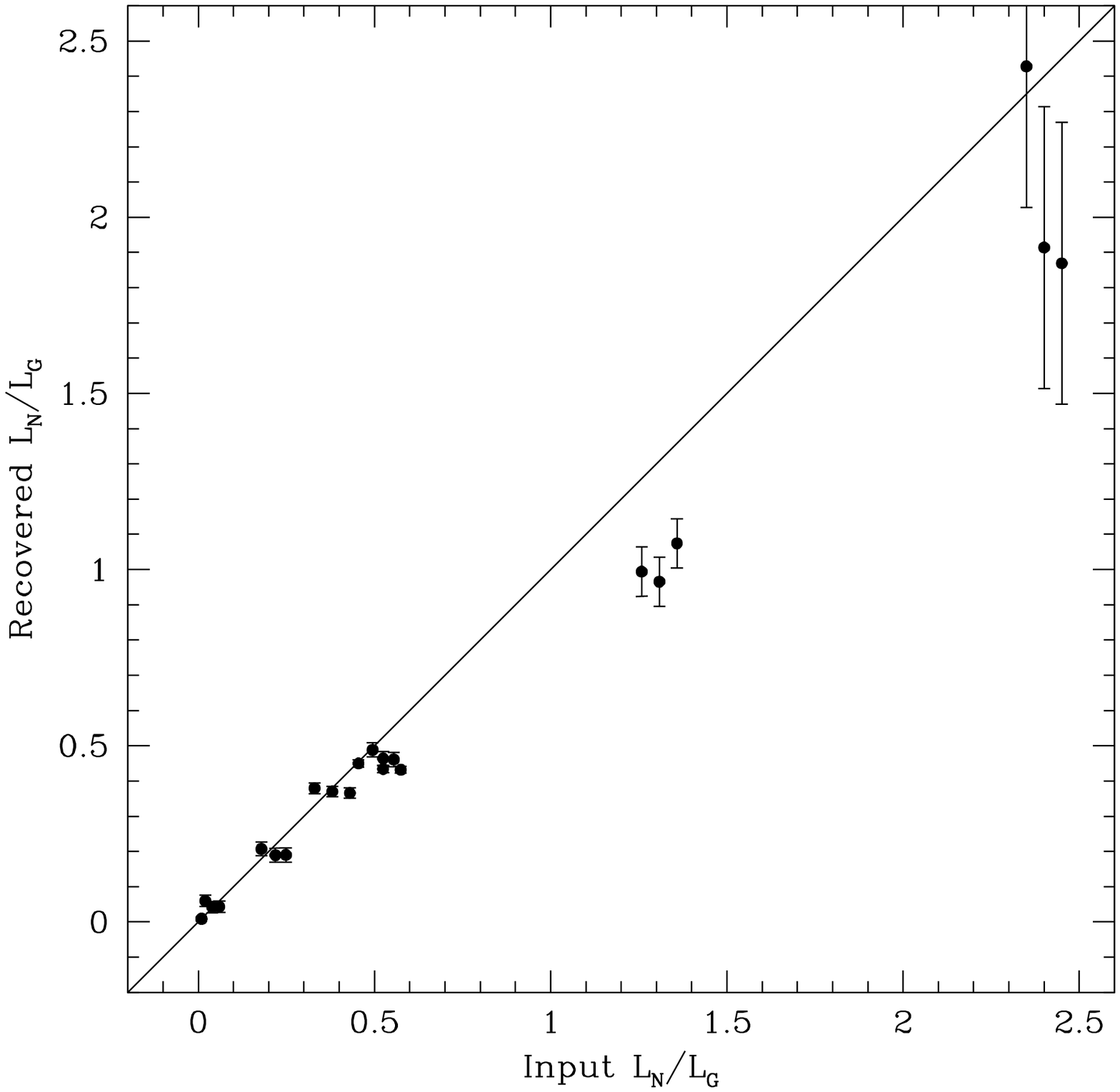,width=3.2in}\break
\noindent\bf Figure 4 \rm Recovered nuclear-to-galaxy I-band luminosity ratios
($L_N/L_G$) compared to input values from the simulations. Errors
represent the spread in the recovered $L_N/L_G$ values for a number of
different simulations with the same input values.}

\subsection{Errors}\tx

The errors on the 17 parameters in each individual fit can be
estimated from the correlation matrix, but such errors are unreliable
since there are strong correlations between the errors in different
parameters (e.g. between the amplitudes of the point source and bulge
components). An estimate of the errors can be made using
simulations. The fitted galaxy parameters can be adopted as a starting
point and the measured parameters can be varied to produce a range of
input models. The results given here provide an indication of the
reliability of the fitting results but are not a complete analysis of
the problem. The present work is focussed on estimating the errors on
the point source versus host galaxy luminosities.

Sets of images were produced with identical object parameters (point
source magnitude, galaxy magnitude and morphology) as the selected
galaxies in the sample.  The objects were simulated in the same (two
or three) bands as the observations and convolved with the appropriate
PSF.  We derived the number of counts from the magnitude and
integration times for the real observations. We also set the sky
levels and noise in the simulated frames to values typical of those
measured in blank regions of the HST and ground-based images.  We did,
however, make the simplifying assumption that the simulated data
frames has been perfectly flat-fielded.  Poisson errors were assumed
throughtout.

After the actual measured parameters were simulated, some of the
parameters were varied to produce a range of simulated object
properties. In total 1400 galaxies with 27 different combinations of
parameters and PSFs were simulated and fit. The simulations were
limited to bulge-plus-point source models because many of the objects
are in that class and also because this is a challenging case in terms
of disentangling the two most compact components. The only shortcut
that was adopted was the use of fitting regions of about three
arcseconds as opposed to the six arcsecond regions use for the real
data. This was done to save computing time but simulations with
varying size of the fitting region shows that this may contribute to
systematic errors in the galaxy properties in some cases. A more
complete analysis of the errors would require a larger fitting radius.

A simulation requires an input point-spread function for each
observation. The PSFs for ground-based observations were always
derived from multiple stars on the same frame as the observation
itself and thus are accurate and reliable. On the other hand, the HST
observations rarely had suitable PSF stars on the image itself and so
a single PSF derived from several bright stars on the PC chip was used
for all of the fits. This PSF was used to construct all of the
simulations but the fitting of the simulations used this same PSF and
also two other PSFs that were constructed from the few AGN snapshot
frames where stars were available. Thus we can estimate the
contribution to the errors that is due to an imperfect knowledge of
the PSF.

Fig.\ 3 shows the results of the simulations and the associated errors
in the $M_{B{\rm (AB)}}$(nucleus) -- $M_{B{\rm(AB)}}$(host) plane. The
errors in the photometry produced by the fitting process are dominated
by systematic errors in the shape and normalization of the PSF, rather
than by statistical errors or sky subtraction. This is because the
signal-to-noise ratio of the observations is high. Typical errors near
the centroid of the distribution of actual objects in this plane are
5-10 per cent in the magnitudes of the nucleus and host galaxy. As
expected, errors in the galaxy magnitude are large where the object is
dominated by the nucleus and vice versa. Near [$M_{B{\rm(AB)}}$(nucleus); 
$M_{B{\rm(AB)}}$(host)]
= [--22,--18] the corresponding errors are [2 per cent, 10 per cent]
whereas near [--15,--22] the errors are [50 per cent, 2 per cent]. Even
when nuclear light dominates it is still possible to detect faint
($M_{B{\rm(AB)}} \sim -18$) host galaxies. Conversely, faint nuclei can be
detected in bright ($M_{B{\rm(AB)}} \sim -22$) host galaxies.

\figure{8}{S}{0mm}{
\psfig{figure=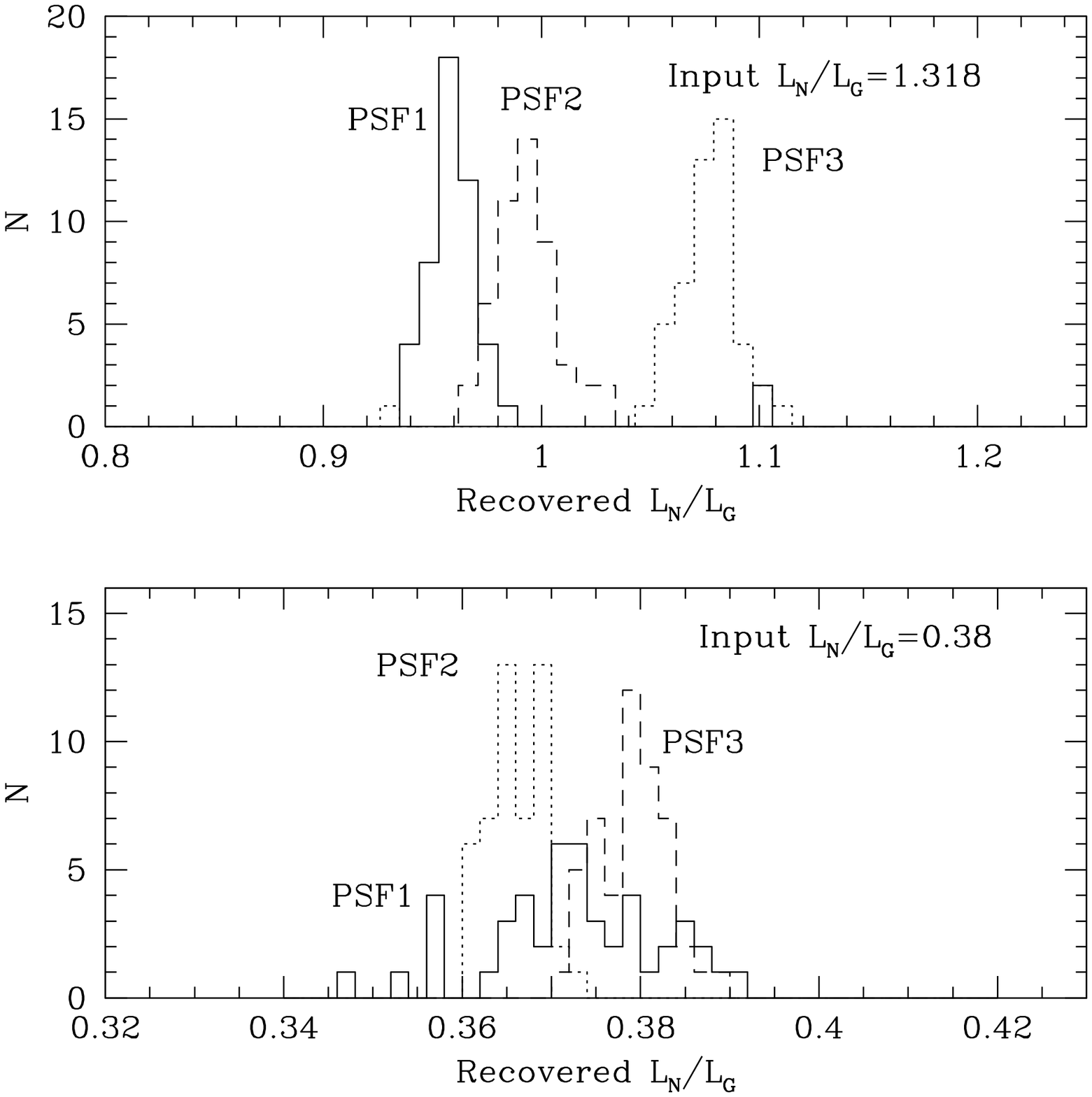,width=3.2in}\break
\noindent\bf Figure 5 \rm Histograms of recovered nuclear-to-host luminosity ratios ($L_N/L_G$) for the three different point-spread-functions used in the
fitting process. Upper panel: Input $L_N/L_G = 1.318$; lower panel:
Input $L_N/L_G = 0.38$.}

Fig.\ 4 shows a comparison of input and recovered values of the ratio
of $I$-band nuclear-to-host galaxy luminosity ($L_N/L_G$). The simulations
indicate that low $L_N/L_G$ values are recovered to within a few per cent
by the fitting method. On the other hand, objects which are dominated
by a strong point source ($L_N/L_G > 1$) may be subject to systematic
errors in the sense that the contribution of the nuclear component may
be underestimated. As demonstrated below, our HST sample contains
relatively few objects with $L_N/L_G > 1$ so that this effect does
significantly affect our results.

Histograms of the recovered values for $L_N/L_G$ for two different
input ratios (0.38 and 1.318) are shown in Fig.\ 5. Separate histograms
are shown for each PSF used in the fitting process. This demonstrates
that the errors due to PSF uncertainty (measured by the shifts between
the individual histograms) are significantly larger than the
statistical errors of the fitting process (measured by the instrinsic
width of individual histograms). For high input values of $L_N/L_G$, the
systematic errors are approximately 25 per cent; for lower values of $L_N/L_G$ 
the systematic errors reduce to 2 -- 3 per cent.

\figure{9}{S}{-35mm}{
\psfig{figure=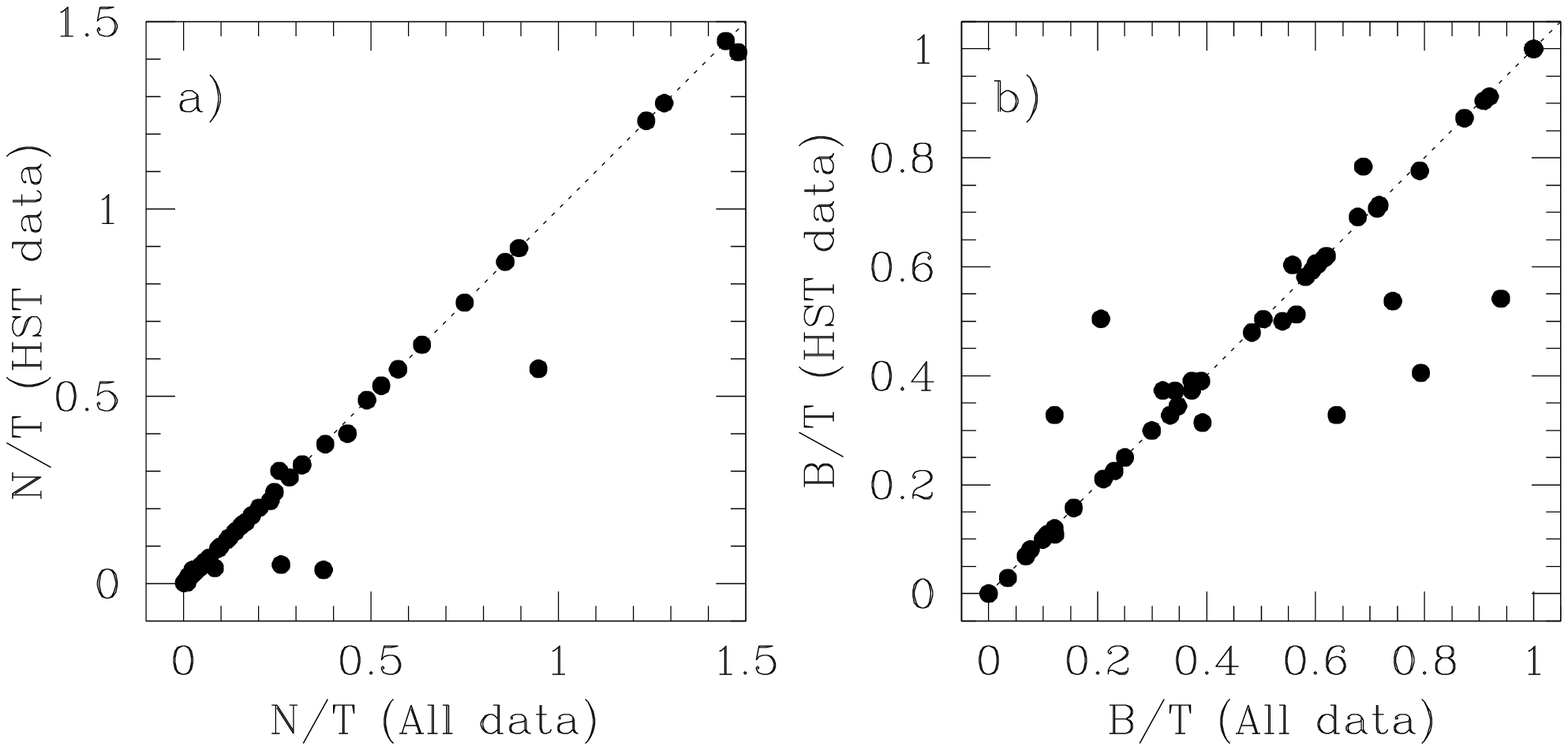,width=3.2in}\break
\noindent\bf Figure 6 \rm a) $I$-band nuclear/host galaxy luminosity
ratios ($N/T$) for fits to the HST images only and fits to both
ground-based and HST images. b) As Fig.\ 6a), for $I$-band
bulge-to-total host galaxy luminosity ratio ($B/T$).}

These simulations indicate that the errors derived from
the correlation matrix are too small by a factor that varies with the
input parameters but is typically a few or more. Those errors are not
reliable for multiple component fits (although they are normally good
for single component fits). The actual errors for data with a very high
signal-to-noise ratio (such as the present case) are dominated by
systematic errors due to uncertainty in the point-spread function used
in the fitting process. Variations in PSF shape and errors in
normalization both contribute to this problem. The actual errors for a
particular galaxy depend on the relative contributions of the galaxy
and nuclear components (see Fig.\ 3).

When combining data of varying image quality, there is also the
concern that the inclusion of lower (ground-based) resolution data may
degrade the fit at the smallest scales, in particular compromising
measurement of the nuclear (point source) and/or bulge component.  For
the 69 AGN in the sample with both HST and ground-based imaging, we
therefore compared the nuclear and bulge $I$-band magnitudes 
derived from fits to the full (HST + ground-based)
data set and to the HST data alone.  The comparison between the
nuclear/total galaxy $I$ band flux ratio ($N/T$) obtained from the
fits to these two different data-sets is shown in Fig.\ 6a).  A
similar comparison for the $I$ band bulge/total galaxy flux ratio
($B/T$) is shown in Fig.\ 6b).

For the vast majority of the sample, the derived $I$-band magnitudes
for both nuclear and bulge components are largely unaffected by
inclusion of the ground-based data in the fit.  A least squares fit to
the relation in Fig.\ 6a) gives a slope of 1.000 with an rms deviation
of 0.067. There are only three cases (MS0721.2+6904, MS1217.0+0700 and
MS1306.1--0115) where $N/T$ varies between the fits by greater than
this value.  Removal of these three points reduces the rms to 0.015,
smaller that systematic errors inherent in the fitting process due to 
the PSF.

The relation between the different $B/T$ estimates shows a larger
scatter, $\sigma(B/T)=0.092$, dominated by six AGN whose $B/T$
values which differ by greater than 0.1 between the fits.  Removal of
these objects from the comparison reduces the observed scatter to
$\sigma(B/T)=0.02$, again below the level of the systematic errors
introduced by the PSF fitting.

However, even the presence of small numbers of objects in our sample
with potentially large uncertainties in their $B/T$ values
($\Delta(B/T) > 0.1$) have little effect on the results presented
below.  In this paper, we use the $B/T$ ratio largely to conduct an
quantitative (albeit crude) morphological classification of the host
galaxy.  Independant visual classification of the host galaxies
confirm that the $B/T$ values derived from the joint HST/ground-based
data-set yield accurate morphological types.  If we were to use the $B/T$
values derived from the HST data alone to carry out the morphological
classification this would only change the type assigned to four AGN
host galaxies -- a net change over the entire sample of 1 less
elliptical, 3 more Sab and 2 less Sbc galaxies.

We conclude that the overall effect of simultaneously fitting to the
full imaging data-set provides useful additional constraints on the
overall parameters of the fits, without systematically biasing the
estimates of nuclear or bulge properties.  In a small fraction of AGN
(5 -- 10 per of the total sample) there are differences between the
derived nuclear/bulge luminosities with/without the inclusion of
ground-based data.  However, these are at level which do not
significantly affect any of our conclusions drawn below. Indeed, there
is no reason to believe that fits to the HST data alone necessarily
produce more accurate estimates of the bulge and/or nuclear
properties.  By neglecting the ground-based data we may be
poorly fitting the low surface brightness disk, biassing the derived
properties for the bulge and/or point source component.

\vfill\eject

\section{Results}\tx

\subsection{Observed Properties of the sample}\tx

The results of the fitting procedure are listed in Table 2. This table
lists 11/17 free parameters in the fit including the fitted $B$, $R$
and $I$ AB magnitudes for the point source, bulge and disk components
for each AGN. The ratio of bulge-to-total light in the $I$ passband
($B/T$) is also given, together with the bulge ($R_e$) and disk ($h$)
radii. We also give the reduced $\chi^2$ for the fit and the
probability ($P_{\rm PS}$) that a point source is not required by the
fit. 

\table{3}{D}
{\noindent\bf Table 2.\ \rm Fitted parameters for AGN/Host Galaxies}
{\tabskip=1em plus 2em minus .7em 
\halign to\hsize{
\hfil#&\hfil#&\hfil#&\hfil#&\hfil#&\hfil#&
\hfil#&\hfil#&\hfil#&\hfil#&\hfil#&\hfil#&\hfil#&\hfil#&\hfil#\cr
\noalign{\medskip} 
\noalign{\hrule}
\noalign{\smallskip} 
&\multispan3\hfil Point\hfil
&\multispan3\hfil Bulge\hfil
&\multispan3\hfil Disk\hfil&&\hfil R$_{\rm e}$\hfil &\hfil h\hfil\cr
\hfil Name \hfil&\hfil $B(AB)$ \hfil & \hfil $R(AB)$ \hfil & \hfil $I(AB)$ \hfil&
\hfil $B(AB)$ \hfil & \hfil $R(AB)$ \hfil & \hfil $I(AB)$ \hfil&
\hfil $B(AB)$ \hfil & \hfil $R(AB)$ \hfil & \hfil $I(AB)$ \hfil&\hfil B/T\hfil &
\hfil ($''$)\hfil &\hfil ($''$) \hfil &\hfil $\chi^2$ \hfil&\hfil $P_{\rm PS}$\cr
\noalign{\smallskip} \noalign{\hrule}  \noalign{\medskip}
MS0007.1$-$0231& 19.00&  --- & 19.01& 19.75&  --- & 16.59& 17.53&  --- & 15.84&  0.33&  0.68&  1.92&  0.74&  0.00\cr
MS0039.0$-$0145&  --- &  --- &  --- & 19.10&  --- & 17.07& 20.60&  --- & 17.63&  0.63&  0.70&  3.13&  0.75&  1.00\cr
MS0048.8$+$2907& 16.78& 17.21& 20.01& 16.79& 14.79& 14.11& 17.89& 16.42& 16.14&  0.87&  3.02&  2.26&  0.58&  0.00\cr
MS0111.9$-$0132& 18.76&  --- & 18.45& 18.30&  --- & 16.52&  --- &  --- &  --- &  1.00&  8.42&  --- &  0.80&  0.00\cr
MS0135.4$+$0256&  --- &  --- &  --- & 18.32& 18.00& 16.81& 18.92& 17.88& 16.98&  0.54&  1.66&  4.35&  0.90&  1.00\cr
MS0144.2$-$0055&  --- &  --- & 18.04&  --- &  --- & 19.13&  --- &  --- & 15.30&  0.03&  0.37&  2.57&  0.62&  0.00\cr
MS0321.5$-$6657&  --- &  --- &  --- & 18.62&  --- & 16.87& 19.29&  --- & 17.15&  0.56&  0.82&  2.64&  0.68&  1.00\cr
MS0330.8$+$0606& 20.10& 19.86& 19.69& 26.70& 18.68& 17.55& 18.38& 17.13& 16.73&  0.32&  1.50&  3.05&  0.76&  0.00\cr
MS0340.3$+$0455& 28.17& 27.08& 20.75& 21.95& 20.04& 19.25& 18.21& 17.15& 16.42&  0.07&  0.45&  2.59&  0.67&  0.56\cr
MS0412.4$-$0802& 15.84& 15.11& 14.89& 16.79& 15.48& 15.16&  --- &  --- &  --- &  1.00&  3.35&  --- &  1.59&  0.00\cr
MS0444.9$-$1000& 20.71& 21.92& 20.40& 19.89& 18.23& 17.82& 18.43& 17.30& 17.10&  0.34&  0.63&  2.75&  0.80&  0.41\cr
MS0457.9$+$0141& 19.47&  --- & 19.39& 18.08&  --- & 16.19&  --- &  --- &  --- &  1.00&  4.94&  --- &  0.78&  0.00\cr
MS0516.6$-$4609&  --- &  --- &  --- &  --- &  --- & 15.00&  --- &  --- & 14.30&  0.34&  5.69& 17.06&  0.36&  1.00\cr
MS0713.4$+$3700& 19.77& 19.59& 20.88& 17.32& 16.17& 15.47&  --- &  --- &  --- &  1.00&  2.57&  --- &  0.52&  0.00\cr
MS0719.9$+$7100& 17.85& 18.20& 17.56& 18.60& 17.29& 16.86&  --- &  --- &  --- &  1.00&  1.44&  --- &  0.83&  0.00\cr
MS0721.2$+$6904& 18.11& 17.26& 17.31& 19.13& 19.19& 17.32& 20.17& 18.28& 20.30&  0.94&  1.49&  1.96&  0.85&  0.00\cr
MS0731.6$+$8011& 27.46& 19.67& 19.07& 17.46& 16.06& 15.63&  --- &  --- &  --- &  1.00&  2.56&  --- &  0.62&  0.00\cr
MS0754.6$+$3928& 15.35& 17.17& 14.22& 16.71& 18.44& 15.56&  --- &  --- &  --- &  1.00&  1.59&  --- &  4.22&  0.00\cr
MS0801.9$+$2129& 16.24& 17.30& 16.46& 18.86& 17.00& 17.60& 18.08& 16.82& 16.68&  0.30&  0.64&  1.79&  1.00&  0.00\cr
MS0803.3$+$7557& 18.32& 17.95& 17.40& 17.28& 15.86& 15.71& 18.64& 18.26& 17.15&  0.79&  6.27&  1.37&  0.60&  0.00\cr
MS0818.8$+$5428& 18.38& 17.52& 18.66& 17.67& 17.67& 16.52& 21.85& 17.86& 18.61&  0.87&  0.65&  1.67&  0.81&  0.00\cr
MS0841.7$+$1628& 20.34& 19.31& 18.53& 19.83& 18.52& 17.93&  --- &  --- &  --- &  1.00&  1.32&  --- &  0.64&  0.00\cr
MS0842.7$-$0720& 17.08& 17.33& 17.26& 20.62& 17.50& 17.22& 18.42& 17.09& 16.75&  0.39&  1.61&  3.01&  0.67&  0.00\cr
MS0844.9$+$1836& 19.61& 18.76& 21.06& 17.97& 16.81& 16.43& 19.11& 17.27& 17.42&  0.71&  1.01&  2.32&  0.50&  0.24\cr
MS0849.5$+$0805& 15.52&  --- & 16.39& 14.91&  --- & 15.10&  --- &  --- &  --- &  1.00&  3.85&  --- &  0.58&  0.00\cr
MS0904.4$-$1505& 18.19& 18.84& 18.71& 17.78& 15.86& 15.73& 19.10& 19.31& 18.22&  0.91&  3.20&  1.21&  0.53&  0.00\cr
MS0905.6$-$0817& 18.59& 17.71& 18.50& 18.10& 17.09& 16.46& 17.62& 16.37& 15.90&  0.37&  1.65&  4.48&  0.41&  0.00\cr
MS0942.8$+$0950& 18.57& 18.34& 18.43& 17.88& 16.50& 16.17& 16.45& 15.46& 14.97&  0.25&  3.12&  2.12&  0.73&  0.00\cr
MS0944.1$+$1333& 15.68& 15.70& 16.06& 17.33& 18.98& 16.68& 18.97& 17.25& 16.70&  0.50&  0.25&  1.50&  1.10&  0.00\cr
MS1020.2$+$6850& 20.77& 17.75& 17.48&  --- &  --- &  --- & 17.89& 18.56& 17.90&  0.00&  --- &  2.66&  0.89&  0.00\cr
MS1058.8$+$1003& 18.96& 18.14& 21.03& 16.00& 15.11& 14.49& 22.78& 15.25& 15.02&  0.62&  2.01&  9.63&  0.46&  0.12\cr
MS1059.0$+$7302& 17.57& 17.22& 17.03& 18.09& 17.41& 15.87& 18.16& 17.11& 17.02&  0.74&  3.47&  2.16&  0.55&  0.00\cr
MS1108.3$+$3530&  --- &  --- &  --- & 17.24& 16.07& 16.03& 18.89& 16.29& 15.95&  0.48&  1.06&  2.52&  0.54&  1.00\cr
MS1110.3$+$2210& 22.12& 19.59& 20.41& 17.41& 15.65& 15.50& 18.70& 18.86& 18.13&  0.92&  2.86&  0.74&  0.52&  0.64\cr
MS1114.4$+$1801&  --- &  --- &  --- & 18.06& 16.96& 16.57& 19.93& 17.88& 17.58&  0.72&  1.03&  2.74&  0.68&  1.00\cr
MS1136.5$+$3413& 17.23& 16.86& 16.82& 16.81& 17.15& 16.16& 17.87& 15.75& 15.59&  0.37&  3.83&  4.18&  0.95&  0.00\cr
MS1138.1$+$0400& 19.77& 19.40& 19.33& 18.18& 17.33& 16.81&  --- &  --- &  --- &  1.00&  0.96&  --- &  0.87&  0.00\cr
MS1139.7$+$1040& 19.00& 19.45& 19.29& 22.42& 21.24& 18.31& 18.46& 17.28& 16.85&  0.21&  1.23&  3.12&  0.74&  0.00\cr
MS1143.5$-$0411&  --- &  --- & 17.71&  --- &  --- & 17.49&  --- &  --- & 16.67&  0.32&  2.63&  2.06&  0.72&  0.00\cr
MS1158.6$-$0323& 20.19& 16.16& 17.72& 16.74& 16.92& 15.98& 16.35& 14.59& 14.67&  0.23&  2.03&  2.51&  0.35&  0.00\cr
MS1200.1$-$0330& 27.70& 19.59& 19.65& 17.70& 16.83& 16.01& 24.69& 16.62& 16.37&  0.58&  1.51&  2.05&  0.69&  0.78\cr
MS1205.7$+$6427& 18.70& 18.79& 18.31& 19.47& 18.72& 17.62& 18.93& 18.41& 18.04&  0.59&  1.98&  0.83&  0.75&  0.00\cr
MS1214.3$+$3811& 20.95& 18.45& 20.47& 18.25& 19.09& 17.92& 17.32& 16.02& 15.63&  0.11&  0.25&  3.49&  0.82&  0.49\cr
MS1217.0$+$0700& 17.79& 17.57& 16.95& 19.85& 17.95& 18.17& 17.35& 16.25& 16.02&  0.12&  1.76&  1.38&  0.65&  0.00\cr
MS1219.6$+$7535& 19.83& 18.42& 15.74& 16.36& 14.96& 15.49& 15.13& 13.72& 16.00&  0.62&  2.30&  3.40&  1.71&  0.00\cr
MS1220.9$+$1601& 21.39& 18.58& 19.95& 17.82& 16.91& 16.40& 17.81& 17.27& 17.02&  0.64&  2.41&  1.19&  0.49&  0.00\cr
MS1232.4$+$1550& 17.00& 16.84& 17.72& 19.89& 16.69& 16.35& 14.55& 12.90& 12.76&  0.04&  0.72& 15.90&  0.58&  0.00\cr
MS1233.3$+$7426&  --- &  --- &  --- & 19.01& 18.43& 17.88& 17.41& 15.93& 15.48&  0.10&  0.25&  3.21&  2.02&  1.00\cr
MS1239.2$+$3219& 18.97& 18.13& 18.34& 20.51& 20.33& 18.03& 18.21& 17.01& 16.60&  0.21&  0.93&  1.96&  0.84&  0.00\cr
MS1242.2$+$1632&  --- &  --- &  --- & 19.48& 18.74& 18.25& 17.48& 16.27& 16.10&  0.12&  0.25&  3.15&  0.62&  1.00\cr
MS1306.1$-$0115& 19.87& 19.75& 18.85& 19.49& 17.66& 17.39&  --- &  --- &  --- &  1.00&  0.95&  --- &  0.83&  0.00\cr
MS1322.3$+$2925& 19.44&  --- & 20.97& 18.31&  --- & 17.12&  --- &  --- &  --- &  1.00&  0.85&  --- &  0.63&  0.33\cr
MS1327.4$+$3209& 18.51& 18.90& 18.00& 17.94& 16.56& 16.38& 17.87& 17.08& 16.91&  0.62&  2.02&  4.85&  1.18&  0.00\cr
MS1333.9$+$5500& 18.13& 18.28& 17.48& 20.76& 17.63& 17.13& 18.53& 17.34& 16.65&  0.39&  2.32&  2.17&  0.76&  0.00\cr
MS1334.6$+$0351& 20.12& 19.44& 19.85& 18.49& 17.11& 16.46&  --- &  --- &  --- &  1.00&  3.17&  --- &  0.72&  0.00\cr
MS1335.1$-$3128& 21.01& 18.88& 19.51& 18.83& 17.81& 16.93&  --- &  --- &  --- &  1.00&  1.63&  --- &  0.95&  0.00\cr
MS1351.6$+$4005& 18.91&  --- & 20.30& 23.40&  --- & 15.56& 17.35&  --- & 17.01&  0.79&  2.97&  1.66&  0.44&  0.66\cr
MS1403.5$+$5439& 21.02& 19.86& 19.90& 18.16& 17.25& 16.64&  --- &  --- &  --- &  1.00&  1.82&  --- &  0.75&  0.00\cr
\noalign{\medskip}
\noalign{\hrule}
\noalign{\medskip}
}}

\table{4}{D}
{\noindent\bf Table 2 contd.\ \rm Fitted parameters for AGN/Host Galaxies}
{\tabskip=1em plus 2em minus .7em 
\halign to\hsize{
\hfil#&\hfil#&\hfil#&\hfil#&\hfil#&\hfil#&
\hfil#&\hfil#&\hfil#&\hfil#&\hfil#&\hfil#&\hfil#&\hfil#&\hfil#\cr
\noalign{\medskip} 
\noalign{\hrule}
\noalign{\smallskip} 
&\multispan3\hfil Point\hfil
&\multispan3\hfil Bulge\hfil
&\multispan3\hfil Disk\hfil&&\hfil R$_{\rm e}$\hfil &\hfil h\hfil\cr
\hfil Name \hfil&\hfil $B(AB)$ \hfil & \hfil $R(AB)$ \hfil & \hfil $I(AB)$ \hfil&
\hfil $B(AB)$ \hfil & \hfil $R(AB)$ \hfil & \hfil $I(AB)$ \hfil&
\hfil $B(AB)$ \hfil & \hfil $R(AB)$ \hfil & \hfil $I(AB)$ \hfil&\hfil B/T\hfil &
\hfil ($''$)\hfil &\hfil ($''$) \hfil &\hfil $\chi^2$ \hfil&\hfil $P_{\rm PS}$\cr
\noalign{\smallskip} \noalign{\hrule}  \noalign{\medskip}
MS1408.1$+$2617& 18.25& 18.16& 18.30& 18.65& 17.49& 17.06&  --- &  --- &  --- &  1.00&  1.68&  --- &  0.83&  0.00\cr
MS1414.0$+$0130& 19.42&  --- & 19.59& 19.83&  --- & 17.74&  --- &  --- &  --- &  1.00&  1.04&  --- &  0.81&  0.01\cr
MS1414.9$+$1337&  --- &  --- &  --- & 19.82& 17.97& 17.49& 19.21& 17.24& 16.80&  0.35&  0.77&  2.75&  0.77&  1.00\cr
MS1416.3$-$1257& 17.46&  --- & 16.56& 19.24&  --- & 17.33& 20.18&  --- & 17.79&  0.60&  0.81&  3.29&  0.89&  0.00\cr
MS1420.1$+$2956& 17.63& 17.24& 17.97& 19.78& 19.70& 17.36& 16.71& 15.57& 15.21&  0.12&  0.39&  3.36&  0.57&  0.00\cr
MS1426.5$+$0130& 18.99&  --- & 14.67& 22.14&  --- & 15.08&  --- &  --- &  --- &  0.54&  3.04&  --- &  1.69&  0.00\cr
MS1455.7$+$2121& 19.08& 18.54& 18.30& 17.55& 16.57& 15.98&  --- &  --- &  --- &  1.00&  2.86&  --- &  0.58&  0.00\cr
MS1456.4$+$2147& 16.88&  --- & 16.65& 17.80&  --- & 16.34&  --- &  --- &  --- &  1.00&  1.93&  --- &  0.57&  0.00\cr
MS1519.8$-$0633& 16.82& 16.64& 16.53& 17.17& 16.13& 16.27& 18.21& 18.04& 16.52&  0.56&  5.89&  2.00&  0.96&  0.00\cr
MS1545.3$+$0305& 18.29&  --- & 17.92& 18.41&  --- & 16.43& 18.64&  --- & 16.88&  0.60&  2.33&  2.04&  0.66&  0.00\cr
MS1846.5$-$7857& 18.77& 16.99& 18.10& 15.88& 14.49& 13.79& 14.98& 13.60& 14.65&  0.69&  6.13& 20.60&  0.77&  0.00\cr
MS2039.5$-$0107& 20.19&  --- & 18.75& 18.34&  --- & 17.22&  --- &  --- &  --- &  1.00&  2.16&  --- &  1.02&  0.00\cr
MS2128.3$+$0349& 16.83&  --- & 16.50& 17.89&  --- & 16.01&  --- &  --- &  --- &  1.00&  3.05&  --- &  1.10&  0.00\cr
MS2144.9$-$2012& 18.94&  --- & 18.84& 18.16&  --- & 16.23& 18.58&  --- & 17.03&  0.68&  1.27&  2.12&  0.67&  0.00\cr
MS2159.5$-$5713&  --- &  --- & 17.67&  --- &  --- & 15.47&  --- &  --- & 18.61&  0.95&  2.38&  2.27&  0.49&  0.00\cr
MS2210.2$+$1827& 20.79& 18.27& 17.48& 17.66& 19.23& 18.68& 17.53& 16.50& 15.98&  0.08&  0.28&  2.27&  0.58&  0.00\cr
MS2348.3$+$3250& 19.41&  --- & 18.85& 20.79&  --- & 17.32&  --- &  --- &  --- &  1.00&  0.97&  --- &  0.93&  0.15\cr
MS2348.6$+$1956&  --- &  --- &  --- & 18.46& 17.53& 16.93& 16.72& 15.52& 15.10&  0.16&  0.23&  2.45&  1.00&  1.00\cr
\noalign{\medskip}
\noalign{\hrule}
\noalign{\medskip}
}}

Ten objects imaged in this survey show little evidence for a point
source component: $P_{\rm PS} = 1$. Of these, only two (MS0039.0--0145
and MS1114.4+1801) are `ambiguous' AGN as identified by Stocke et al.\
(1991). This leaves eight objects, or approximately ten per cent of
the sample, which have been classified as broad emission-line AGN but
have no detectable nuclear component. It is possible that these
objects were incorrectly classified as broad-emission line AGN in the
EMSS, despite the care taken to flag all potentially ambiguous
cases. Although no cases were found of an AGN without a detectable
point source in any of the HST imaging survey of bright ($M_B < -23$) AGN 
(Bachall et al.\ 1997, Boyce et al.\ 1997, McLure et al.\ 1999),
in an HST imaging study of 91 Seyfert 1 galaxies, Malkan et al.\ (1998)
find an even greater percentage ($\sim $35 per cent) of broad emission line
AGN that exhibit no evidence of any point source component. Malkan et
al.\ (1998) ascribe this to dust absorption of the central
source. It could be argued that the amount of dust required
to obscure the central regions under these circumstances would also
extinguish the broad line region, the basis on which these objects
were classified as AGN. Of course, the obscuration may be patchy and
the nucleus may have become obscured since its spectroscopic
classification as a broad lined AGN. Equally, these objects may not
even harbour a compact point source; the broad lines created by
intense star formation in the central regions of the galaxy (see
e.g. Terlevich et al.\ 1992). Whatever the origin, the results from the
current HST surveys appear to indicate a trend for an increasing
fraction of AGN with no point source component with decreasing AGN
luminosity.

At the opposite extreme, we do not find any cases where there is no
evidence for a host galaxy. In the case of MS1020.2+6850, HST imaging
shows evidence for a weak disk only but both $B$ and $R$-band images show
a luminosity profile that is significantly more extended than the PSF.

Good fits ($\chi^2\le2$) were obtained for the vast majority of
the AGN in this analysis.  Although the large fitting radius may
bias estimates of $\chi^2$ towards low values (see above), visual 
examination of all residual (data $-$ model) images confirmed that no
significant systematic effects remained after the model fitting process.
The largest reduced $\chi^2$ residual ($\chi^2=4.2$) was found
for the fit to MS0754.6+3928. Visual inspection of
this object clearly reveals a strong point source component and a low
surface brightness disk.

\figure{10}{S}{0mm}{
\psfig{figure=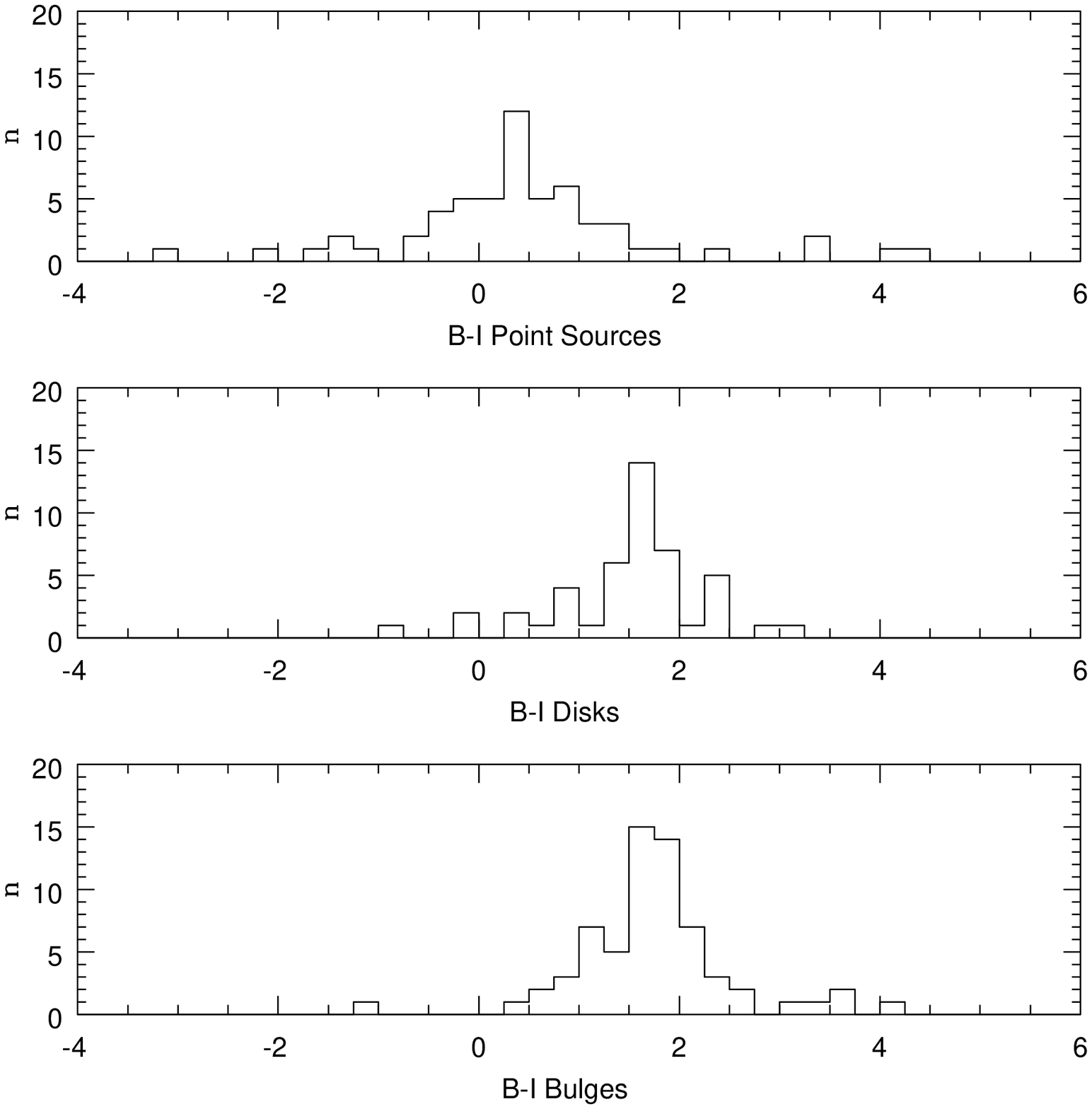,width=3.2in}\break
\noindent\bf Figure 7 \rm $(B-I)_{\rm AB}$ histogram for the 
point source, bulge and disk components.}

Fig.\ 7 shows the observed $(B-I)_{\rm AB}$ colour histograms for the
point source, bulge and disk components. The mean fitted $(B-I)_{\rm AB}$
colour for the point source component is significantly bluer,
$(B-I)_{\rm AB}=0.2$, than that derived for the disk or bulge components,
$(B-I)_{\rm AB}=1.2$. These colours are consistent with previous
observations of QSOs and galaxies, and thus provide a useful
consistency check on the fitting procedure since no a priori
assumptions were fed into the fit relating to the colour of the
components.

Although the $(B-I)_{\rm AB}$ colour distribution for the galaxy
components are reasonably tight ($\sigma = 0.3\,$mag), there is a long
tail to both blue and red in the $(B-I)_{\rm AB}$ colour distribution
for point sources. This is an artifact caused by the fitting procedure.
Where the point source is weak and/or the ground-based $B$ data is
poor, the $B$ fit is poorly constrained, resulting in large
errors. For this reason, we chose not to use $B$ band magnitudes
obtained from the fit to compute derived rest-frame absolute
magnitudes for point sources in the $M_{B{\rm (AB)}}$ band. Instead we
used the mean point source $(B-I)_{\rm AB}$ colour to transform the
$I$-band HST fits to the $B$ passband. For the galaxy components we
used the fitted $B$- or $R$-band magnitudes, unless there were no
ground-based data in the relevant band. In that case, the median
colour for the component was used.

\table{5}{D}
{\noindent\bf Table 3.\ \rm Derived parameters for AGN/Host Galaxies}
{\tabskip=1em plus 2em minus .7em 
\halign to\hsize{
\hfil#\hfil&\hfil#\hfil&\hfil#\hfil&\hfil#\hfil&\hfil#\hfil&\hfil#\hfil&
\hfil#\hfil&\hfil#&\hfil#\cr
\noalign{\medskip} 
\noalign{\hrule}
\noalign{\smallskip} 
&&&&\multispan3\hfil Point \hfil&\cr
Name&$M_{B{\rm(AB)}}$(Point)&$M_{B{\rm(AB)}}$(Bulge)&$M_{B{\rm(AB)}}$(Disk)&
$\log(L_{\rm 2keV})$&
$\hfil \log(L_{\rm 5GHz})$\hfil&
$\log(L_{\rm 2500A})$&$\log[R_{\rm e} {\rm (kpc)}]$&$\log[h{\rm (kpc)}$]\cr
\noalign{\smallskip} \noalign{\hrule}  \noalign{\medskip}
MS0007.1$-$0231&$-$19.10&$-$19.58&$-$21.52& 25.62& 29.64& 28.20&  0.17&  0.62\cr
MS0039.0$-$0145&  --- &$-$20.73&$-$20.15& 25.21& 29.64&  --- &  0.27&  0.92\cr
MS0048.8$+$2907&$-$16.17&$-$20.20&$-$19.01& 25.66& 29.03& 27.03&  0.47&  0.35\cr
MS0111.9$-$0132&$-$20.36&$-$21.58&  --- & 25.62& 30.14& 28.71&  1.38&  --- \cr
MS0135.4$+$0256&  --- &$-$22.00&$-$21.62& 25.77& 29.86&  --- &  0.75&  1.17\cr
MS0144.2$-$0055&$-$19.88&$-$17.96&$-$21.77& 25.36& 29.36& 28.52& $-$0.13&  0.72\cr
MS0321.5$-$6657&  --- &$-$20.61&$-$20.11& 25.29&  --- &  --- &  0.28&  0.79\cr
MS0330.8$+$0606&$-$18.83&$-$13.01&$-$21.11& 25.62& 29.54& 28.09&  0.58&  0.89\cr
MS0340.3$+$0455&$-$17.60&$-$18.27&$-$21.08& 25.05& 29.53& 27.60&  0.03&  0.79\cr
MS0412.4$-$0802&$-$21.35&$-$20.15&  --- & 25.32& 29.06& 29.10&  0.53&  --- \cr
MS0444.9$-$1000&$-$17.90&$-$19.54&$-$20.66& 25.12& 29.39& 27.72&  0.17&  0.81\cr
MS0457.9$+$0141&$-$19.56&$-$22.02&  --- & 26.12& 29.78& 28.39&  1.17&  --- \cr
MS0516.6$-$4609&  --- &$-$20.94&$-$21.62& 25.08& 31.00&  --- &  0.86&  1.34\cr
MS0713.4$+$3700&$-$17.97&$-$22.64&  --- & 25.70& 29.78& 27.75&  0.87&  --- \cr
MS0719.9$+$7100&$-$21.35&$-$21.37&  --- & 25.85& 29.75& 29.10&  0.63&  --- \cr
MS0721.2$+$6904&$-$21.33&$-$20.57&$-$18.94& 26.05& 29.65& 29.10&  0.60&  0.72\cr
MS0731.6$+$8011&$-$19.04&$-$21.65&  --- & 25.99& 29.48& 28.18&  0.75&  --- \cr
MS0754.6$+$3928&$-$24.10&$-$22.36&  --- & 25.99& 30.15& 30.20&  0.58&  --- \cr
MS0801.9$+$2129&$-$22.32&$-$20.75&$-$21.58& 25.85& 29.80& 29.49&  0.25&  0.70\cr
MS0803.3$+$7557&$-$20.88&$-$21.90&$-$20.50& 25.85& 29.50& 28.92&  1.17&  0.51\cr
MS0818.8$+$5428&$-$19.42&$-$21.13&$-$17.44& 25.73& 29.60& 28.33&  0.15&  0.56\cr
MS0841.7$+$1628&$-$20.77&$-$20.69&  --- & 25.87& 29.92& 28.87&  0.65&  --- \cr
MS0842.7$-$0720&$-$21.95&$-$20.13&$-$21.88& 25.88& 29.88& 29.35&  0.73&  1.00\cr
MS0844.9$+$1836&$-$17.02&$-$20.99&$-$19.91& 25.60& 29.30& 27.37&  0.34&  0.70\cr
MS0849.5$+$0805&$-$20.97&$-$21.42&  --- & 26.35& 28.93& 28.95&  0.80&  --- \cr
MS0904.4$-$1505&$-$18.35&$-$20.20&$-$18.58& 25.24& 29.06& 27.90&  0.66&  0.24\cr
MS0905.6$-$0817&$-$19.16&$-$20.42&$-$20.93& 25.30& 29.13& 28.23&  0.48&  0.91\cr
MS0942.8$+$0950&$-$15.52&$-$16.66&$-$18.07& 24.60& 27.64& 26.77&  0.06& $-$0.11\cr
MS0944.1$+$1333&$-$22.95&$-$22.28&$-$21.37& 26.60& 30.34& 29.74& $-$0.12&  0.66\cr
MS1020.2$+$6850&$-$20.39&  --- &$-$20.47& 25.00& 29.21& 28.72&  --- &  0.72\cr
MS1058.8$+$1003&$-$14.60&$-$20.26&$-$13.51& 24.31& 29.30& 26.40&  0.19&  0.87\cr
MS1059.0$+$7302&$-$21.13&$-$21.22&$-$20.73& 25.79& 29.45& 29.01&  0.89&  0.68\cr
MS1108.3$+$3530&  --- &$-$20.76&$-$19.46& 25.06& 29.75&  --- &  0.23&  0.61\cr
MS1110.3$+$2210&$-$15.37&$-$19.08&$-$17.60& 24.42& 28.71& 26.71&  0.38& $-$0.22\cr
MS1114.4$+$1801&  --- &$-$21.03&$-$19.51& 24.70& 29.36&  --- &  0.37&  0.80\cr
MS1136.5$+$3413&$-$19.10&$-$19.64&$-$18.81& 25.45& 28.43& 28.20&  0.53&  0.56\cr
MS1138.1$+$0400&$-$19.04&$-$21.01&  --- & 25.20& 29.48& 28.18&  0.37&  --- \cr
MS1139.7$+$1040&$-$20.02&$-$18.33&$-$21.91& 26.21& 30.27& 28.57&  0.62&  1.03\cr
MS1143.5$-$0411&$-$21.33&$-$20.75&$-$21.54& 25.98& 29.86& 29.10&  0.91&  0.81\cr
MS1158.6$-$0323&$-$17.17&$-$18.69&$-$19.16& 25.16& 28.35& 27.43&  0.06&  0.15\cr
MS1200.1$-$0330&$-$17.82&$-$20.62&$-$13.70& 24.76& 29.61& 27.69&  0.41&  0.54\cr
MS1205.7$+$6427&$-$20.21&$-$20.11&$-$20.28& 25.58& 30.00& 28.65&  0.71&  0.33\cr
MS1214.3$+$3811&$-$16.89&$-$19.62&$-$20.89& 24.99& 29.18& 27.32& $-$0.39&  0.75\cr
MS1217.0$+$0700&$-$20.97&$-$18.98&$-$21.34& 25.51& 29.36& 28.95&  0.55&  0.45\cr
MS1219.6$+$7535&$-$21.92&$-$21.94&$-$23.01& 26.49& 29.43& 29.33&  0.62&  0.79\cr
MS1220.9$+$1601&$-$18.00&$-$20.93&$-$20.77& 25.41& 29.47& 27.77&  0.69&  0.39\cr
MS1232.4$+$1550&$-$18.99&$-$17.67&$-$22.96& 25.33& 28.87& 28.16& $-$0.05&  1.29\cr
MS1233.3$+$7426&  --- &$-$19.74&$-$21.65& 25.07& 29.40&  --- & $-$0.28&  0.83\cr
MS1239.2$+$3219&$-$18.68&$-$17.48&$-$19.58& 24.87& 29.00& 28.03&  0.12&  0.44\cr
MS1242.2$+$1632&  --- &$-$19.37&$-$21.43& 25.85& 29.71&  --- & $-$0.26&  0.84\cr
MS1306.1$-$0115&$-$19.79&$-$20.34&  --- & 25.48& 29.70& 28.48&  0.41&  --- \cr
MS1322.3$+$2925&$-$16.72&$-$20.08&  --- & 25.09& 29.27& 27.25&  0.20&  --- \cr
MS1327.4$+$3209&$-$20.18&$-$21.13&$-$21.01& 25.53& 30.30& 28.64&  0.66&  1.04\cr
MS1333.9$+$5500&$-$21.08&$-$19.07&$-$21.11& 25.73& 29.71& 29.00&  0.78&  0.75\cr
MS1334.6$+$0351&$-$19.24&$-$21.84&  --- & 25.50& 29.77& 28.26&  1.00&  --- \cr
MS1335.1$-$3128&$-$18.46&$-$20.16&  --- & 25.07& 29.32& 27.95&  0.53&  --- \cr
MS1351.6$+$4005&$-$17.07&$-$14.87&$-$20.51& 25.25& 29.23& 27.39&  0.68&  0.43\cr
MS1403.5$+$5439&$-$18.08&$-$20.67&  --- & 25.26& 29.38& 27.80&  0.58&  --- \cr
\noalign{\medskip}
\noalign{\hrule}
\noalign{\medskip}
}}
\table{6}{D}
{\noindent\bf Table 3 contd.\ \rm Derived parameters for AGN/Host Galaxies}
{\tabskip=1em plus 2em minus .7em 
\halign to\hsize{
\hfil#\hfil&\hfil#\hfil&\hfil#\hfil&\hfil#\hfil&\hfil#\hfil&\hfil#\hfil&
\hfil#\hfil&\hfil#&\hfil#\cr
\noalign{\medskip} 
\noalign{\hrule}
\noalign{\smallskip} 
&&&&\multispan3\hfil Point \hfil&\cr
Name&$M_{B{\rm(AB)}}$(Point)&$M_{B{\rm(AB)}}$(Bulge)&$M_{B{\rm(AB)}}$(Disk)&
$\log(L_{\rm 2keV})$&
$\hfil \log(L_{\rm 5GHz})$\hfil&
$\log(L_{\rm 2500A})$&$\log[R_{\rm e} {\rm (kpc)}]$&$\log[h{\rm (kpc)}$]\cr
\noalign{\smallskip} \noalign{\hrule}  \noalign{\medskip}
MS1408.1$+$2617&$-$19.39&$-$19.89&  --- & 25.19& 29.27& 28.32&  0.49&  --- \cr
MS1414.0$+$0130&$-$19.60&$-$20.64&  --- & 25.91& 29.97& 28.40&  0.53&  --- \cr
MS1414.9$+$1337&  --- &$-$19.49&$-$20.12& 25.05& 29.44&  --- &  0.23&  0.78\cr
MS1416.3$-$1257&$-$22.42&$-$20.89&$-$20.16& 26.76& 29.78& 29.53&  0.39&  1.00\cr
MS1420.1$+$2956&$-$19.05&$-$18.19&$-$21.05& 25.06& 29.05& 28.18& $-$0.26&  0.68\cr
MS1426.5$+$0130&$-$23.41&$-$22.18&  --- & 26.36& 29.67& 29.93&  0.82&  --- \cr
MS1455.7$+$2121&$-$19.62&$-$21.25&  --- & 25.64& 29.41& 28.41&  0.76&  --- \cr
MS1456.4$+$2147&$-$20.71&$-$20.33&  --- & 25.44& 29.23& 28.85&  0.50&  --- \cr
MS1519.8$-$0633&$-$21.47&$-$21.51&$-$20.72& 25.95& 30.24& 29.15&  1.09&  0.62\cr
MS1545.3$+$0305&$-$20.45&$-$21.06&$-$20.73& 25.58& 29.54& 28.74&  0.75&  0.69\cr
MS1846.5$-$7857&$-$17.61&$-$20.55&$-$21.22& 24.79&  --- & 27.61&  0.69&  1.22\cr
MS2039.5$-$0107&$-$20.43&$-$21.70&  --- & 25.63& 29.81& 28.74&  0.85&  --- \cr
MS2128.3$+$0349&$-$21.78&$-$21.43&  --- & 26.09& 29.50& 29.28&  0.85&  --- \cr
MS2144.9$-$2012&$-$19.61&$-$21.38&$-$20.79& 25.54& 29.62& 28.41&  0.50&  0.73\cr
MS2159.5$-$5713&$-$20.34&$-$21.71&$-$18.54& 25.66& 31.48& 28.70&  0.70&  0.68\cr
MS2210.2$+$1827&$-$20.42&$-$20.73&$-$21.23& 25.75& 29.35& 28.73& $-$0.25&  0.66\cr
MS2348.3$+$3250&$-$19.33&$-$20.03&  --- & 25.54& 29.56& 28.29&  0.34&  --- \cr
MS2348.6$+$1956&  --- &$-$18.77&$-$20.58& 24.82& 29.25&  --- & $-$0.57&  0.45\cr
\noalign{\medskip}
\noalign{\hrule}
\noalign{\medskip}
}}

%
%

The absolute $M_{B{\rm (AB)}}$ magnitudes for all three components
derived in this manner are given in Table 3, along with the physical
sizes for the disk and bulge components. For completeness, the
monochromatic X-ray, radio and optical fluxes (at 2keV, 5GHz and
2500\AA\ respectively) of the point source component are also given. To
derive the radio and X-ray luminosities we have assumed spectral
indices of $\alpha_{\rm R}=0.5$ and $\alpha_{\rm X}=1$ in the radio
and X-ray regimes. We have also assumed that all the radio and X-ray
flux comes from the central component.

\figure{11}{S}{0mm}{
\psfig{figure=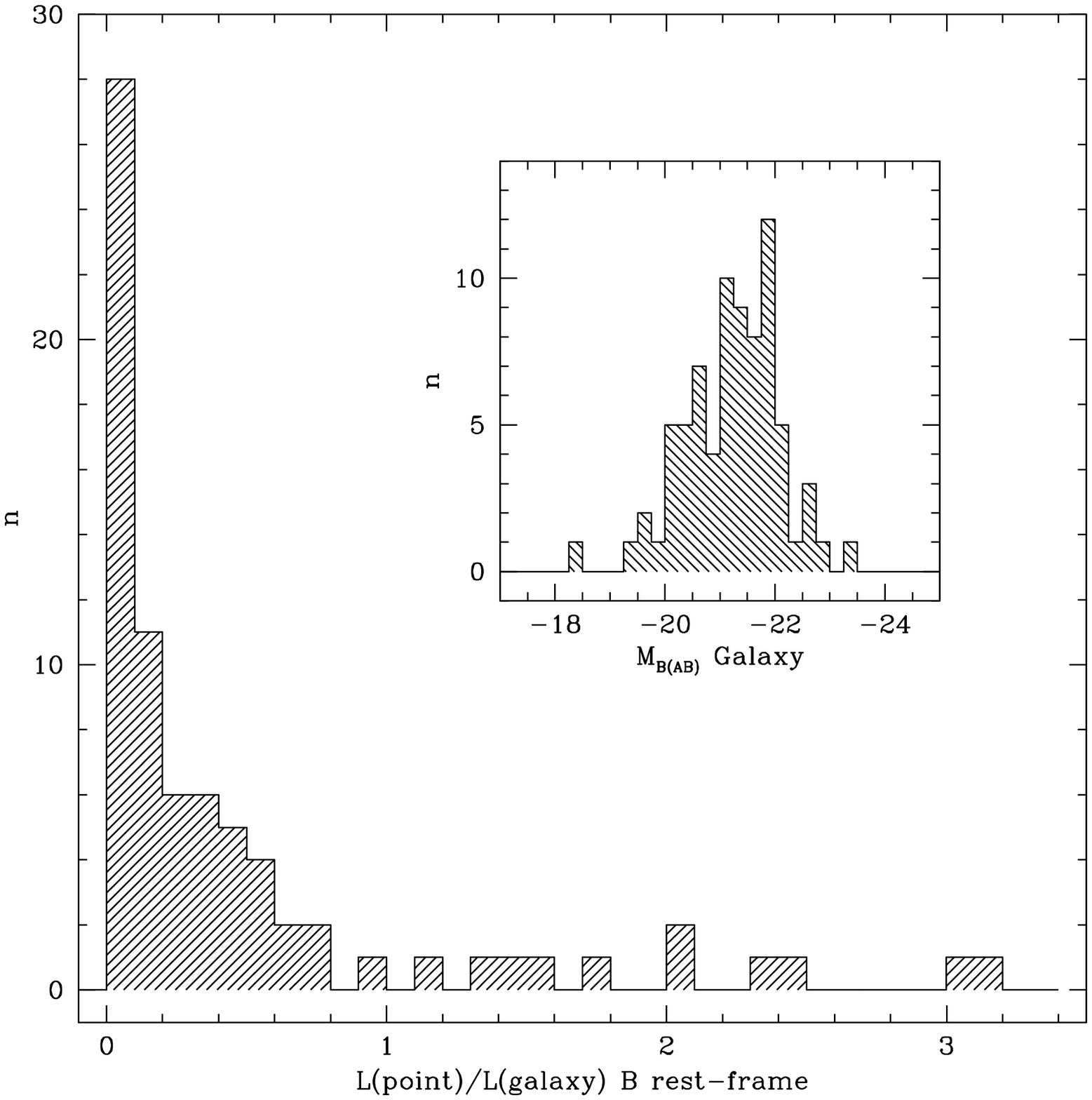,width=3.2in}\break
\noindent\bf Figure 8 \rm Histogram of nuclear-to-host luminosity ratio for 
the sample.  (Inset: Host galaxy luminosity histogram).}

\subsection{Host Galaxy Properties}\tx

\subsubsection{Luminosity}\tx

The histogram of host galaxy luminosities (corrected to the rest-frame
$B({\rm AB})$ pass-band) is plotted in Fig.\ 8. The luminosity range of
the host galaxies is $-23.1 < M_{B{\rm (AB)}} < -18.3$ with a median
value of $M_{B{\rm (AB)}} = -21.1$. The mean value for the $I$-band
nuclear-to-host luminosity ratio (also plotted in Fig.\ 8) 
is $L_N/L_G = 0.2$, lower than that
observed in previous samples of bright AGN. Over 75 per cent of our
sample exhibit $L_N/L_G < 0.5$. In contrast, McLure et al.\ (1999)
obtain a median $R$-band value $L_N/L_G = 1.5$ from their sample of
nine radio-quiet AGN. At the low $L_N/L_G$ measured in this sample,
any systematic effects introduced by the fitting procedure are, at
most, at the 2 -- 3 per cent level (see section 3.2). We conclude,
therefore the low values for $L_N/L_G$ found here are unlikely to be
artifact of the fitting procedure.

\figure{12}{D}{0mm}{
\psfig{figure=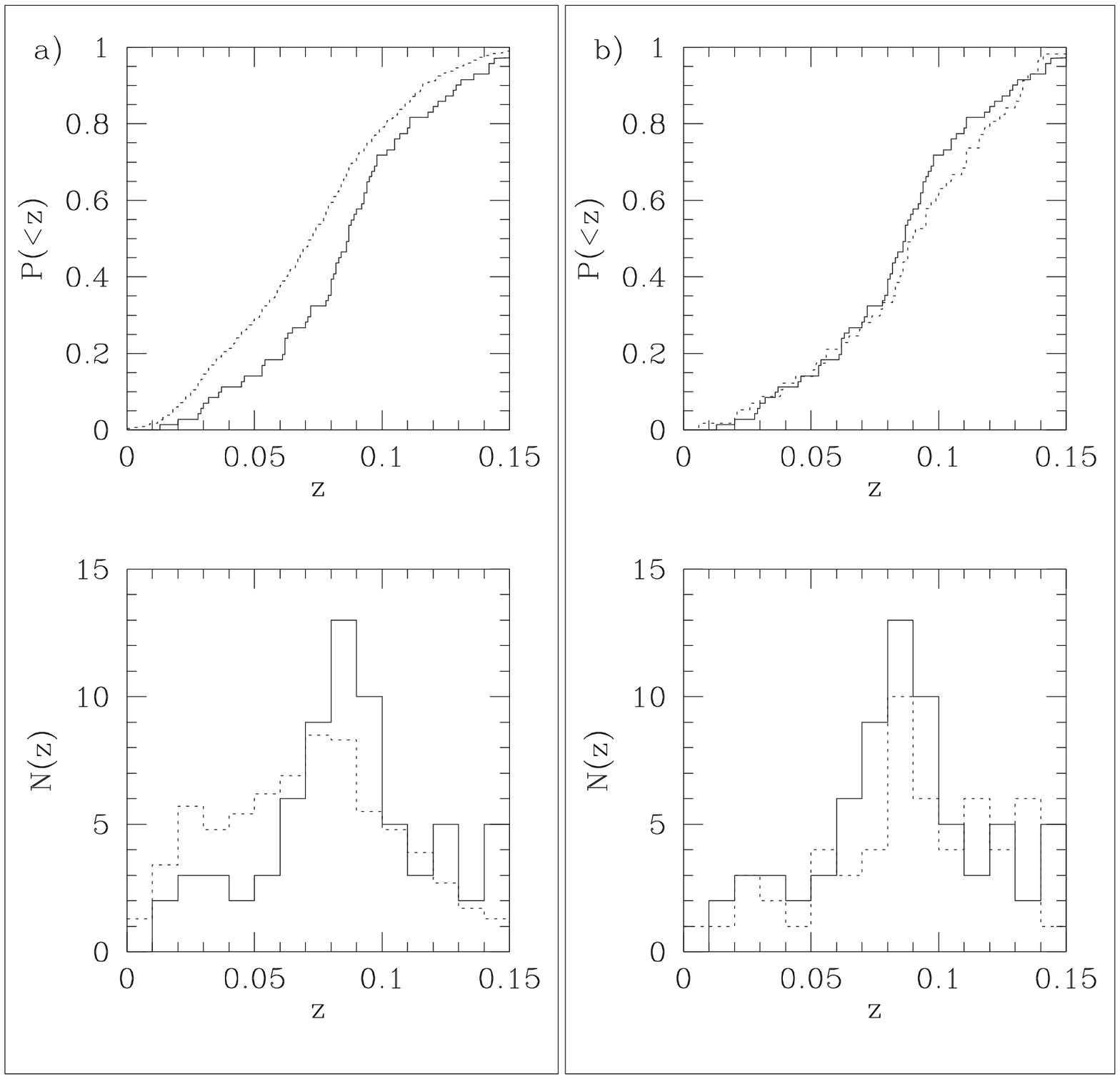,width=6.4in}\break
\noindent\bf Figure 9 \rm Comparison of redshift distributions
for the EMSS sample (solid line) with random-sample drawn from 
Autofib sample (dotted line) a) Autofib random-sample drawn to match
apparent magnitude distribution of EMSS sample b) random-sample 
drawn to match apparent magnitude and morphological distribution
of the EMSS sample.  Upper panel: cumulative redshift
distribution.  Lower panel: redshift histograms. }

To establish whether the properties of the AGN host galaxies were
representative of the field galaxy population, we tested the
luminosity distribution of the host galaxies in this sample against a
control sample of galaxies from the Autofib redshift survey (Ellis et
al.\ 1996). For each AGN host galaxy, ten galaxies
with the same apparent magnitude ($\pm 0.05\,$mag) were chosen at
random and with replacement from the Autofib sample. A small random offset ($-0.01 <
\delta z < 0.01$) was applied to each redshift in the Autofib sample to
minimise the effects of clustering in this sample. If the luminosity
distributions of the randomly-drawn Autofib sample and the AGN host
sample are identical then we would expect the redshift distributions
of the two samples to match one another. A Kolmogorov-Smirnoff (K-S)
test shows that the distributions are different at greater than the
99.9 per cent significance level (see Fig.\ 9a). The sense of the
difference is that the AGN hosts are displaced toward higher
redshifts, implying the hosts are more luminous than typical field
galaxies as represented by the Autofib sample. By applying increasingly
large magnitude offsets to the galaxies drawn at random from the
Autofib sample, we were able to establish that the AGN hosts were
brighter by $0.75 \pm 0.25\,$mag at the 95 per cent confidence level
than the Autofib galaxies.

\subsubsection{Morphology}\tx

\figure{13}{S}{0mm}{
\psfig{figure=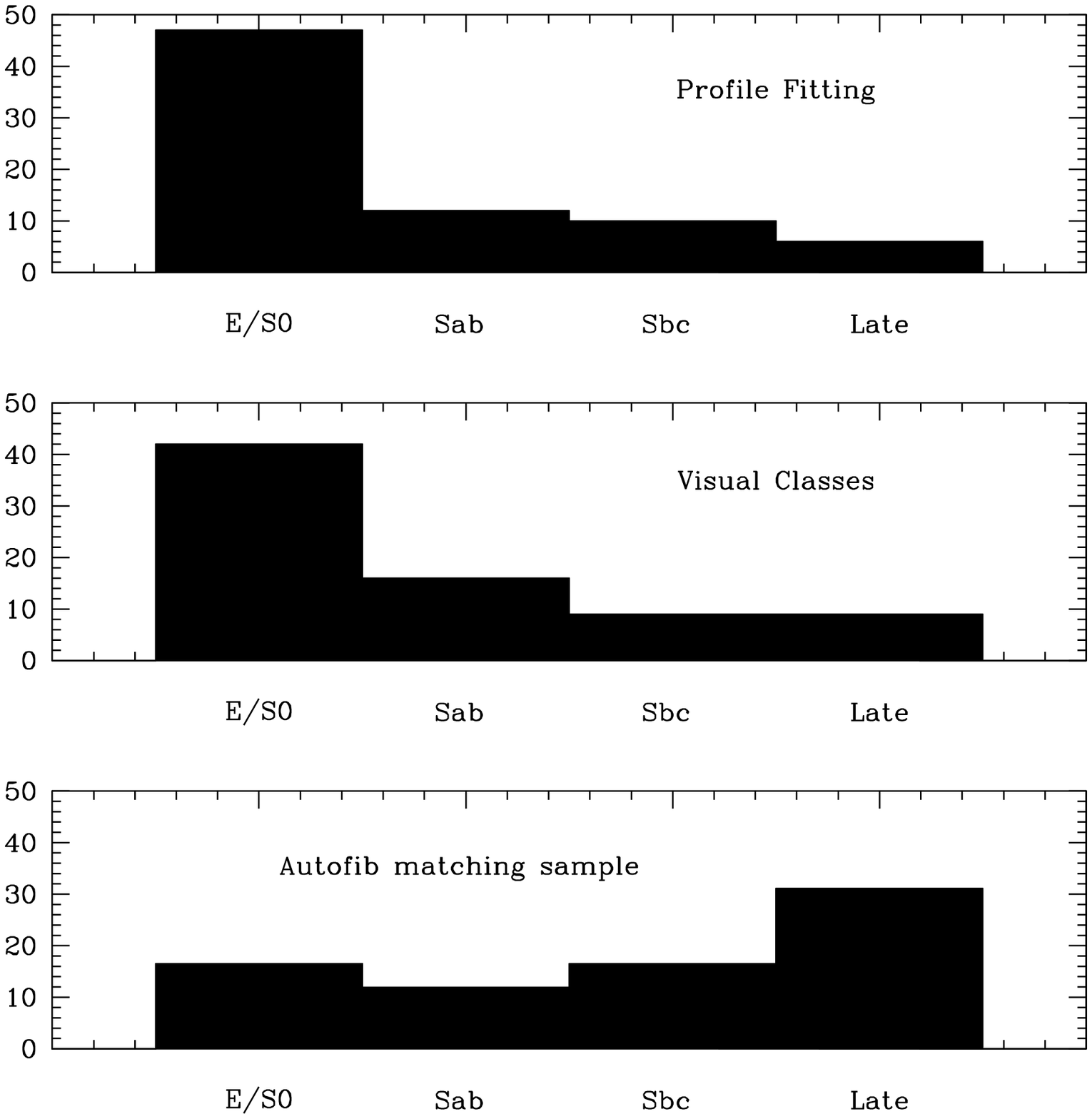,width=3.2in}\break
\noindent\bf Figure 10 \rm Distribution of host galaxy morphological types.
a) EMSS survey: fit parameters, b) EMSS survey: visual inspection
c) Autofib survey: randomly-drawn sample (see text).}

The morphological types can be characterised according to the output
parameters of the fitting procedure. In this scheme the fractional
bulge luminosity $B/T$ is the primary
classification parameter. We approximately follow Simien \& de
Vaucouleurs (1986) and define the E/S0 class with $0.5 \le B/T < 1.0$,
Sab; $0.3 \le B/T < 0.5$, Sbc; $0.1 \le B/T < 0.3$, and we define our
own `Late' class as $B/T < 0.1$. Fig.\ 10a) plots the histogram of the
rest-frame B-band values of $B/T$ computed using the median 
$(B-I)_{\rm AB}$ galaxy colors.

An alternative method is to visually classify the images roughly
according to the Hubble classification system. We used the Hubble
Atlas (Sandage 1961) as a reference. One difficulty with this approach
is that some of the objects in this sample are dominated by a nuclear
component so that estimating the contribution of the bulge is
problematical. The bulge and nuclear light are easily confused. This
problem was dealt with by subtracting the point source component from
the best-fit model and then re-evaluating the classifications. The
affect of this re-evaluation was negligible. However, nine galaxies
were very compact and/or dominated by the nuclear contribution so that
it was not possible to classify them with any degree of
confidence. All of these objects were classified as `Late' for the
purposes of the comparisons below. A comparison of the distributions
of the profile-fitting and visual classification (Figs 10a and b) shows
no significant difference.

To test whether these distributions are characteristic of the field
galaxy population at these magnitudes, we compared our visual
classifications against those from 10 random samples generated from
the Autofib survey using the method described above. The resulting
histogram of morphological types for the Autofib sample is shown in
Fig. 10c). We adopted our visual classifications for the purpose of
this comparison since these are likely to be derived in a similar way
to those of the Autofib survey. Using only the 4 rough classes defined
above, the comparison between the Autofib and AGN host galaxies
samples yielded a $\chi^2 = 69$ for 4 degrees of freedom . Clearly the
AGN host galaxies are drawn from a different parent population than
the general field population. AGN host galaxies in this sample tend to
be of earlier type than the field. Remarkably, 55 per cent of the AGN
host galaxies are E/S0 type.

This percentage is similar to the fraction of early-type
hosts identified amongst bright radio-quiet QSOs ($M_B < -23$) by both
Bahcall et al.\ (1997) and McLure et al.\ (1999). Bahcall et
al. (1997) identified 7 out of 14 of their radio-quiet AGN to have
bulge luminosity profiles, while McLure et al.\ (1999) found
elliptical galaxy fits were favoured over disk galaxy fits in seven
out of the nine radio-quiet AGN they studied. Because the disk host
galaxies were found preferentially around the low luminosity AGN in
their sample, McLure et al.\ (1999) postulated that early-type hosts
may be more prevalent amongst bright AGN. Within the statistical
errors, our analysis would suggest that this is not the case; the
frequency of early type hosts is almost as high amongst our fainter sample as
in the McLure et al.\ (1999) sample. This observation is also
internally consistent within our own sample.  Using the
Spearman rank test, we find no significant correlation between $B/T$ and 
point source luminosity (see Fig.\ 11).  A least squares fit to the data
points in Fig.\ 11 formally gives a slope of $0.0$. 
At the very brightest nuclear magnitudes, our statisitics are too poor to 
determine whether AGN inhabit {\em exclusively} bulge-dominant 
systems as suggested by McLure et al.  We have only two AGN with
$M_{B{\rm (AB)}}{\rm(nucleus)}<-23$ in our sample, both of which
have $B/T>0.5$.  

In contrast, although Malkan et al.\ (1998) report that Seyfert 1s
have earlier-type host galaxies than Seyfert 2s, the overall fraction
of Seyfert 1 galaxies in E/S0 hosts in their HST imaging survey is
much lower ($\sim 20$ per cent) than observed in this analysis. Thus
the high incidence of early-type hosts for radio quiet AGN may break
down at the very lowest AGN luminosities ($M_B > -20$).

\figure{14}{S}{0mm}{
\psfig{figure=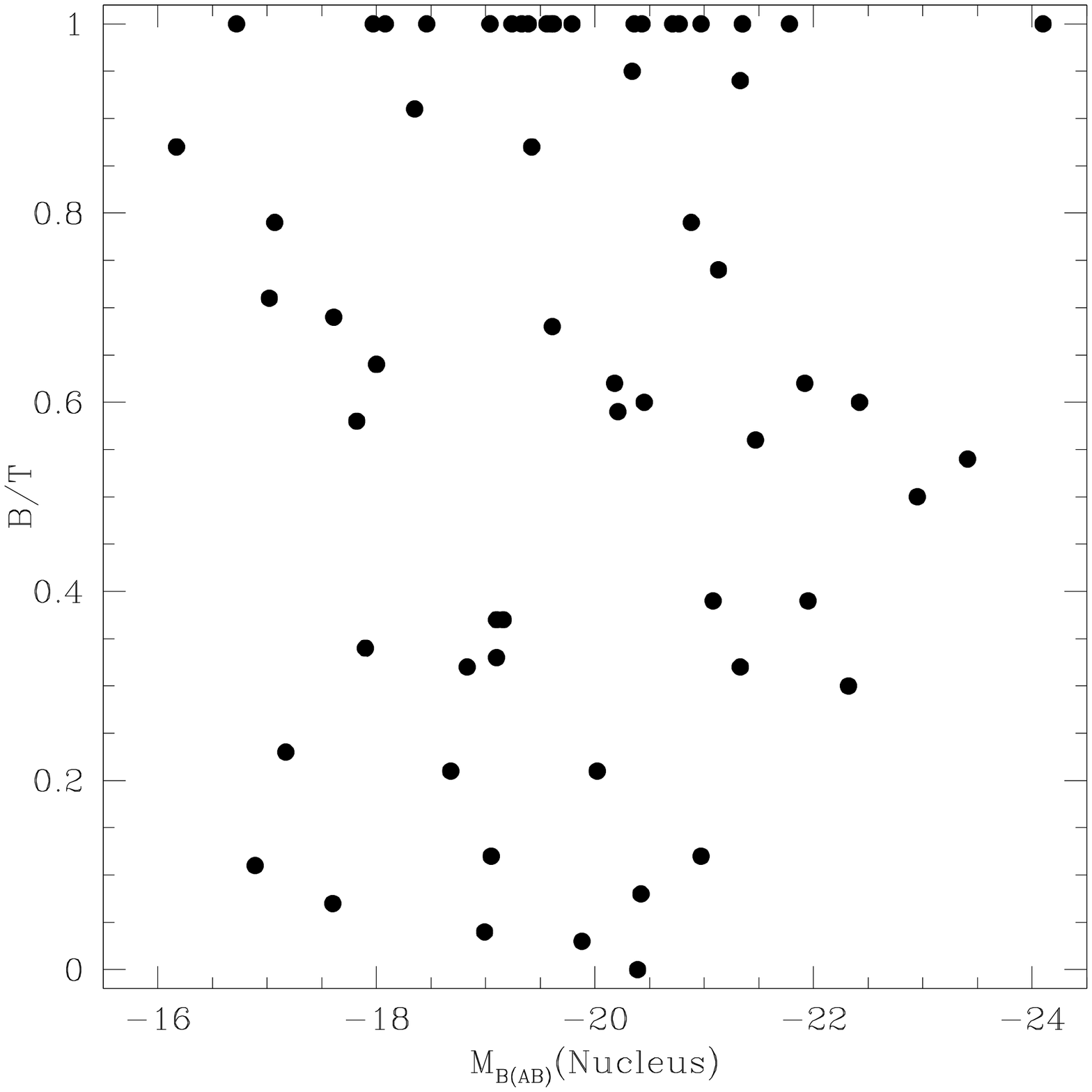,width=3.2in}\break
\noindent\bf Figure 11 \rm $I$ band bulge-to-total luminosity ratio ($B/T$) 
plotted as a function of nuclear $M_{B({\rm AB})}$ magnitude.}
 
The observation that the AGN host galaxies are biased towards earlier
types is also consistent with our observation that the absolute
magnitudes of the host galaxies are brighter than the field
population. Folkes et al.\ (1999) have recently derived the field
galaxy luminosity function for different spectral types in the 2dF
galaxy redshift survey. Based on almost 6000 galaxies, they obtain an
$M^*_{B{\rm (AB)}} = -21.2$ for early type galaxies, $0.7\,$mag
brighter that the $M^*_{B{\rm (AB)}}$ for late-type galaxies. This is
close to the median luminosity of the AGN host galaxies in this
sample. Furthermore, the difference between the $M^*_{B{\rm (AB)}}$
derived for early and late-type galaxies in the 2dF survey is close to
the observed luminosity difference between the AGN host galaxies and
the random field sample.

We performed a variant of the earlier test with the Autofib sample to
see whether the luminosity difference is consistent with the galaxies
being biased toward earlier spectral types. This time we selected galaxies
at
random from the Autofib sample with identical apparent
magnitudes ($\pm 0.05\,$mag) and spectral types. Since we were much
more restricted in our choice of galaxy from the Autofib sample we
were only able to do this test with the same number of objects in the
randomly-selected Autofib sample as in the EMSS sample (typically only
1--3 objects had the same apparent magnitude and morphology) as the
host galaxies in the EMSS sample. We computed the KS probability for
the two resultant redshift distributions being drawn from the same
sample (see Fig.\ 9b). In this case the KS probability was $P_{\rm KS}
= 0.75$, i.e. there is no evidence that the AGN host galaxies in this
sample have a different luminosity distribution when compared to the
same morphological type distribution in the field. The difference in
luminosity between the AGN host galaxies and the random field galaxy
population is therefore a natural consequence of the bias towards
earlier type galaxies in this population.

\subsubsection{Sizes}\tx

\figure{15}{D}{0mm}{
\psfig{figure=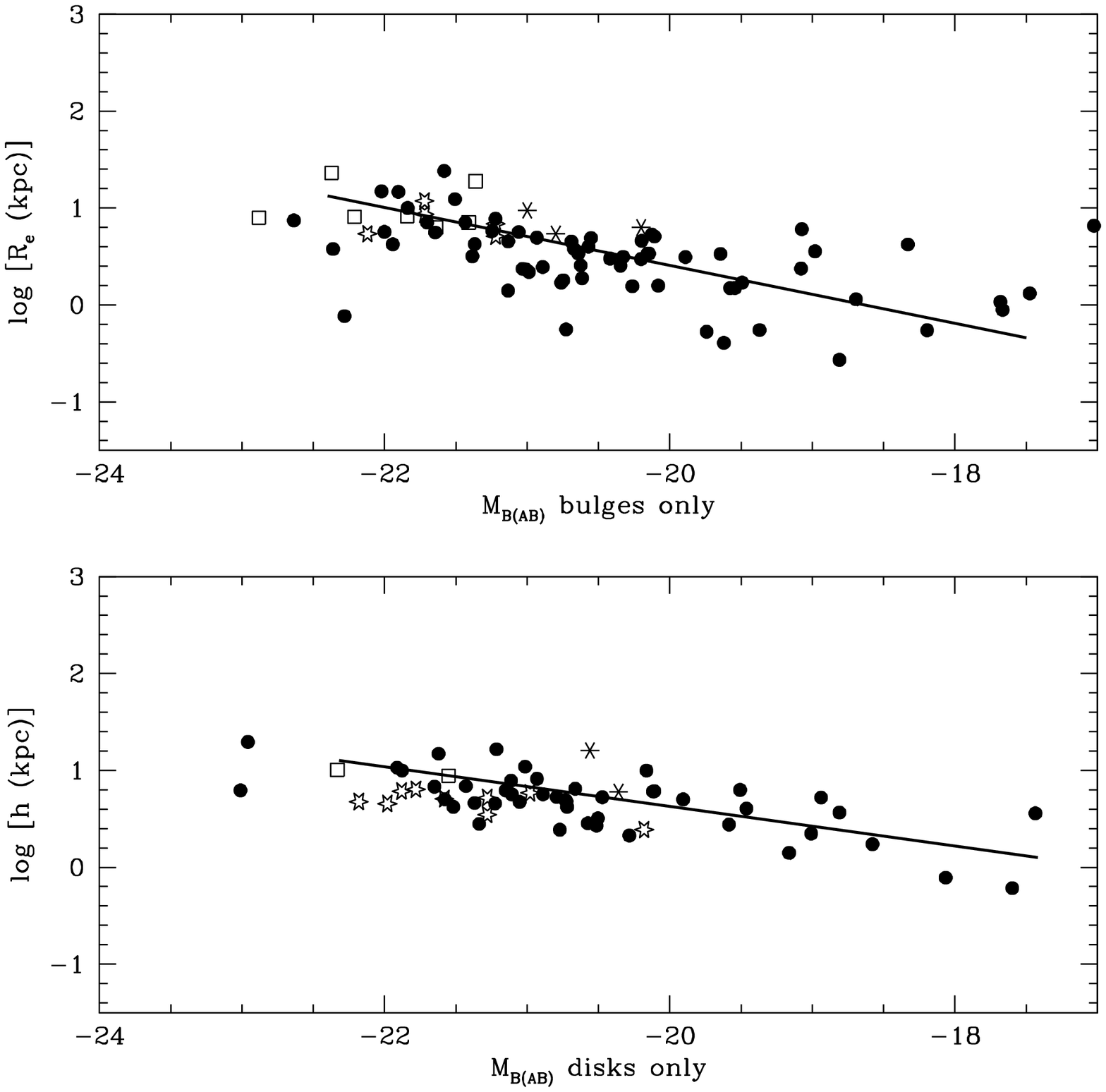,width=6.4in}\break
\noindent\bf Figure 12 \rm Scale-length {\it v.} absolute magnitude for a) 
bulges b) disks for the host galaxies of the AGN in the sample.  The
solid lines denote the observed relations for elliptical bulges
(Schade, Barrientos, \& Lopez-Cruz 1997)
and spiral disks (Freeman 1970).  Other symbols represent different
HST AGN imaging surveys. Open squares: McLure et al.\ (1999); stars:
Bahcall et al.\ (1997); asterisks: Boyce et al.\ (1998).}

In Fig.\ 12 we have plotted the fitted disk and bulge scale lengths
against galaxy luminosity. We have also plotted in this diagram the
observed size/luminosity relation for ellipticals from Schade,
Barrientos, \& Lopez-Cruz (1997) and spirals (Freeman 1970). The AGN
host galaxies follow these relations surprisingly well, the large
scatter caused, in part, by the errors on the parameters in the
fitting process.

We can straightforwardly compare the sizes of these host galaxies with
those identified with HST by other authors. McLure et al.\ (1999),
Boyce et al.\ (1997) and Bahcall et al.\ (1997) all give absolute
magnitudes and effective radii or scale heights for their favoured fit
(bulge or disk) to the AGN host galaxies. In the comparison, we have
only considered properties of the radio-quiet AGN observed by these
authors. To minimise possible discrepancies arising from different
fitting procedures, we used the results of the 2D-fitting process
employed by all authors. Bahcall et al.\ (1997) and Boyce et al.\ (1997)
both give host galaxy magnitudes in the V passband. To convert this
into the $B({\rm AB})$ band we used the following relations:

$$B({\rm AB}) = V + 0.78\qquad\qquad {\rm (bulge)}$$
$$B({\rm AB}) = V + 0.42\qquad\qquad {\rm (disk)}$$

For McLure et al.\ (1999), we adopted the following transformations
between their $R$ passband and the $B({\rm AB})$ band.

$$B({\rm AB}) = R + 1.41\qquad\qquad {\rm (bulge)}$$
$$B({\rm AB}) = R + 0.99\qquad\qquad {\rm (disk)}$$

Note that the absolute magnitudes derived by these authors correspond
to total galaxy luminosity which is fit by a single component
i.e. either bulge or disk but not both as in this analysis. Thus the
bulge or disk luminosities quoted by these authors will be
systematically higher than the similar luminosities derived in this
analysis where a bulge plus disk model is fit simultaneously. In
general, however, one or other of the components is likely to be
dominant (particularly true for bulges) and so the offset will be
small.

We have plotted the size-absolute magnitude distribution for these
host galaxies alongside those for the EMSS sample in Fig.\ 12. Although
the galaxies are clearly larger and more luminous on average than the
EMSS sample, they follow the identical relation to the EMSS AGN and
exhibit a large overlap in their properties. From this diagram we
conclude that AGN host galaxies exhibit a continuous range of
properties, broadly correlated with their nuclear luminosity. To
investigate this further, we now consider the detailed correlation
between host galaxy and nuclear luminosity.

\subsection{Host and Nuclear properties}\tx

We have plotted in Fig.\ 13a) the rest-frame $M_{B({\rm AB})}$ host galaxy
absolute magnitude as a function of point source  $M_{B({\rm AB})}$. Comparison
with Fig.\ 3 gives an indication of the errors associated with the
determination of the host galaxy and nuclear absolute magnitudes at
various points in this diagram. There is a weak but significant
correlation between the magnitude of the galaxy and the point source
AGN component. A Spearman rank test yields a positive correlation at
greater the $3\sigma$ level. The least squares fit (slope=0.21) to the
points with point source detections (filled circles) is also shown.

\figure{16}{D}{0mm}{
\psfig{figure=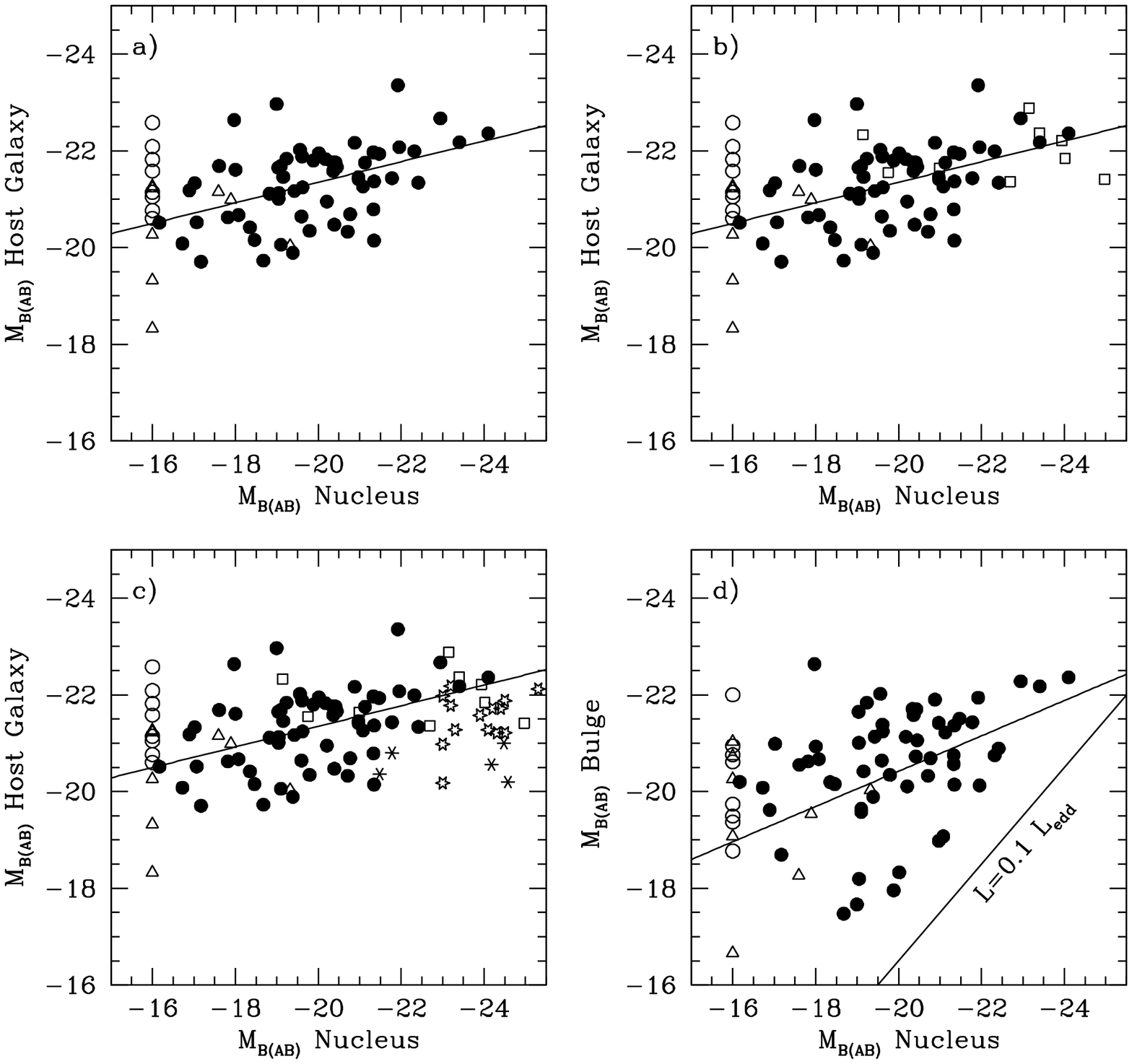,width=6.4in}\break
\noindent\bf Figure 13 \rm a) Nuclear luminosity v.\ host galaxy luminosity for
all objects imaged in the survey.  Open symbols represent those
objects with no detectable or weak ($M_{B({\rm AB})}>-16$) point
source components.  Triangles denote `ambiguous' AGN.  b) as figure a)
with data-points (open squares) from radio-quiet AGN in the McLure et
al.\ (1999) survey.  c) as b) with additional data points from Bahcall
et al.\ (1997)\ (stars) and Boyce et al.\ (1998) (asterisks). d)
Nuclear luminosity {\it v.} bulge luminosity for all objects imaged in
the survey.  Symbols as for a).  Also shown is the predicted relation
for AGN with $L=0.1L_{\rm Edd}$ (see text).}

In Fig.\ 13b)  we have included the data points from the radio-quiet
AGN observed by McLure et al.\ (1999). In this case the $R$ magnitude for
the nuclear component was transformed to the $B$ band by:

$$B({\rm AB}) = R + 0.55$$

Although at the high luminosity end of the distribution, the data
points of McLure et al.\ (1999) are consistent with the trend seen in
the EMSS sample.

We have also added the results of the Bahcall et al.\ (1997) and Boyce
et al.\ (1997) analysis in Fig 13c). These results are treated
separately because the nuclear components in these studies was
strongly saturated in the HST images, leading to some uncertainties in
the photometry of the point source component. In this case the
following relation was used to convert nuclear V band magnitudes into
the $B({\rm AB})$ band.

$$B({\rm AB}) = V + 0.19$$

Again, these points lie at the high luminosity end of the
distribution, although in this case all the points appear
systematically shifted toward lower host galaxy luminosities than the
trend apparent in this analysis or in the observations of McLure et
al.\ (1999). This could be caused by a strongly saturated nuclear image
making it difficult to detect all the galaxy light, or simply the fact
that any weak correlation between nuclear and host galaxy luminosity
breaks down at the highest luminosities.

However, similar weak correlations have also been found by a number of
other authors (e.g.\ Bahcall et al.\ 1997, McLeod et al.\ 1999). This
has been potentially ascribed to an underlying correlation between the
bulge mass ($M_{\rm bulge}$) and black hole mass ($M_{\rm BH}$) where
$M_{\rm BH} = 0.006M_{\rm bulge}$ based on the observations of
Magorrian et al.\ (1998). Translating this into a correlation between
bulge and nuclear absolute magnitudes, the approximate relation can be
obtained (see McLeod et al.\ 1999):

$$M_{B(AB)_{\rm AGN}} = M_{B(AB)_{\rm Bulge}} - 6.0 \hfil\break
-2.5[\log\epsilon + \log({\Upsilon_{B(AB)}\over{10M/L}}) + \log({f\over{0.006}}) 
-\log({{\rm BC}\over{10}})]$$

\noindent
where $\epsilon$ is the ratio to Eddington luminosity, $\Upsilon_{B(AB)}$ is
the mass-to-light ratio in the $B(AB)$ band, BC is the bolometric
correction from $B(AB)$ band luminosity to total luminosity for the AGN,
and $f$ is the fraction of the spheroid mass in the black hole. The
normalisation constants are the typical observed values for each of
these parameters.

Substituting these default values for $\Upsilon_{B(AB)}$, BC and $f$, the
correlation expected for a constant Eddington ratio (in this case 
$L = 0.1L_{\rm Edd}$) is shown in Fig.\ 13d), where we have now plotted bulge
luminosity against nuclear luminosity for the EMSS sample. Again we
find a correlation which is significant at the 99 per cent confidence
level (based on the Spearman rank test) with a least squares slope of
0.36. Based on the black hole model described above, the lower bound
of the correlation is consistent with an inferred Eddington ratio of $\sim$ 
5 per cent, with most AGN radiating significantly below this limit.

The correlation observed between the host galaxy and nuclear
components is, of course, very much flatter than that given by a
single Eddington ratio. Such a flat correlation could be explained by
appealing to the fact that lower luminosity AGN preferentially radiate
at lower Eddington ratios. This would naturally explain the inference
that brighter AGN appear to radiate at Eddington ratios up to 20 per
cent (McLure et al.\ 1999, McLeod et al.\ 1999), see also Fig.\ 13b) and
c).

Perhaps the greatest concern over any correlation is that it may
simply be an artifact of the detection limits of our analysis
procedure. For example, bright galaxies (in particular bright bulges)
might mask the existence of a weak point source. Equally bright point
source components might hide faint galaxies (in particular small
bulges). Thus the areas of Fig.\ 13 which might be selected against are
precisely those areas in which no data points are seen.

The simulations (section 3.2) that were done to estimate the errors
are indicative rather than comprehensive. Nevertheless, they show that
relatively faint galaxies can be detected in the presence of a strong
point source and that relatively weak point sources can be detected in
the presence of bright galaxies (see Fig.\ 3). These results suggest
that the correlation between host galaxy and nuclear luminosity is
unlikely to be due to selection effects in the fitting process.

\subsection{Interactions}\tx

In marked contrast to previous studies of bright AGN (Bahcall et
al.\ 1997, Boyce et al.\ 1997), few, if any, of the AGN in this study
show evidence for interaction or a strong excess of close
companions. The latter result is hardly surprising, since Smith et
al.\ (1995) have already demonstrated that the excess number of
galaxies around $z < 0.3$ AGN in the EMSS is consistent with clustering
strength of field galaxies. Similar results are also reported for
Seyfert 1 galaxies (Dultzin-Hacyan et al.\ 1999).

The frequency of mergers in this sample is harder to put on a
quantitative basis. Nevertheless the fit residuals (see Table 2) show
little evidence for significant postmerger/interaction activity
(e.g. disrupted morphologies, tidal tails etc.). Inner bars and weak
spiral structure are the most common residual features seen. As noted
by McLure et al.\ (1999) a definitive measure of the extent to which
AGN activity is accompanied by evidence for interactions awaits a
detailed study of the level of activity in otherwise `normal'
galaxies. A low incidence of tidal tails/multiple nuclei ($< 10$ per
cent) was also noted by Malkan et al.\ (1998) in their imaging study of
lower luminosity Seyfert galaxies.

It is certainly true that the limited depth of our 600-sec HST
exposures could lead us to miss low level residuals implying
post-merger activity. Nevertheless, the level of strong interactions
seen in this low-luminosity AGN sample ($< 5$ per cent) is much less
than has been seen in similar studies of brighter AGN. One possible
interpretation is that interactions do not play as strong a role in
fuelling lower luminosity AGN ($M_B > -23$). Another explanation might
be that lower luminosity AGN represent a more advanced stage of the
AGN evolutionary process, i.e.\ the AGN declines in luminosity with
time from the merger event which initially fuelled the AGN.

Unfortunately with the absence of a similarly detailed morphological
study of `normal' galaxies it is impossible to determine whether AGN
do, in fact, show any strong evidence for any enhanced
merger/interaction activity compared to the field galaxy
population. From this study, the indication is that there is little,
if any, evidence for such activity.

\section{Conclusions}\tx

We have carried out a systematic ground- and HST-based imaging study of
a large sample of nearby AGN. The X-ray selection of the initial
sample minimises any optical morphological bias. Although on average ten 
time fainter than
many previous samples of nearby AGN imaged with HST, the objects
studied here comprise the bulk of local AGN, with space densities up
to 100 times higher than their more luminous counterparts. As such
they are responsible for the vast majority of the AGN luminosity
density in the local Universe.

We find that the properties of the host galaxies of these AGN are much
more `normal' compared to those of more luminous AGN/QSOs. The host
galaxies follow the observed size-luminosity relations for bulges and
disks, with sizes typically $10h_{50}\,$kpc. The host galaxies span a wide
range in luminosity, with a median luminosity of $M_{B{\rm (AB)}} = -21.5$. 
All but one of the host galaxies are detected with $M_{B{\rm (AB)}} > -18$.

Compared to a random sample of field galaxies at these redshifts, the
host galaxies are biased towards early morphological types
(E, S0). This is consistent with the observation that the host galaxies
are also $0.75\pm0.25\,$mag brighter than field galaxies at $z < 0.15$. 
The median luminosity of the sample is also consistent with the
most recent estimates of $L^*$ for early spectral types.

There is a weak correlation between the host galaxy and nuclear
luminosity, the origin of which may be due to the underlying energy
generation mechanism. Assuming the standard black hole model for
energy generation in AGN and the derived relation between spheroid and
black hole mass, the AGN in this study typically radiate at or below a
few per cent of their Eddington luminosity.

There is no evidence for any enhanced merger activity/interactions in
this sample of objects. The host galaxies of these AGN thus appear to
represent a rather typical subset of `normal' galaxies in the local
Universe, albeit biased towards bulge-dominated objects.

When combined with HST imaging studies of brighter AGN, it is clear
that the properties of AGN host galaxies form a continuous
distribution, over all sizes and luminosities. The host galaxies of
AGN are not unusual with respect to the overall galaxy
population. Galaxies with luminosities $L^*$ and fainter are capable
of harbouring an AGN. Indeed the correlation which leads the brighter
AGN to be found in the large, more luminous galaxies also reveals that
the fainter AGN that comprise the bulk of the population in the local
Universe will be found in normal galaxies. The underlying parameter
driving this correlation may be bulge mass and/or energy generation
efficiency.

The HST continues to provide a wealth of information on AGN host
galaxies at low redshifts. However, the vast majority of low redshift
AGN imaged to date are only of moderate luminosity ($-24 < M_B <
-18$).  Even the most luminous of low redshift AGN ($M_B \sim -25$) are
still significantly fainter than the typical `break'-luminosity QSOs
($M_B = -26$) at $z \sim 2$, where QSO activity reaches its peak. One
of the next major observational steps will therefore be the extension
of similarly comprehensive AGN imaging studies to high redshift. It is
only by considering unbiased samples over as wide a luminosity as
possible that we can hope to disentangle the relationship between the
large scale (the host galaxy and its environment) and the small scale
(the nucleus and energy generation mechanism) phenomena in AGNs. It is
to be hoped that the combination of large aperture and outstading
image quality provided by new groud-based telescopes such as Gemini
will yield major advances in this field in the near future.

\section*{Acknowledgements}\tx

\noindent   
The observations were obtained with the Jacobus Kapteyn Telescope at
the Observatorio del Roque de los Muchacos operated by the Royal
Greenwich Observatory, the 40-inch telescope at Siding Spring operated by
the Research School of Astronomy and Astrophysics, Australian National
University and with the Hubble Space Telescope operated by
STScI. BJB acknowledges the hospitality of Dominion Astrophysical
Observatory. We are indebted to Matthew Colless for supplying the
Autofib survey galaxy catalogue in digital format. We thank Nicholas
Ross and Danielle Frenette for their work on the morphological
classifications.

\section*{References}

\bibitem 
Adams T.F., 1977, ApJS, 33, 19 

\bibitem
Avni Y., Tanenbaum H., 1986, ApJ, 305, 83 

\bibitem
Bahcall J.N., Kirakos, S., Saxe D.H., Schneider D.P., 1997, ApJ, 479, 658 
\bibitem
Boyce P.J. et al., 1997, MNRAS, 298, 121 
\bibitem
Boyle B.J., McMahon R.G., Wilkes B.J., Elvis M. 1995, MNRAS, 276, 315 
\bibitem
Boyle B.J., Shanks T., Peterson B.A., 1988, MNRAS, 235, 935 
\bibitem
Dultzin-Hacyan D., Krongold Y., Fuentes-Guridi I., Marziani 1999, ApJ., 513, 111 
\bibitem
Dunlop J.S., Taylor G.L., Hughes D.H., Robson E.I., 1993, MNRAS 264, 455 
\bibitem
Ellis R.S., Colless M., Broadhurst T., Heyl J., Glazebrook K., 1996, MNRAS, 280, 235 
\bibitem
Folkes S. et al., 1999, MNRAS, 308, 459 
\bibitem
Freeman K. 1970, ApJ, 160, 811 
\bibitem
Kotilainen J., Ward M.J., 1994, MNRAS, 266, 953 
\bibitem
Green R., Schmidt M., Liebert J., 1986, ApJS, 61, 305 
\bibitem
Landolt 1992, AJ, 104, 320 
\bibitem
Maccacaro T., Wolter A., McLean B., Gioia I., Stocke J.T., Della Ceca R., Burg R., Faccini R., 
1994, Ap. Lett. \& Comm., 29, 267 
\bibitem
Magorrian J.\ et al., 1998, AJ, 115, 2285 
\bibitem
MacKenty J.W., 1990, ApJS, 72, 231 
\bibitem
Malkan M.A., Margon B., Chanan G.A., 1984, ApJ, 280, 66 
\bibitem
Malkan M.A., Gorjian V., Tam R., 1998, ApJS, 117, 25 
\bibitem
McLeod K.K., Reike G.H., 1994, ApJ, 431, 137 
\bibitem
McLeod K.K., Reike G.H., Storrie-Lombardi L.J., 1999, ApJ, 511, 67 
\bibitem
McLure R.J., Dunlop J.S., Kukula M.J., Baum S.A., O'Dea C.P., Hughes D.H., 1999, MNRAS, 308, 377
\bibitem
Peacock J.A., Miller L., Mead A.R.G., 1986, MNRAS, 218, 265 
\bibitem
Piccinotti et al.\ 1982, ApJ, 253, 485 
\bibitem
Sandage A., 1961, The Hubble Atlas of Galaxies, (Carnegie Institution: Washington) 
\bibitem
Schade D.J., Barrientos L., Lopez-Cruz O. 1997, ApJ, 477, 17 
\bibitem
Schade D.J., Lilly S.J., Le F\`evre O., Hammer F., Crampton D. 1996, ApJ, 464, 79 
\bibitem
Simien, de Vaucouleurs G. 1986, ApJ, 302, 564 
\bibitem
Simkin S.M., Su H.J., Schwarz M.P., 1980, ApJ, 237, 404 
\bibitem
Smith E.P., Heckman T.M., Bothun G.D., Romanshin W., Balick B. 1986, ApJ, 306, 64 
\bibitem
Smith E.P., Heckman T.M., 1990, ApJ, 348, 38 
\bibitem
Smith R.J., Boyle B.J., Maddox S.J. 1995, MNRAS, 277, 270 
\bibitem
Stetson, P., 1987, PASP, 99, 191 
\bibitem
Stocke J.T. et al., 1991, ApJS, 76, 813 
\bibitem
Taylor G.L., Dunlop J.S., Hughes D.H., Robson E.I., 1996, MNRAS, 283, 930 
\bibitem
Terlevich R., Tenorio-Tagle G., Franco J., Melnick J., 1992, MNRAS, 255, 713 
\bibitem
V\`eron-Cetty M.P, Woltjer L., 1990, A\&A., 236, 69 
\bibitem
V\`eron-Cetty M.P., V\`eron P., 1997, A Catalogue of Active Galactic Nuclei, 7th Edition 
\bibitem
Zitelli V., Granato G.L., Mandolesi N., Wade R., Danese L., 1993, ApJS, 84, 185
\bye